\documentclass[twocolumn]{aastex701}

\usepackage{makecell}
\usepackage{subfig}
\usepackage{siunitx}
\usepackage{caption}
\usepackage{subcaption}
\usepackage{xcolor}

\begin{document}

\title{A Systematic Study of Type Ia Supernova Remnants: Using Nucleosynthesis to Probe their Supernova Progenitors}

\author[orcid=0009-0005-5637-6538,gname='Cole',sname='Treyturik']{Cole Treyturik}
\affiliation{Department of Physics and Astronomy, University of Manitoba, Winnipeg, MB R3T 2N2, Canada}
\email[show]{treyturc@myumanitoba.ca}

\author[orcid=0000-0001-6189-7665]{Samar Safi-Harb}
\affiliation{Department of Physics and Astronomy, University of Manitoba, Winnipeg, MB R3T 2N2, Canada}
\email[show]{\\samar.safi-harb@umanitoba.ca}

\author[0000-0002-4231-8717]{Gilles Ferrand}
\affiliation{Department of Physics and Astronomy, University of Manitoba, Winnipeg, MB R3T 2N2, Canada}
\affiliation{RIKEN Center for Interdisciplinary Theoretical and Mathematical Sciences (iTHEMS), Wak\={o}, Saitama 351-0198 Japan}
\email[show]{\\gilles.ferrand@umanitoba.ca}

\begin{abstract}
We present the first systematic, spatially resolved X-ray spectroscopic study of a largely thermonuclear (Type Ia) sample of supernova remnants (SNRs), aimed at probing the explosion properties and progenitors through a comparison to a suite of nucleosynthesis models available in the literature. Our sample focuses on Galactic and LMC ejecta-dominated SNRs believed to be, or otherwise assumed to be, of thermonuclear origin. Using archival \textit{XMM–Newton} observations (and \textit{Chandra} for G1.9+0.3), we extract spectra from adaptively binned regions across each remnant and model the emission to constrain the plasma temperature, ionization timescale, and ejecta abundances. We then compare abundance ratios (relative to Si) to a library of 335 individual models spanning 11 commonly-used supernova nucleosynthesis simulation sets from the literature including seven thermonuclear and four core-collapse sets. Across the sample, we find that individual remnants can be well matched by subsets of models, but no single model reproduces all measured elemental ratios at once. As a result, the best fit model for a given object is typically set by a selection of well-fitted abundance ratios, highlighting both the strength and limitations in yield-based model determination. For some SNRs, the abundance comparisons show better agreement with particular families of Type Ia SN explosions, including near-Chandrasekhar-mass delayed detonations, sub-Chandrasekhar-mass explosions, and dynamically driven double detonations, although these interpretations are not unique. Finally, we outline the need for model improvements, including refined nuclear reaction rates, higher dimensional treatment of mixing and turbulence, expanded metallicity coverage, and the exploration of non-standard supernova explosion energies. 
\end{abstract}


\section{Introduction}
\label{sec:introduction}
Supernova (SN) explosions play a key role in the chemical enrichment of galaxies, and in the production of the heavy elements. The remnants of these explosions -- expanding clouds of hot gas known as supernova remnants (SNRs) -- offer us a window into the past, allowing us to probe the products of explosive nucleosynthesis even without having borne witness to the supernova event itself.

While supernovae themselves come in two broad classifications -- known typically as core-collapse (CC) and thermonuclear (Type Ia) -- the classification of SNRs is seldom so simple. Type Ia SNRs are commonly believed to result from thermonuclear explosions of carbon-oxygen (CO) white dwarfs, encompassing multiple progenitor channels and explosion mechanisms (for a recent review, see \citealt{2025A&ARv..33....1R} and references therein). A key distinction is whether ignition occurs near the Chandrasekhar mass (near-$M_{Ch}$) or at sub-Chandrasekhar mass (sub-$M_{Ch}$). Within near-$M_{Ch}$ channels, delayed detonations, i.e. deflagrations that transition to detonations, span a broad range of outcomes depending on ignition conditions, while pure deflagrations may produce increased mixing and distinct yield patterns. Another key distinction is the binary channel that brings the system to ignition: single-degenerate (SD) accretion from a non-degenerate companion versus double-degenerate (DD) white-dwarf binaries. Within the sub-$M_{Ch}$ systems, double detonation models, where a helium-shell detonation triggers a secondary detonation in the CO core, have emerged as a leading mechanism. Furthermore, DD systems involve additional pathways, including violent mergers and helium-triggered detonations, with the dynamically-driven double-detonation models providing sub-$M_{Ch}$ channels with distinct abundance diagnostics (see \S4). 

Core-collapse supernovae, on the other hand, result from the gravitational collapse of massive ($\gtrsim 8$ M$_{\odot}$) stars that have exhausted their nuclear fuel and thus reached the end of their lives. A broad range of outcomes arise based largely on diversity in the progenitors, with aspects such as mass, metallicity, and evolutionary history determining the extent of mass loss and the internal configuration of the star at the time of collapse. Similarly, the explosion mechanism itself can lead to different outcomes, with scenarios such as a spherical explosion leading to results that can notably differ from, for example, a bi-polar explosion (see e.g., \citealt{2012ARNPS..62..407J, 2009ARA&A..47...63S, 2008arXiv0811.4648F} for reviews). The differences in these scenarios -- as well as those that occur in Type Ia supernovae -- are often significant, but without direct observation of the supernova itself, it can be difficult to determine exactly which scenario was responsible for a given event.

SNRs, which glow in X-rays for thousands of years after the explosion, provide a nearby laboratory to probe this diversity of SN explosion mechanisms and progenitors. By revealing shock-heated, often spatially resolved ejecta whose X-ray spectra yield elemental ratios directly linked to the explosion and progenitor properties, SNRs help directly probe the parent supernova. This motivates a homogeneous, spatially resolved analysis across a sample of (believed or possible) Type Ia SNRs, coupled to uniform comparisons with a broad suite of published nucleosynthesis models (thermonuclear and core-collapse) to identify which abundance ratios provide the strongest constraints and where the models succeed or fail.

The goals of this work are therefore to: (1) perform a systematic spatially-resolved spectroscopic study on a selection of Type Ia SNRs in order to determine their plasma temperatures, ionization timescales, and chemical abundances; (2) compare these properties to a suite of nucleosynthesis yield models, thus inferring properties of both the supernova explosion as well as the progenitor system; (3) to draw any further conclusions regarding these objects and overall the tested models of supernova nucleosynthesis, so as to provide feedback to the nucleosynthesis modeling community.

This paper is organized as follows: in Section \ref{sec:targets}, we introduce the criteria by which our targets were selected, and present a brief summary of each of the chosen targets. In Section \ref{sec:data}, we summarize the observations, the data preparation, and the method of region generation. In Section \ref{sec:nucsyn_models}, we present information about the suite of nucleosynthesis models used in this work. In Section \ref{sec:results}, we describe the result of both our spectroscopic study and our derived findings, presenting these results on an object-by-object basis, while in Section \ref{sec:discussion}, we discuss some of the broader findings that these results may imply. Finally, in Section \ref{sec:conclusion}, we summarize our findings, and include some possible avenues of future work.

\section{Targets}
\label{sec:targets}
\begin{figure*}
    \gridline{
        \fig{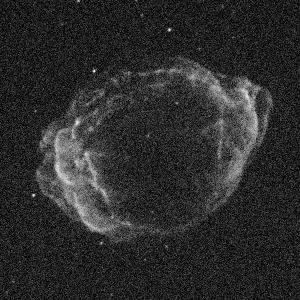}{0.24\textwidth}{(a) G1.9+0.3}
        \fig{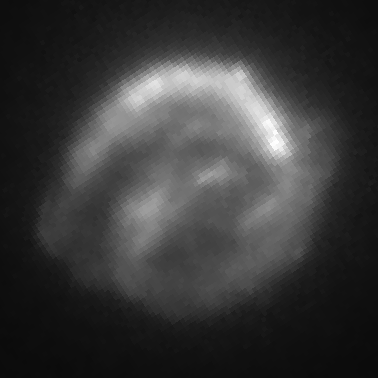}{0.24\textwidth}{(b) G4.5+6.8}
        \fig{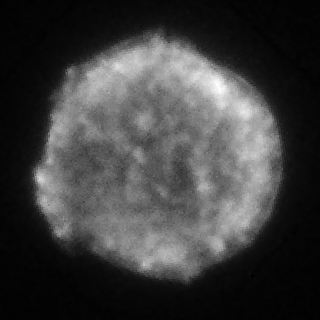}{0.24\textwidth}{(c) G120.1+1.4}
        \fig{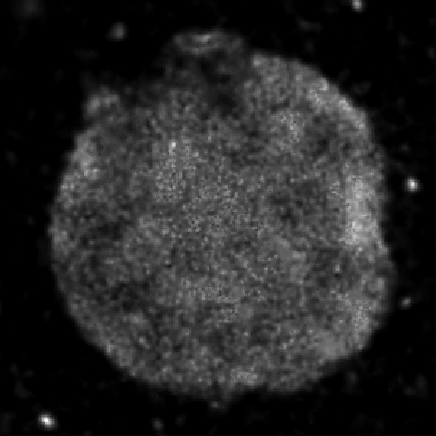}{0.24\textwidth}{(d) G272.2-3.2}
    }
    \gridline{
        \fig{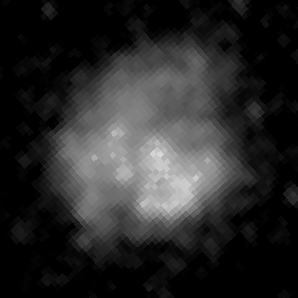}{0.24\textwidth}{(e) G337.2--0.7}
        \fig{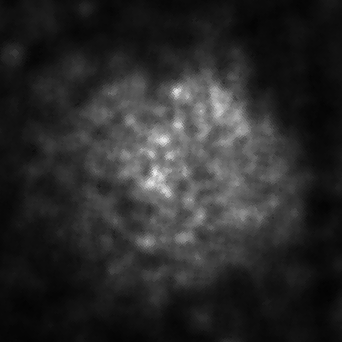}{0.24\textwidth}{(f) G344.7--0.1}
        \fig{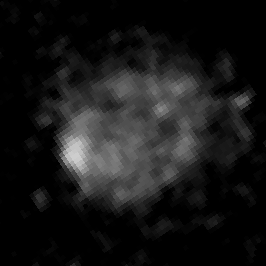}{0.24\textwidth}{(g) G352.7--0.1}
        \fig{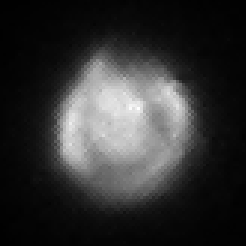}{0.24\textwidth}{(h) 0505--67.9}
    }
    \gridline{
        \fig{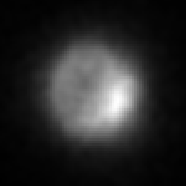}{0.24\textwidth}{(i) 0509--67.5}
        \fig{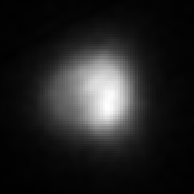}{0.24\textwidth}{(j) 0509--68.7}
        \fig{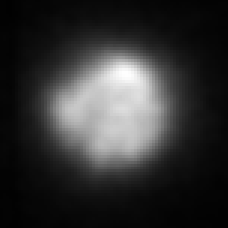}{0.24\textwidth}{(k) 0519--69.0}
        \fig{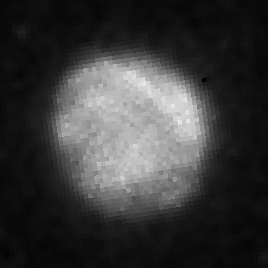}{0.24\textwidth}{(l) 0534--69.9}
    }
    \gridline{
        \leftfig{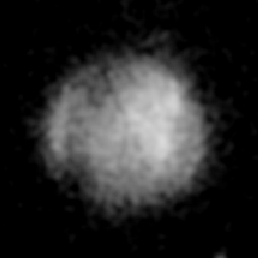}{0.24\textwidth}{(m) 0548--70.4}
    }
    \caption{Greyscale intensity images for all objects involved in this study, produced using the observations listed in Table~\ref{tab:info_table}. The first seven objects (those with a G-name) are the selected Galactic SNRs, while the remaining six are in the LMC.}
    \label{fig:intensity_maps}
\end{figure*}

In choosing targets for this work, we aimed to include SNRs that: (a) are believed to be, or otherwise have been assumed to be of a Type Ia origin; (b) could be considered young, in that they are less than $\sim 10$ kyr old and thus are young enough to still be in the ejecta-dominated phase so that they (c) show evidence of thermal X-ray emission attributed to shock-heated ejecta; and (d) have been observed and fully covered by at least one of the \textit{XMM-Newton} and \textit{Chandra} telescopes. To that end, 15 targets\footnote{Two further targets -- SN1006 and RCW86 -- were strongly considered, but were ultimately excluded from this study due to their weak thermal X-ray emission combined with their large size prohibiting a full coverage of the SNR.} were selected -- 9 Galactic SNRs and 6 from the Large Magellanic Cloud (LMC) -- 13 of which are detailed below. Intensity images for all objects covered in this text can be seen in Figure~\ref{fig:intensity_maps}. 
For information on the remaining two objects whose nature as Type Ia continue to be debated -- G43.3--0.2 (also known as W49B) and G41.1--0.3 (also known as 3C~397) -- we direct the reader to a dedicated recent study \citep{paper_i}. For all citations pertaining to the Galactic objects covered within this work, we refer the reader to SNRcat\footnote{\url{http://snrcat.physics.umanitoba.ca}}, the online catalogue of supernova remnants hosted at the University of Manitoba \citep{2012AdSpR..49.1313F}.

\subsection{Galactic SNRs}
\label{ssec:targets_galactic_snrs}

\textbf{G1.9+0.3}: At an estimated 120 years old \citep{reynolds_2008, borkowski_2013}, G1.9+0.3 is the youngest known SNR in our galaxy, located nearby to the Galactic center at an estimated distance of 8.5 kpc. Its young age and moderate distance corresponds to a small angular size, appearing with a diameter of approximately 1$\arcmin$.7. Although its supernova was never observed, its radio and X-ray spectra suggest a Type Ia origin, although a core-collapse nature has not been ruled out primarily based on its location close to the Galactic center. In the X-ray regime, G1.9+0.3 is a synchrotron-dominated SNR \citep{reynolds_2009}, which imposes some difficulty when determining chemical composition and possible abundances. However, evidence of thermal emission in several regions of the SNR was detected by \cite{borkowski_2010}.

\textbf{G4.5+6.8}: More commonly known as Kepler's SNR, G4.5+6.8 holds historical significance as the last Galactic SNR whose progenitor explosion was visible to the naked eye, having occurred in 1604 CE. The remnant itself was first identified by \cite{baade_1943}, and is a roughly circular shell with distinct ``ears" at its northwest and southeast edges, measuring about 3$\arcmin$.6 in diameter. As the SN was observed, the SNR age is precisely known, though the distance remains debated, with recent estimates ranging from 4.8 to 7 kpc, with a commonly adopted value of 5 kpc \citep{reynoso_goss_1999, aharonian_2008, chiotellis_2012, patnaude_2012}.

Though it is one of the most studied SNRs, the nature of its supernova origin was for a time unclear. Initially, debates focused on whether it originated from a Type Ia or core-collapse (CC) event, but X-ray observations -- particularly those of \cite{reynolds_2007} -- largely settled the debate with the conclusion of a Type Ia origin. The exact mechanism -- single-degenerate (SD) or double-degenerate (DD) -- remains unclear, however. In the X-ray regime, the remnant is well-described by non-equilibrium ionization (NEI) models, with studies indicating complex emission from elements like Si, S, Fe, and sub-solar abundances of O and Mg \citep{cassam-chenai_2004, katsuda_2015, sato_2017, sun_2019, kasuga_2021, holland-ashford_2023}. Evidence of nonthermal emission, particularly at the outer rims, suggests particle acceleration at shock fronts \citep{tsuji_2021}.

\textbf{G120.1+1.4}: Generally referred to as Tycho's SNR, G120.1+1.4 is associated with the supernova event observed in 1572 CE. With its classification as the remnant of a Type Ia event being confirmed through the observation of light echoes \citep{kraus_2008}, Tycho's SNR is one of the few remnants for which both its age and supernova origin are well-established \citep{2023FrASS..1012880R}. The remnant itself is a limb-brightened, nearly circular shell with an angular diameter of around 8$\arcmin$, assuming a distance of 3.6 kpc \citep{hayato_2010}. Its sharp forward shock contrasts with the more diffuse interior, a structure likely caused by Rayleigh-Taylor instabilities at the boundary between the ejecta and interstellar medium (ISM) \citep{warren_blondin_2013}.

As a young and relatively bright remnant, G120.1+1.4 is one of the most extensively studied SNRs. Its X-ray spectrum is dominated by emission from intermediate-mass elements such as Mg, Si, S, Ar, and Ca, along with significant Fe L and Fe K emission. Stratification of the elements is observed, with lighter elements found at the outer regions and Fe concentrated towards the center \citep{hwang_1997, aschenbach_2002}. Thermal emission from the ejecta is consistent with non-equilibrium ionization (NEI) models, showing super-solar abundances of Si, S, Ar, Ca, and Fe \citep{guo_2017}, while nonthermal synchrotron emission from the forward shock suggests active cosmic ray acceleration \citep{warren_2005, cassam-chenai_2007}. Several bright knots of enhanced emission, particularly in the southeast, exhibit variations in Si and Fe abundances, hinting at incomplete mixing between the layers \citep{decourchelle_2001}.

\textbf{G272.2-3.2}: First discovered in X-rays by the \textit{ROSAT} All-Sky Survey and later studied in radio, G272.2--3.2 is a faint, roughly circular SNR with a diameter of 15$\arcmin$. Its X-ray emission is predominantly thermal and exhibits a centrally-brightened morphology, while its radio spectrum shows a steep index of $-0.55 \pm 0.15$ ($S \propto \nu^{\alpha}$), typical of shell-type SNRs; it was this contrast that lead to its classification as a thermal composite SNR \citep{harrus_2001}. The distance to the remnant is uncertain, with estimates ranging from 2.5 to 10 kpc \cite{greiner_egger_1994, harrus_2001, kamitsukasa_2016}, and its age is estimated between 3,600 to 6,000 years \cite{sezer_gok_2012, sanchez-ayaso_2013, kamitsukasa_2016}. The SNR's X-ray spectrum has been modeled with a non-equilibrium ionization (NEI) plasma, while abundances in the remnant show a mix of super-solar (Si, S, Fe) and sub-solar (Ne, Mg) values \citep{mcentaffer_2013, kamitsukasa_2016}. Studies suggest the presence of both NEI and collisionally ionised equilibrium (CIE) plasmas, the latter likely resulting from shock-heated interstellar medium (ISM). Based on its abundance patterns, lack of a compact remnant, and Fe K$\alpha$ line energy, G272.2--3.2 is considered the remnant of a Type Ia supernova.

\textbf{G337.2--0.7}: Initially identified as an SNR candidate due to its nonthermal radio emission, G337.2--0.7 possesses an angular size of roughly $4\arcmin.5 \times 5\arcmin.5$, and radio observations reveal an incomplete elliptical ring structure in the southern part of the SNR. X-ray observations, meanwhile, show that the remnant is dominated by metal-rich ejecta \citep{rakowski_2001}. Radio HI absorption measurements by \cite{rakowski_2006} place the remnant between $2.0\pm0.5$ and $9.3\pm0.3$ kpc, and \cite{takata_2016} use arguments of a high X-ray absorption column density to place it at the further end of this range, i.e. at $9$ kpc. Similarly, the remnant’s age is debated, with estimates ranging from 750 to 5000 years depending on assumptions of distance and model \citep{rakowski_2006, takata_2016}. The remnant is classified as a Type Ia SNR, based on abundance patterns and ejecta mass.

\textbf{G344.7--0.1}: In the radio regime, G344.7--0.1 appears as an asymmetric and bright shell-like structure with a size of $8\arcmin-10\arcmin$, while in the X-ray regime, it appears as an extended source of thermal emission roughly $6\arcmin$ in diameter. It's assumed to be a somewhat older SNR, with its measured ionization timescale yielding an estimated age between $3000 - 6000$ years \citep{combi_2010, giacani_2011}. The distance to the SNR is similarly poorly constrained, with estimates yielding values between $6.2 - 14$ kpc \citep{dubner_1993, giacani_2011, yamaguchi_2012, fukushima_2020}. G344.7--0.1 was initially categorised as a core-collapse SNR by \cite{lopez_2011} and \cite{giacani_2011}, using arguments of asymmetry and possible association with star-forming regions, respectively; however, the energy of the remnant's Fe K$\alpha$ line, as well as the abundance patterns and distribution, have caused it to be reclassified as a Type Ia SNR \citep{yamaguchi_2012, fukushima_2020}

\textbf{G352.7--0.1}: Though initially classified as a shell-like SNR due to its appearance in the radio regime, which features two loop-like structures with an angular size of $8\arcmin \times 6\arcmin$, G352.7--0.1 is considered a mixed-morphology SNR due to its thermal X-ray emission \citep{giacani_2009}. Age estimations range from 2200 years \citep{kinugasa_1998} to 7600 years \citep{leahy_ranasinghe_2016}, with more recent estimates coming firmly in the middle at $4900 \pm 1300$ years \citep{dang_2024}. The remnant’s classification remains uncertain: \cite{giacani_2009} and \cite{pannuti_2014} suggest a core-collapse origin due to its barrel-shaped morphology and large swept-up mass, while \cite{yamaguchi_2014}, \cite{sezer_2014}, and \cite{fujishige_2023} argue for a Type Ia origin based on elemental abundances and Fe K$\alpha$ line energy.

\subsection{LMC SNRs}
\label{ssec:targets_lmc_snrs}

\textbf{0505--67.9}: Also known as DEM L71, the LMC SNR 0505--67.9 is possessed of a slightly elliptical outer rim roughly $1\arcmin.4 \times 1\arcmin$ across surrounding faint, diffuse emission from its center in an apparent double-shock morphology suggestive of an outer blast wave surrounding a central region of reverse-shock heated ejecta. Its age is moderately well constrained, with estimates ranging from $4360 - 6600$ years \citep{hughes_1998, ghavamian_2003, alan_2022}. Detection of enhanced abundances for Fe, Mg, Si, and S by \cite{hughes_2003} and \cite{rakowski_2003}, alongside an estimated Fe ejecta mass of $1.4$ M$_{\odot}$ \citep{hughes_2003, vanderheyden_2003} have resulted in the remnant being classified as a Type Ia SNR.

\textbf{0509--67.5}: In the optical regime, 0509--67.5 appears as a largely circular SNR. In the X-ray regime, it appears similar, displaying a clearly circular shell with fairly faint emission from its center. The SNR's southwestern half exhibits brighter emission than the northeastern half, giving it a distinct, two-halved appearance roughly $0\arcmin.42$ across. Through the observation of light echoes, it has been constrained to an age roughly $400$ years \citep{rest_2005}. These light echoes -- alongside the elemental abundances observed in the X-ray regime by \cite{warren_2004} -- strongly suggest a Type Ia origin for the SNR.

\textbf{0509--68.7}: Also known as N103B, 0509--68.7 is the fourth brightest X-ray remnant in the LMC. A shell-type remnant with an angular size of roughly $0\arcmin.5$, N103B exhibits a highly asymmetric morphology with a bright enhancement towards the west, where it's believed to be interacting with a nearby molecular cloud \citep{williams_2014}. Through the use of light echoes, \cite{rest_2005} estimated its age at $860$ years, and this young age is supported by the measurements of high average shock velocities that exceed $4 \times 10^{3}$~\si{km.s^{-1}} \citep{williams_2018}. The origin of N103B has not been firmly established: while its proximity to a nearby HII region and the measured abundances of O, Ne, and Mg support a core-collapse origin \citep{vanderheyden_2002, someya_2014}, its substantial Fe mass and morphology have been used to instead argue for a Type Ia origin \citep{lewis_2003, lopez_2011}.

\textbf{0519--69.0}: 0519--69.0 is a somewhat irregular, patchy SNR with clear inner ejecta surrounded by a thin shell of swept-up ISM roughly $0\arcmin.55$ across. Through observation of light echoes, its age was estimated to be to roughly $600$ years \citep{rest_2005}; however, more recent estimates based on measurements of circumstellar density and the radii of the forward and reverse shocks have placed it at approximately $450$ years old instead \citep{kosenko_2010}. Though it hasn't been extensively studied in the X-ray regime, the prominence of the emission lines for the elements between Si and Fe have resulted in it being classified as a Type Ia SNR.

\textbf{0534--69.9}: 0534--69.9 possesses a bright central region surrounded by a limb-brightened shell of swept-up ISM, and is fairly large for an LMC SNR, with an angular size of approximately $2\arcmin$. Based on measurements of shock velocities, this size corresponds to an age of roughly $10100$ years \citep{hendrick_2003}. Early studies of the remnant have been limited, but observations by \textit{Chandra} and \textit{Suzaku} have revealed strong Fe emission, alongside K-shell emission from Mg, Si, and S \citep{hendrick_2003, takeuchu_2016}. The abundances of these elements have resulted in an assumption of a Type Ia origin for the SNR.

\textbf{0548--70.4}: Discovered in X-rays by \cite{long_1981}, 0548--70.4 is an approximately circular object with an angular radius of $1\arcmin.8$. It possesses a bright central region alongside bright limbs that are most prominent on the remnant's eastern and western sides. An estimated age of $7100$ years is based on the measurement of shock velocities by \cite{hendrick_2003}. Like 0534--69.9, 0548--70.4 has not been studied extensively in the X-ray regime. Strong Fe emission has been detected, but the emission from intermediate mass elements is notably lower \citep{hendrick_2003, takeuchu_2016}. These abundances have led to an assumed Type Ia origin for this SNR.

\section{Observations and Data Reduction}
\label{sec:data}
All observations used in this analysis made use of archival data obtained by the \textit{XMM-Newton} X-ray telescope, save for those of G1.9+0.3, which used archival data from the \textit{Chandra} X-ray Observatory instead. Observation IDs and combined exposure times can be found in Table~\ref{tab:info_table}. \textit{XMM-Newton} observations were performed using the European Photon Imaging Camera (EPIC), using both Metal Oxide Semi-conductor (MOS) CCD cameras, as well as the pn camera, which together afford an energy range of 0.15 to 15 keV, with a spectral resolution of roughly 0.1 keV \citep{turner_2001, struder_2001}. The \textit{Chandra} observations were performed using the Advanced CCD Imaging Spectrometer (ACIS), using the ACIS-S array.

For all \textit{XMM-Newton} observations, calibration and filtering were performed using the \textit{XMM-Newton} Science Analysis System\footnote{\url{https://www.cosmos.esa.int/web/xmm-newton/sas}} (SAS) v19.1.0, with the latest calibration files being used. All observation files were reprocessed using the SAS tasks \texttt{emproc} and \texttt{epproc} and filtered for good time intervals (GTI), bad pixels, and out-of-time events, and all data was checked for any potential photon pile-up using the command \texttt{epatplot}. The MOS data was filtered to retain patterns 0 -- 12 in the 200 -- 12000 eV range and using the \texttt{\#XMMEA\_EM} flag, while the pn data was filtered to retain patterns 0 -- 4 in the 200 -- 15000 eV range and using the \texttt{\#XMMEA\_EP} flag.

For the \textit{Chandra} data, data processing was done using the Chandra Interactive Analysis of Observations (CIAO) version 4.16, and followed the standard procedure laid out in the CIAO Analysis Guides.\footnote{\url{https://cxc.harvard.edu/ciao/guides/index.html}} The \texttt{chandra\_repro} command was used to reprocess the data before filtering was applied to limit the energies to the $0.3-10$~keV range, and periods of background flaring were removed through the use of GTI filters which were applied using the command \texttt{dmcopy}. Spectral extraction was performed using the \texttt{specextract} command, and once extracted, spectra were regrouped to a minimum of 20 counts per bin using the FTOOLS task GRPPHA. Sample spectra -- including fitted models and residuals -- for all objects can be seen in Figure~\ref{fig:sample_spectra}.

\begin{deluxetable}{lcccccc}
\centering
\tablecaption{Basic information, observational IDs, and combined exposure times for each object studied in this work. For references, see Section \ref{sec:targets}. \label{tab:info_table}}
\tablehead{
\colhead{Object Name} &
\colhead{\makecell{Unique\\Name}} &
\colhead{\makecell{Distance\\(kpc)}} &
\colhead{\makecell{Size\\(arcmin.)}} &
\colhead{\makecell{Age\\(yrs)}} &
\colhead{Observation ID(s)} &
\colhead{\makecell{Exposure Time\\(ks)}}
}
\startdata
G1.9+0.3 &  & 8.5 & 1.7 & $\sim$120 & \makecell{12689, 12690, 12691,\\12692, 12693, 12694,\\12695, 13407, 13509} & 980.5 \\
G4.5+6.8 & Kepler & 5 & 3.6 & 421 & 0842550101 & 140.5 \\
G120.1+1.4 & Tycho & 3.6 & 8 & 453 & \makecell{0412380101, 0412380201,\\0412380301, 0412380401} & 149.6 \\
G272.2--3.2 &  & 2.0 - 10 & 15 & 3600 - 6000 & 0112930101 & 38.4 \\
G337.2--0.7 &  & 2.0 - 9.3 & $4.5 \times 5.5$ & 750 - 5000 & 0087940101 & 40.2 \\
G344.7--0.1 &  & 6.1 - 14 & 6 & 3000 - 6000 & 0111210101, 0111210401 & 22.2 \\
G352.7--0.1 &  & 7.5 - 11 & $8 \times 6$ & 2200 - 7600 & 0150220101 & 30.7 \\
0505--67.9 & DEM L71 & 50 & $1.4 \times 1$ & 4360 - 6600 & 0884620101 & 129.8 \\
0509--67.5 &  & 50 & 0.42 & 400 & 0111130201 & 44.3 \\
0509--68.7 & N103B & 50 & 0.50 & 860 & 0113000301 & 26.5 \\
0519--69.0 &  & 50 & 0.55 & 450 & 0113000501 & 48.4 \\
0534--69.9 &  & 50 & 2 & 10100 & 0673780101 & 61.9 \\
0548--70.4 &  & 50 & 1.8 & 7100 & 0883390101 & 59.9
\enddata
\end{deluxetable}

\subsection{Region Generation}
\label{ssec:regions}
For this study, region maps for 12 of the 13 objects were generated using the \textit{contbin}\footnote{\url{https://www-xray.ast.cam.ac.uk/papers/contbin/}} algorithm \citep{sanders_2006}, an adaptively smoothed binning program which generates regions based on a set of input parameters. This allowed for the generation of regions which seamlessly encompass the entirety of an SNR while simultaneously maintaining its morphology. Through this method regions are generated based on surface brightness, which is applicable due to the generally-observed correlation between this parameter and spectral properties, and are intended to have the same (or at least, a very similar) number of counts per region.

\begin{figure*}
    \gridline{
        \fig{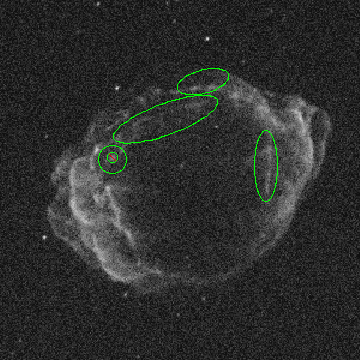}{0.24\textwidth}{(a) G1.9+0.3 (4 regions)}
        \fig{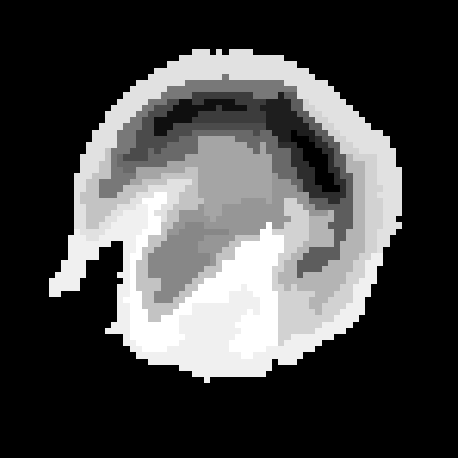}{0.24\textwidth}{(b) G4.5+6.8 (18 regions)}
        \fig{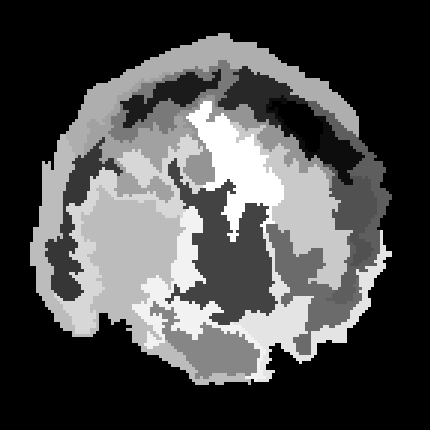}{0.24\textwidth}{(c) G120.1+1.4 (20 regions)}
        \fig{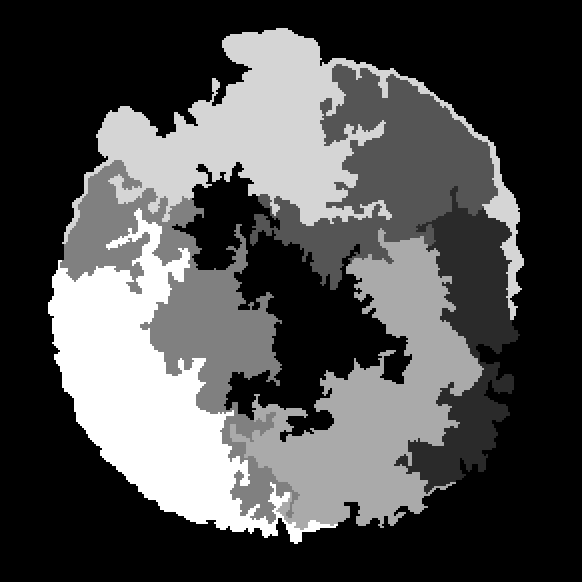}{0.24\textwidth}{(d) G272.2-3.2 (7 regions)}
    }
    \gridline{
        \fig{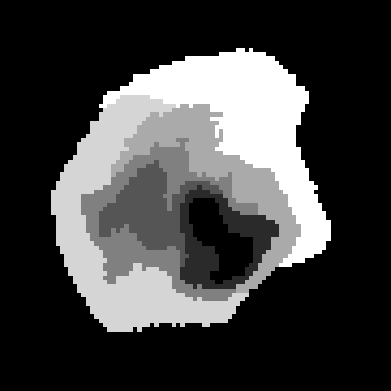}{0.24\textwidth}{(e) G337.2--0.7 (7 regions)}
        \fig{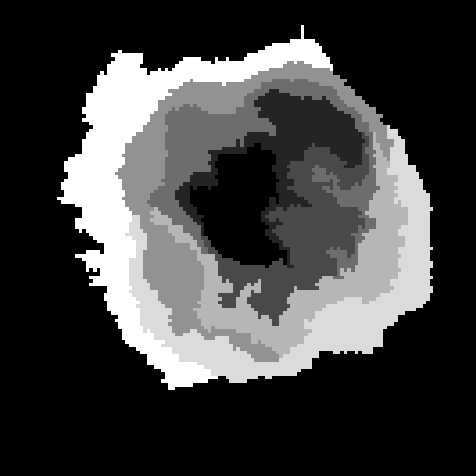}{0.24\textwidth}{(f) G344.7--0.1 (8 regions)}
        \fig{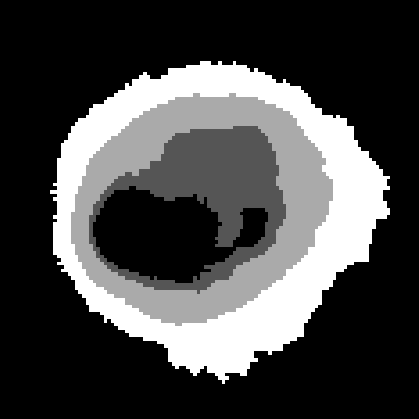}{0.24\textwidth}{(g) G352.7--0.1 (4 regions)}
        \fig{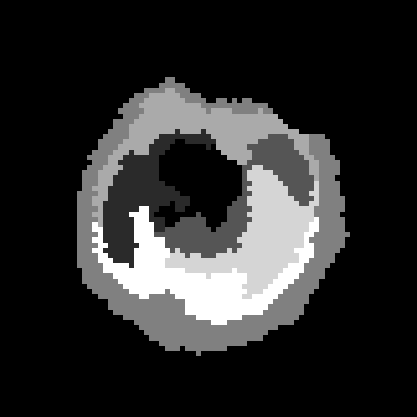}{0.24\textwidth}{(h) 0505--67.9 (7 regions)}
    }
    \gridline{
        \fig{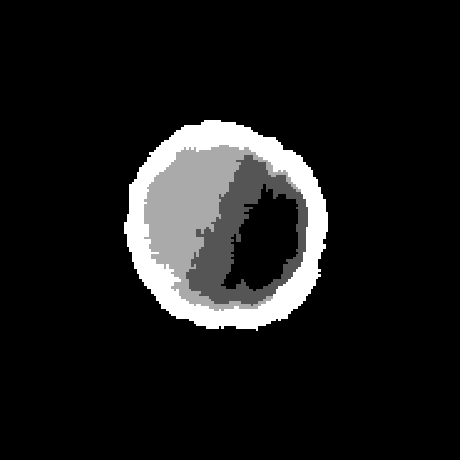}{0.24\textwidth}{(i) 0509--67.5 (4 regions)}
        \fig{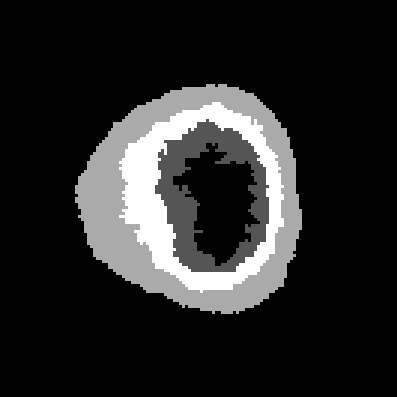}{0.24\textwidth}{(j) 0509--68.7 (4 regions)}
        \fig{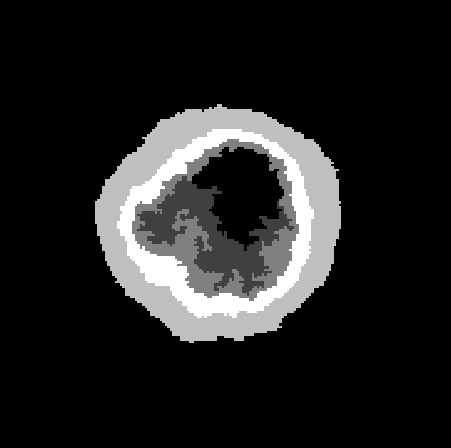}{0.24\textwidth}{(k) 0519--69.0 (5 regions)}
        \fig{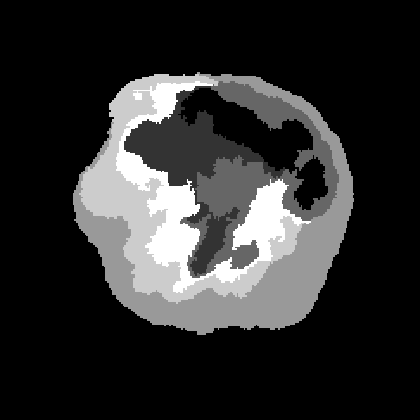}{0.24\textwidth}{(l) 0534--69.9 (6 regions)}
    }
    \gridline{
        \leftfig{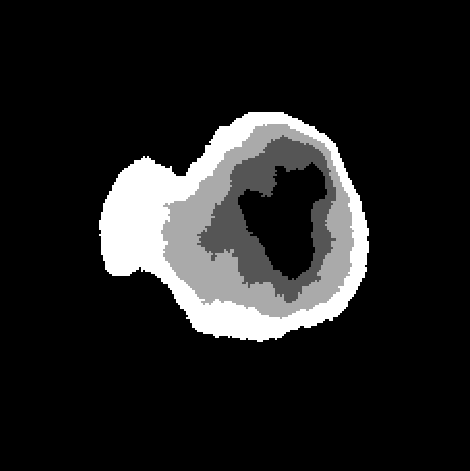}{0.24\textwidth}{(m) 0548--70.4 (4 regions)}
    }
    \caption{Region maps for all objects involved in this study. For G1.9+0.3, the regions were selected by hand, and are indicated by green ellipses. For all other objects, the regions were generated via the \textit{contbin} algorithm (see \S\ref{ssec:regions}), and are indicated by varying shades on a white-black gradient.}
    \label{fig:region_maps}
\end{figure*}

As several of the objects involved in this study made use of multiple observations, we opted not to run the algorithm on each observation individually. Instead, broadband (0.1 -- 10 keV) images from all observations were first merged to create a single image, from which a mask was created to better constrain the algorithm to the borders of the SNR. The results were then binned to an appropriate bin size before running the algorithm. These steps were done to avoid the generated regions being biased towards the surface brightness of any particular observation. The final number of regions was chosen to allow for sufficient statistics for each region during the spectral fitting process, whilst still allowing for the examination of any arcsecond-scale structure within the object. The final regions for each object are shown in Figure~\ref{fig:region_maps}, in which each color represents a different region. In our final results, each region is labeled (e.g. ``R00," ``R01," ``R02," etc.) according to the output of the \textit{contbin} algorithm -- that is to say, in order of descending flux of their highest-flux pixel.

The sole exception to the use of the \textit{contbin} algorithm was for G1.9+0.3. While we initially attempted to use the same method for this object, as G1.9+0.3 is dominated by non-thermal emission \citep{reynolds_2008, reynolds_2009}, we found that the generated regions were similarly dominated, with no notable thermal emission to which we could fit our spectral models. However, the presence of thermal emission in the SNR was noted by \cite{borkowski_2010}, largely concentrated in the SNR's northern rim. This emission was further isolated by \cite{borkowski_2013}, who noted four distinct regions in which thermal emission was detected. For this object, we thus opted to reproduce these four regions for our study.

\begin{figure*}
	\begin{center}
		\subfloat[G1.9+0.3]{\includegraphics[angle=0,width=0.45\textwidth,scale=0.5]{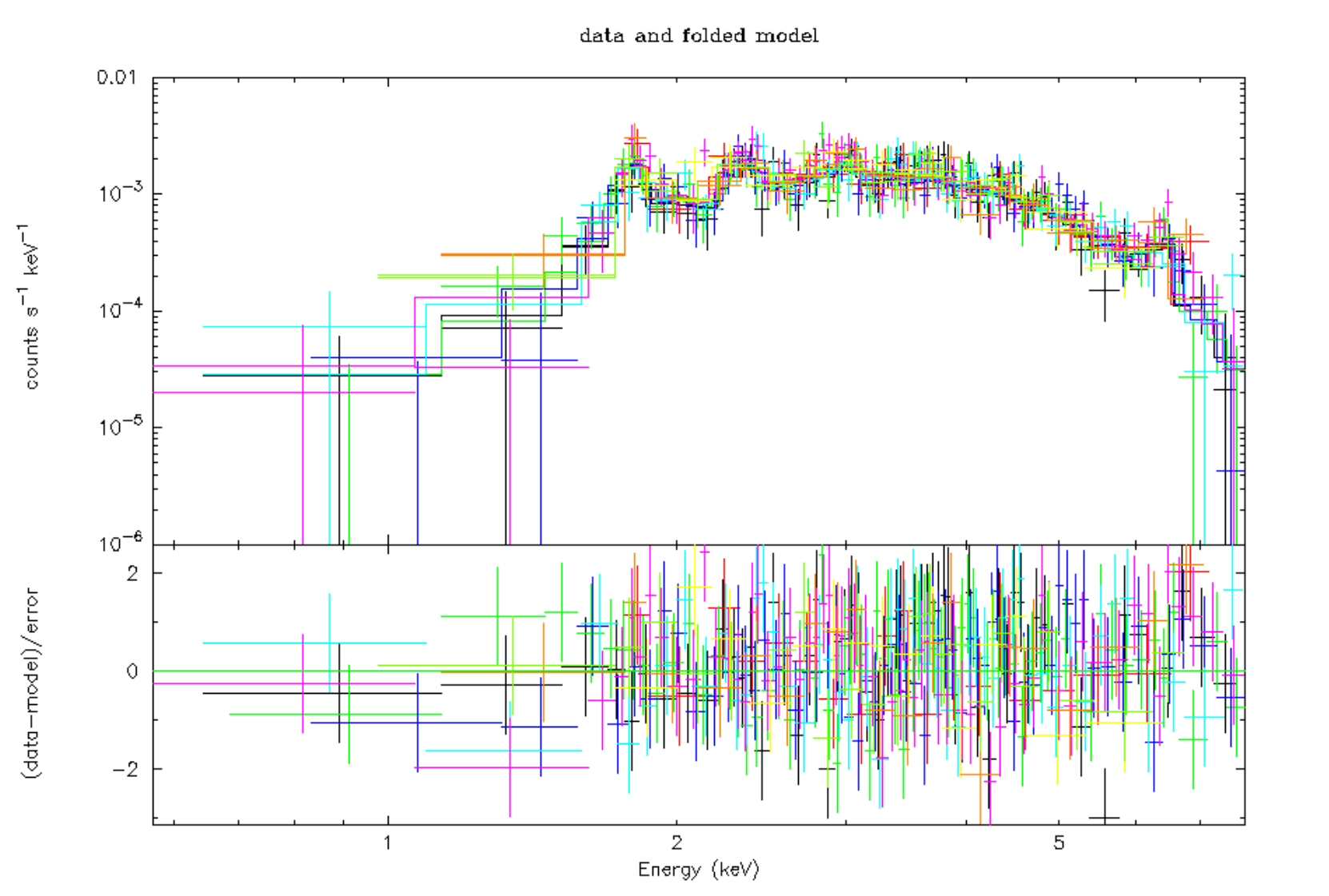}}
		\subfloat[G4.5+6.8]{\includegraphics[angle=0,width=0.45\textwidth,scale=0.5]{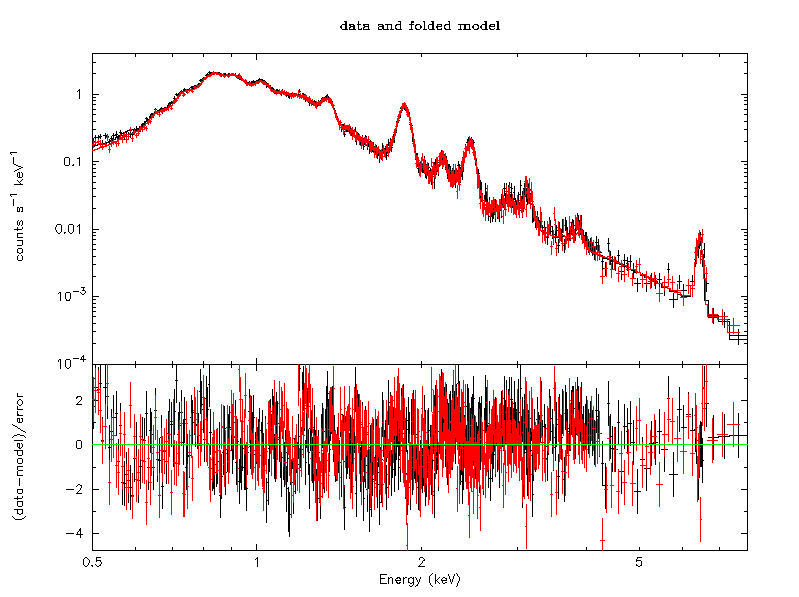}}\\
		\subfloat[G120.1+1.4]{\includegraphics[angle=0,width=0.45\textwidth]{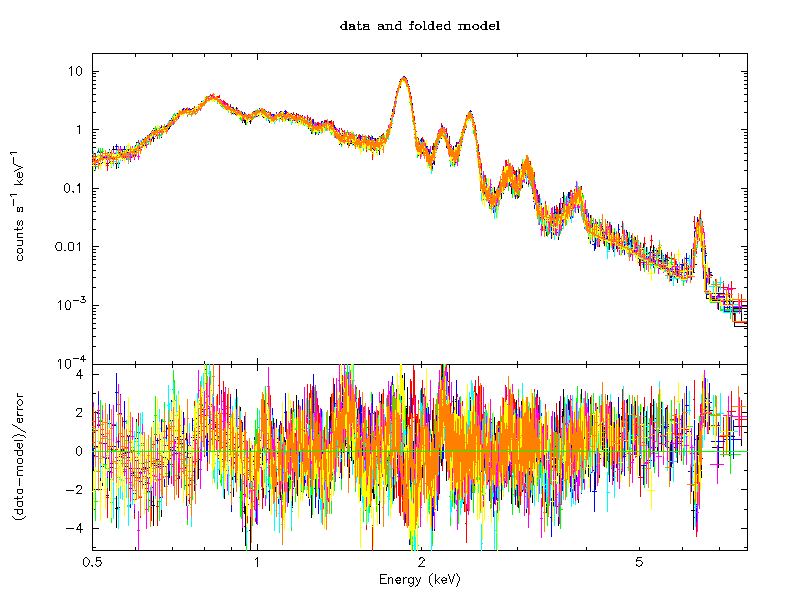}}
		\subfloat[G272.2--3.2]{\includegraphics[angle=0,width=0.45\textwidth]{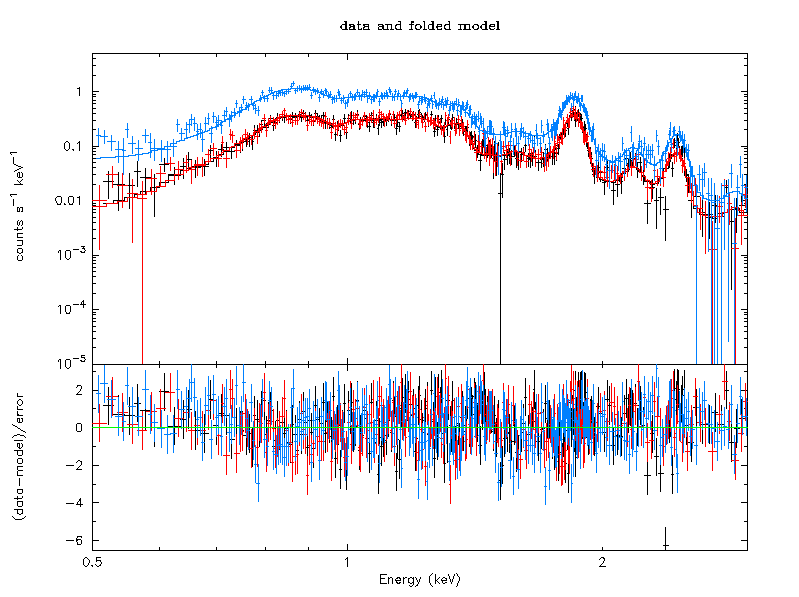}} \\
		\subfloat[G337.2--0.7]{\includegraphics[angle=0,width=0.45\textwidth,scale=0.5]{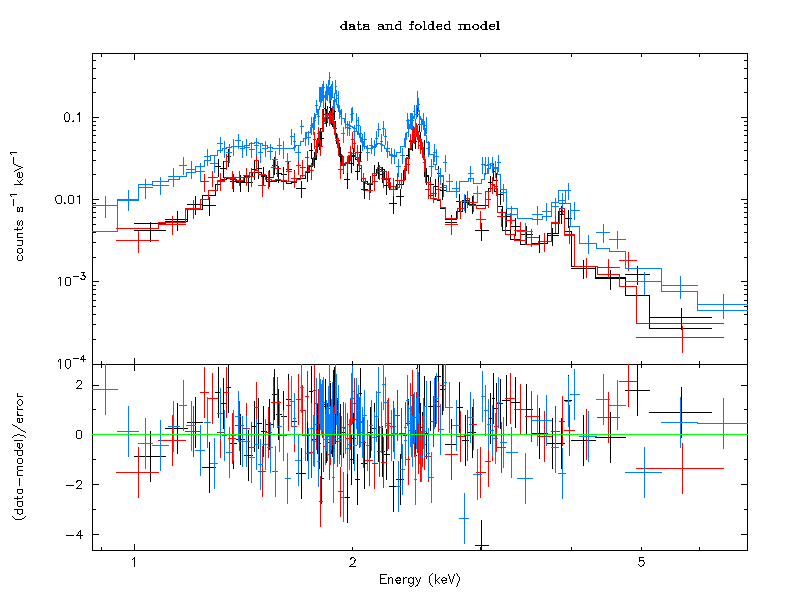}}
		\subfloat[G344.7--0.1]{\includegraphics[angle=0,width=0.45\textwidth]{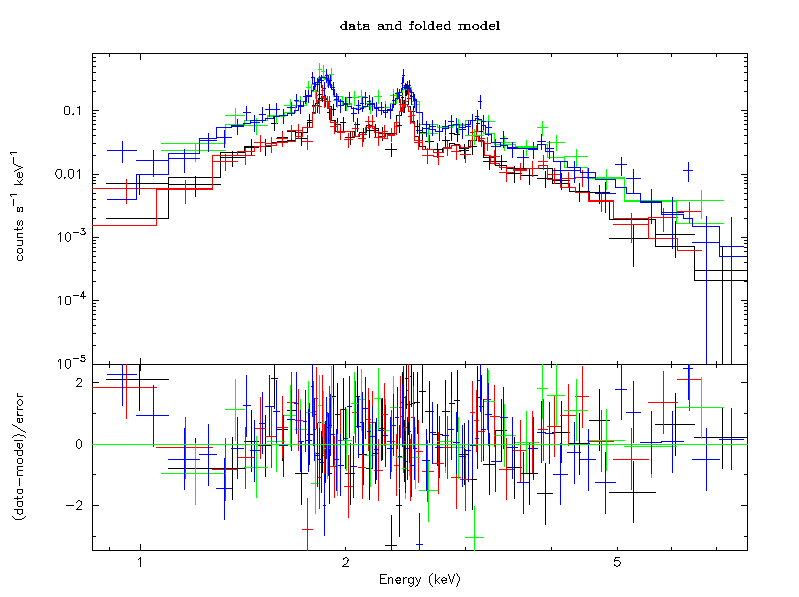}}
	\end{center}
    \caption{Sample spectra for each object. All spectra taken from region labeled `R00' in each object, with the exception of G1.9+0.3, for which the sample spectrum was taken from the region labeled `N'.}
    \label{fig:sample_spectra}
\end{figure*}

\begin{figure*}
    \ContinuedFloat
	\begin{center}
		\subfloat[G352.7--0.1]{\includegraphics[angle=0,width=0.45\textwidth,scale=0.5]{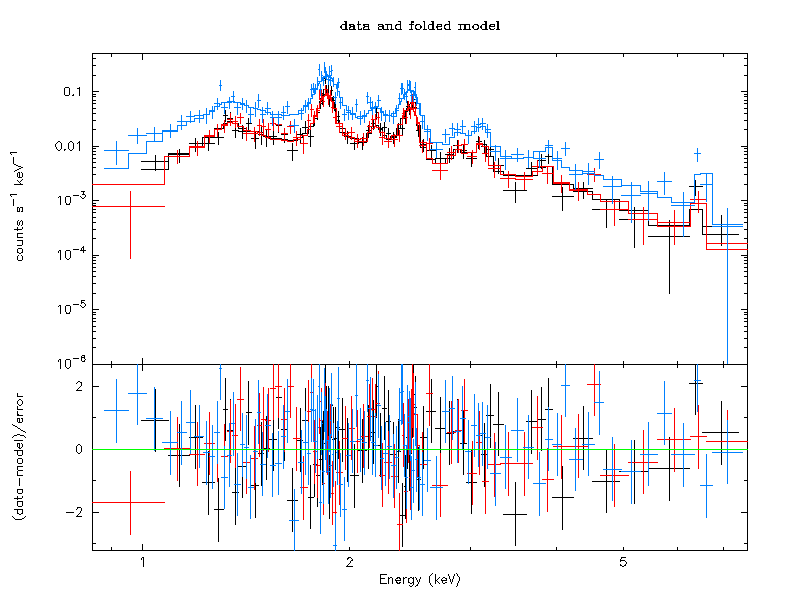}}
		\subfloat[0505--67.9]{\includegraphics[angle=0,width=0.45\textwidth,scale=0.5]{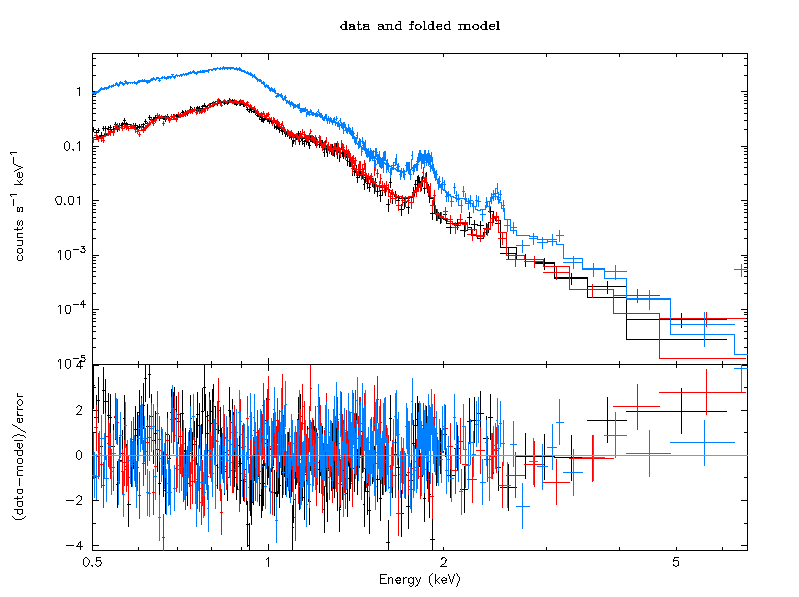}} \\
		\subfloat[0509--67.5]{\includegraphics[angle=0,width=0.45\textwidth]{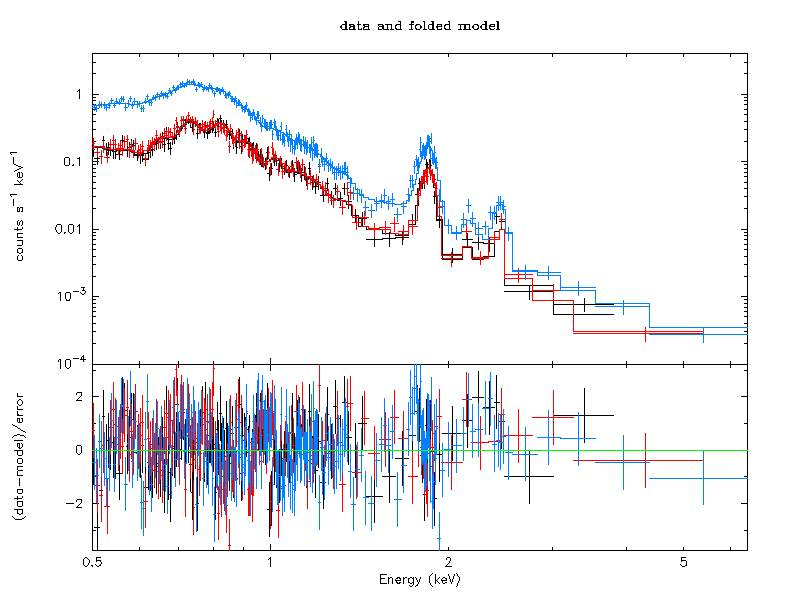}}
		\subfloat[0509--68.7]{\includegraphics[angle=0,width=0.45\textwidth]{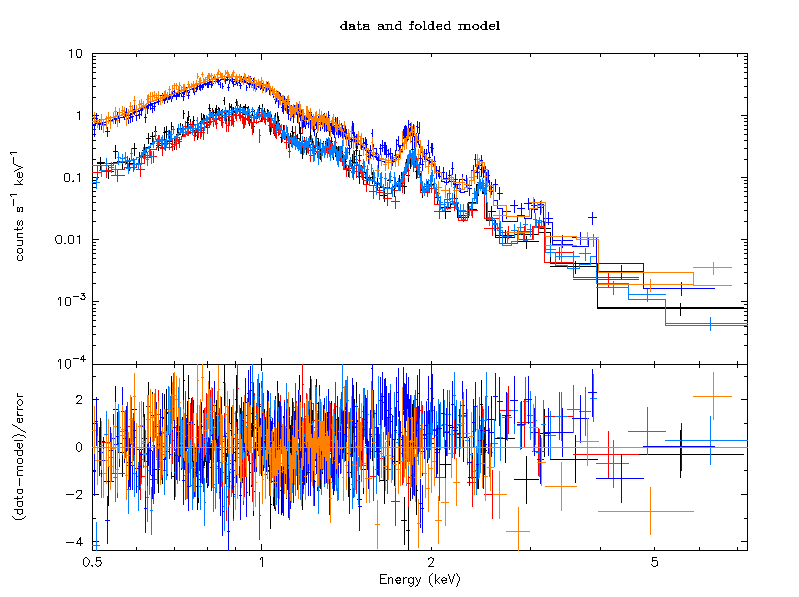}} \\
		\subfloat[0519--69.0]{\includegraphics[angle=0,width=0.45\textwidth,scale=0.5]{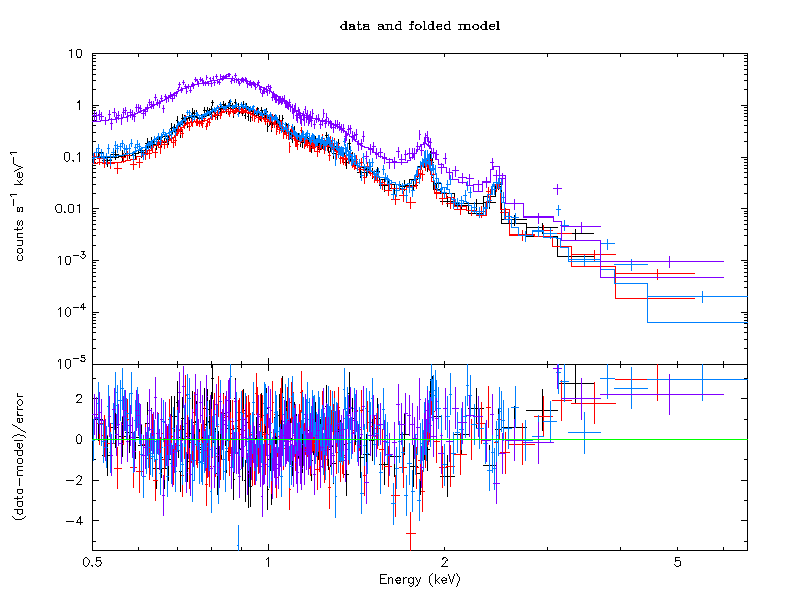}}
		\subfloat[0534--69.9]{\includegraphics[angle=0,width=0.45\textwidth]{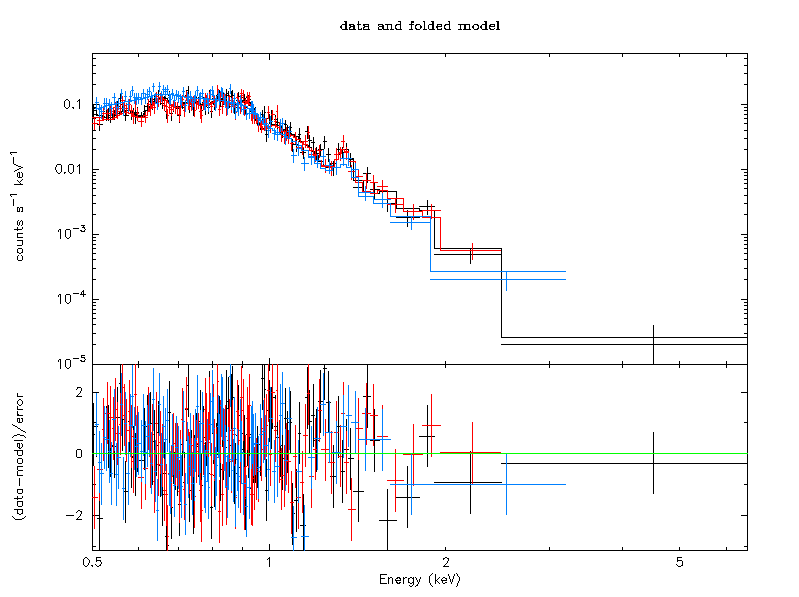}}
	\end{center}
    \caption{Continued from above.}
\end{figure*}

\begin{figure*}
    \ContinuedFloat
	\begin{center}
		\subfloat[0548--70.4]{\includegraphics[angle=0,width=0.45\textwidth,scale=0.5]{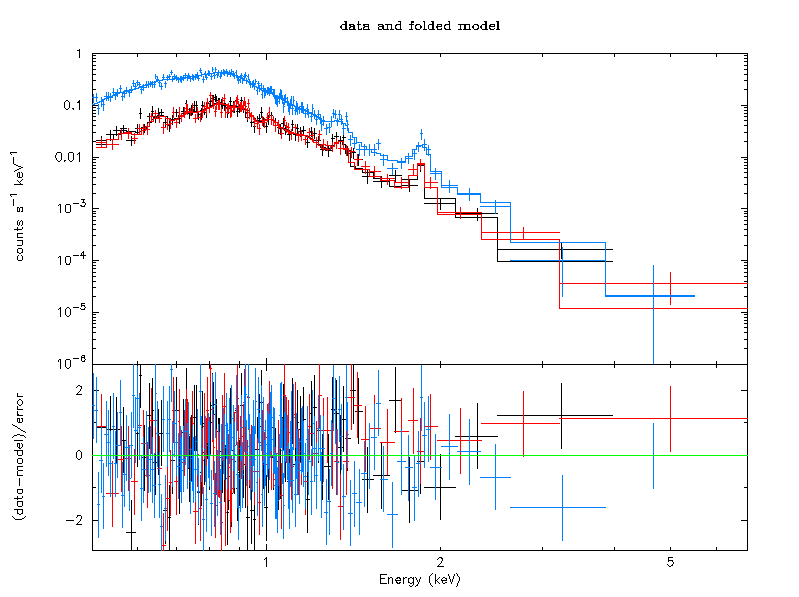}}
	\end{center}
    \caption{Continued from above.}
\end{figure*}

\begin{figure*}
	\begin{center}
		\subfloat[G1.9+0.3]{\includegraphics[angle=0,height=0.48\textwidth,scale=0.5]{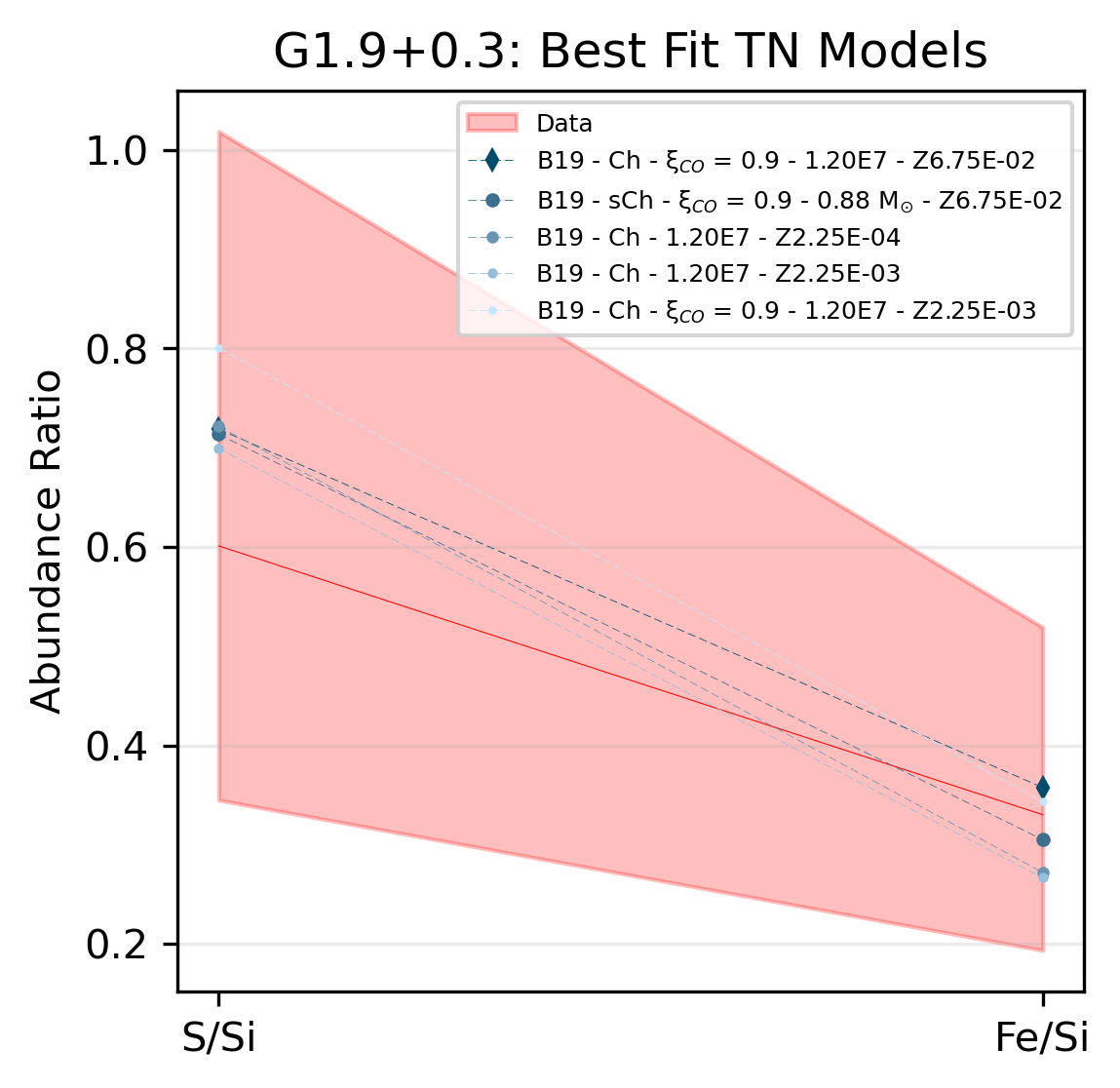}}
		\subfloat[G4.5+6.8]{\includegraphics[angle=0,height=0.48\textwidth,scale=0.5]{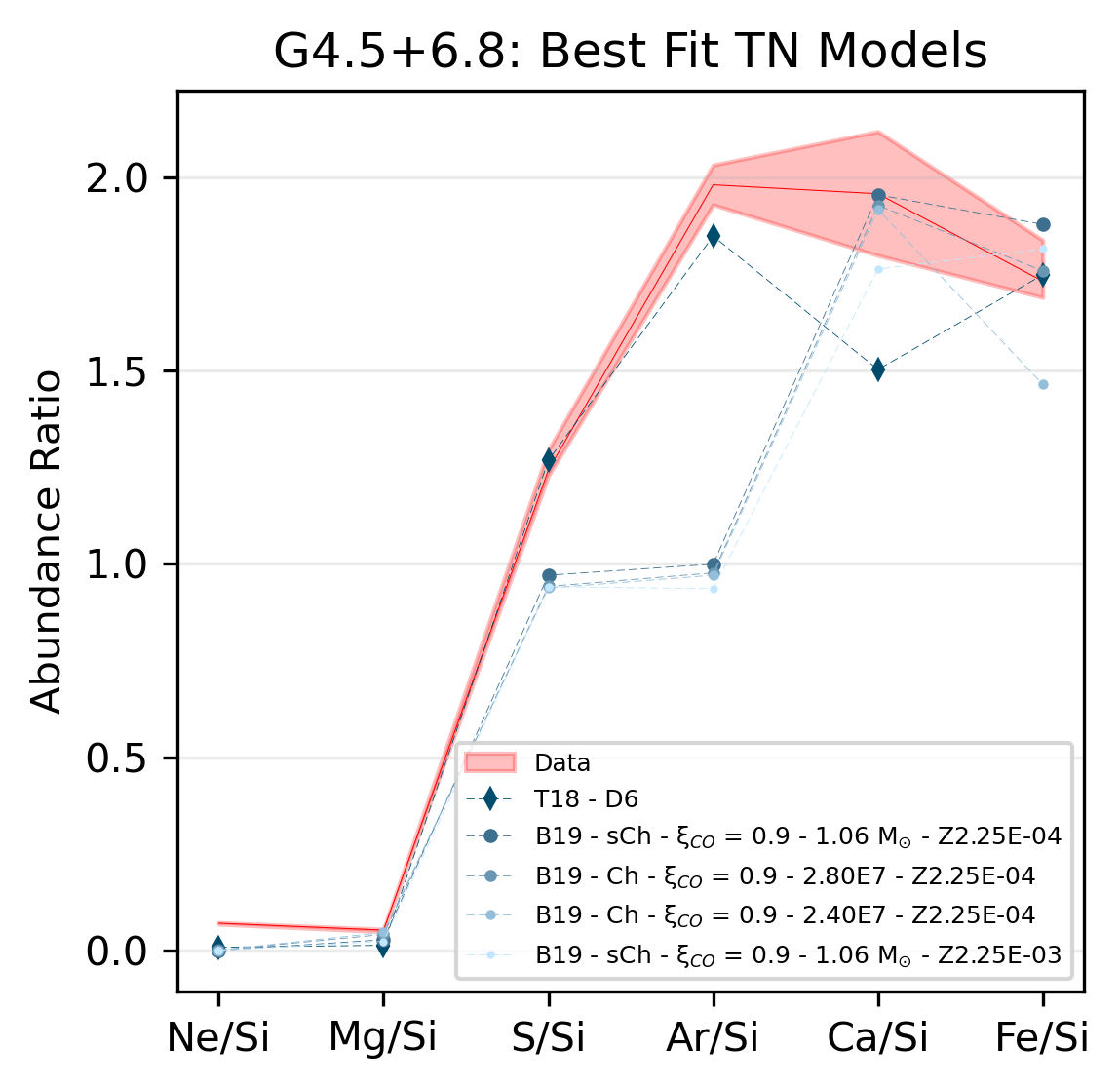}}\\
		\subfloat[G272.2--3.2]{\includegraphics[angle=0,height=0.48\textwidth]{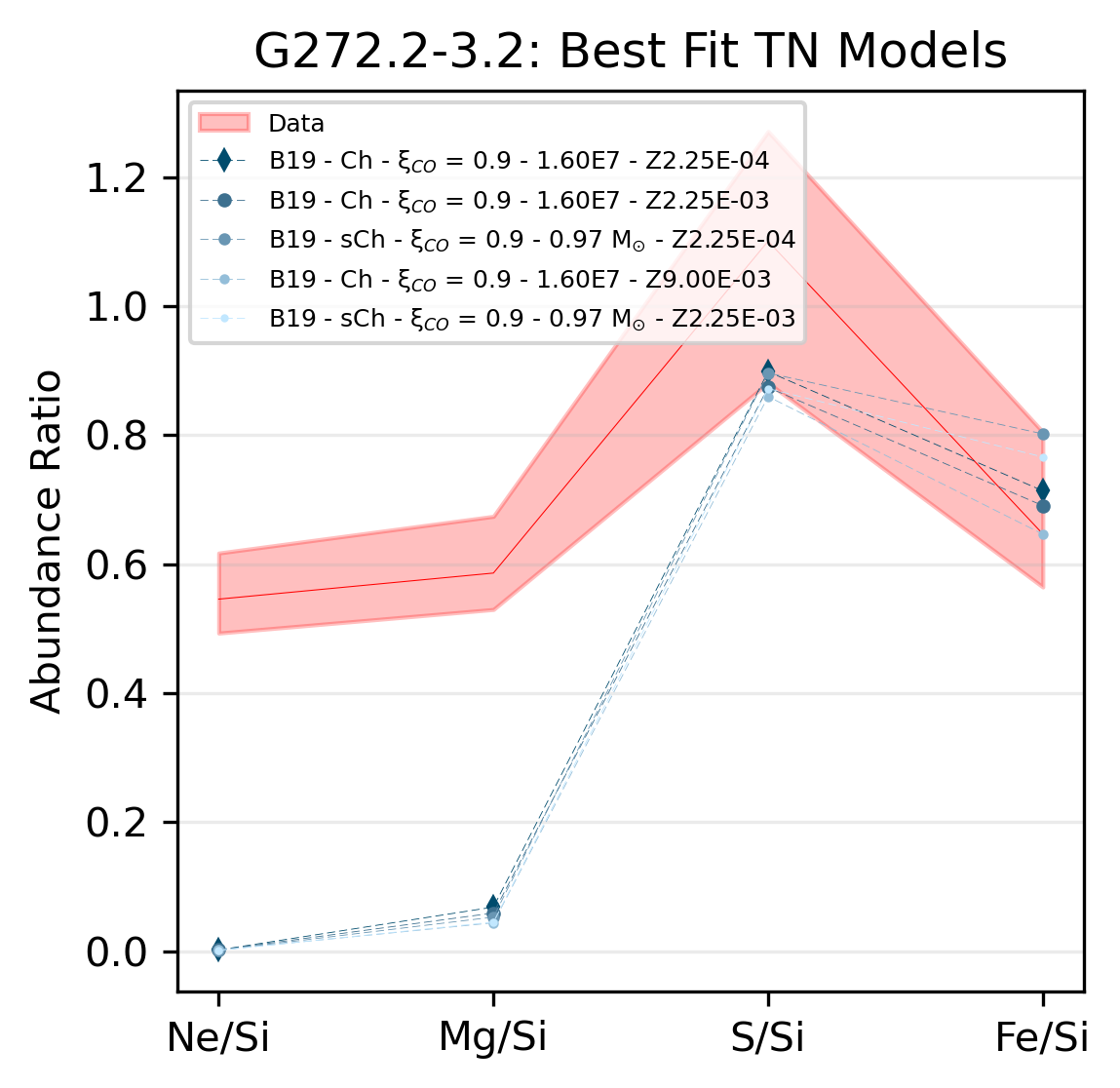}}
		\subfloat[G337.2--0.7]{\includegraphics[angle=0,height=0.48\textwidth,scale=0.5]{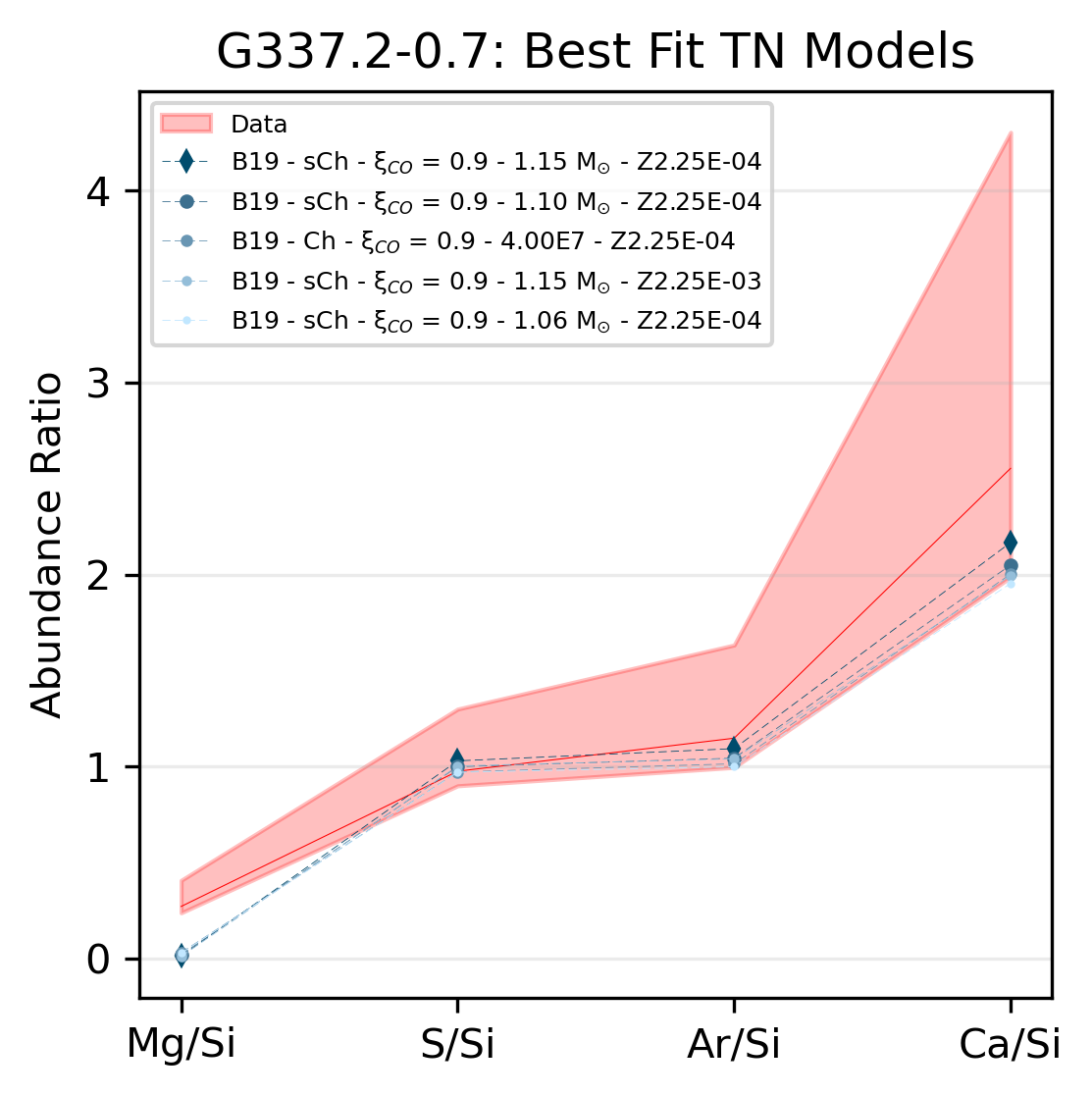}}
	\end{center}
    \caption{Comparisons between observational data and the best-fit models of nucleosynthesis in Type Ia supernovae for all objects. Shaded red areas represent observational data, while colored points represent abundance ratios from a given model and are listed in order of decreasing quality-of-fit. 
    The abundance ratios shown (in this and subsequent figures) are given in the form of $(\text{X}/\text{Si})/(\text{X}/\text{Si})_{\odot}$, where X is the abundance value of the relevant element with respect to the solar values of \cite{wilms_2000}.
    See \S\ref{ssec:thermonuclear} and Table~\ref{tab:model_summary} for details of the models.}
    \label{fig:ia_nucsyn_plots}
\end{figure*}

\begin{figure*}
    \ContinuedFloat
	\begin{center}
		\subfloat[G344.7--0.1]{\includegraphics[angle=0,height=0.48\textwidth]{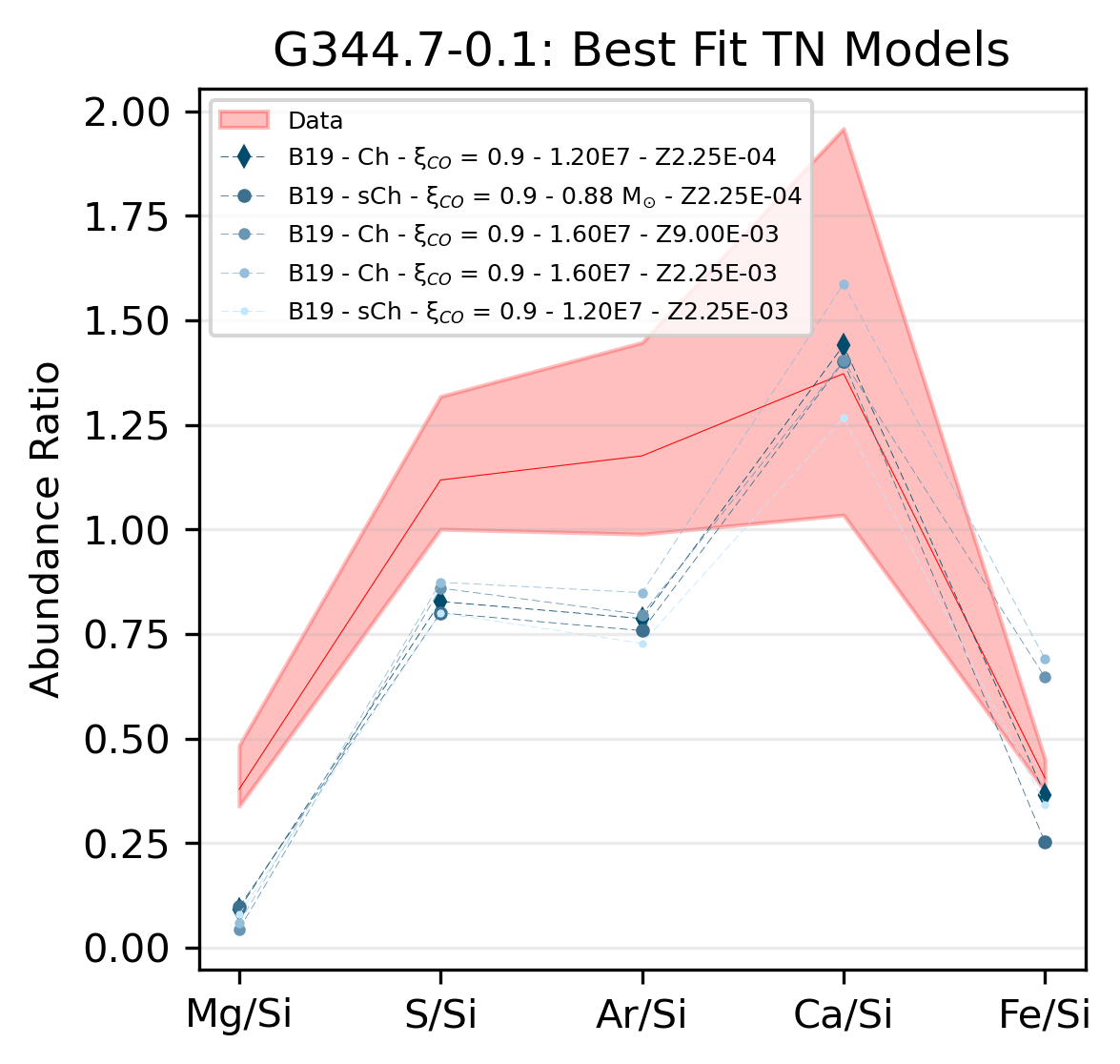}}
		\subfloat[G352.7--0.1]{\includegraphics[angle=0,height=0.48\textwidth,scale=0.5]{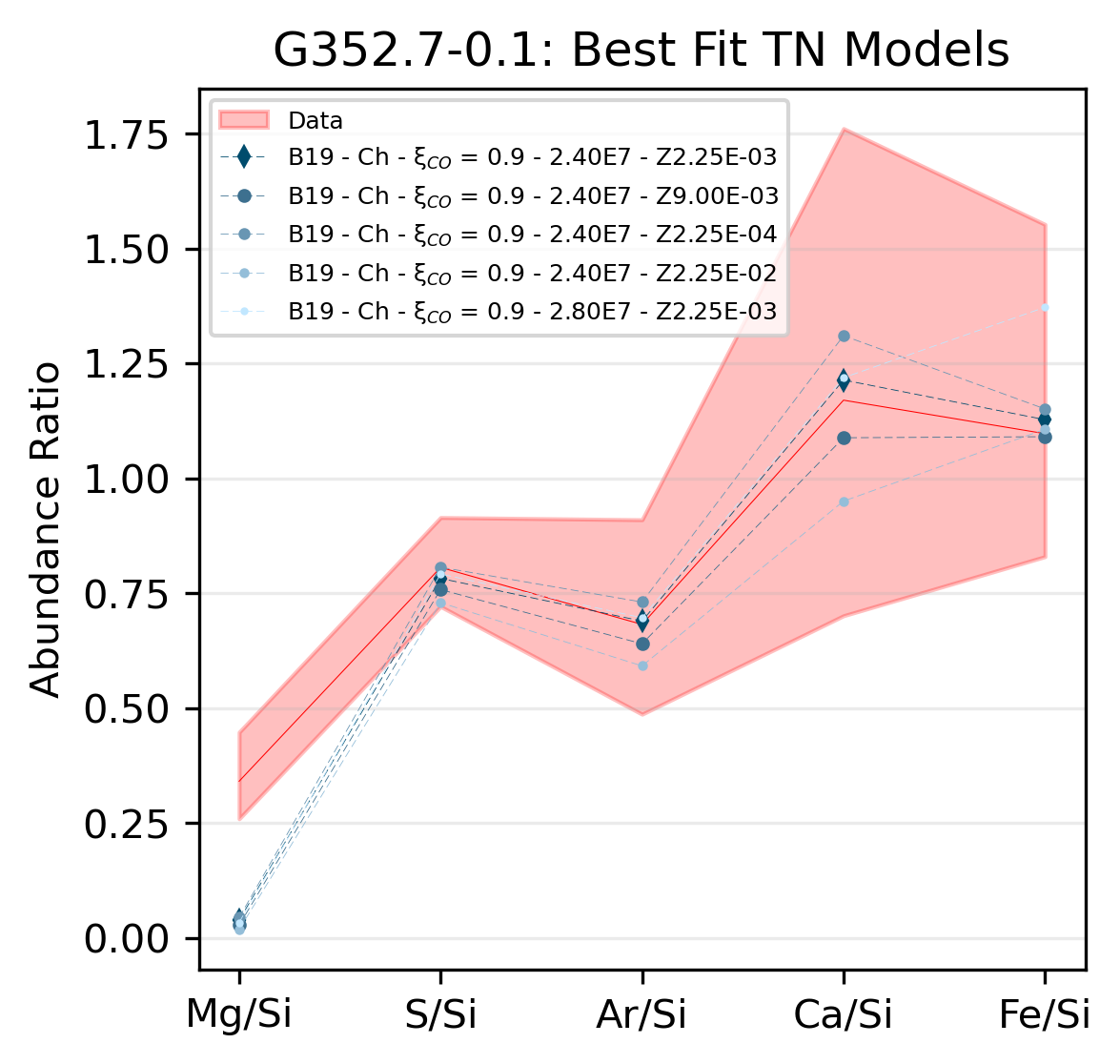}} \\
		\subfloat[0505--67.9]{\includegraphics[angle=0,height=0.48\textwidth,scale=0.5]{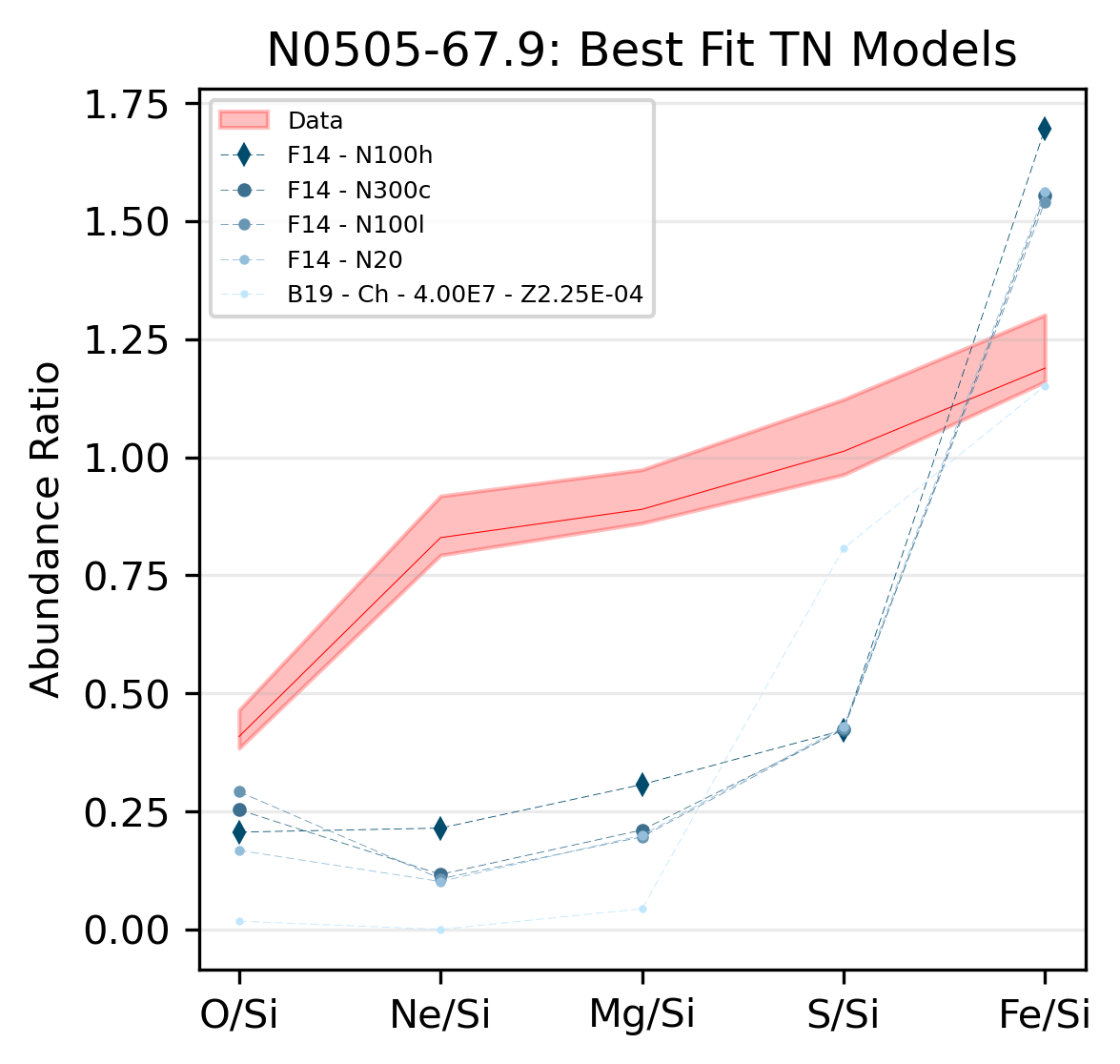}}
		\subfloat[0509--67.5]{\includegraphics[angle=0,height=0.48\textwidth]{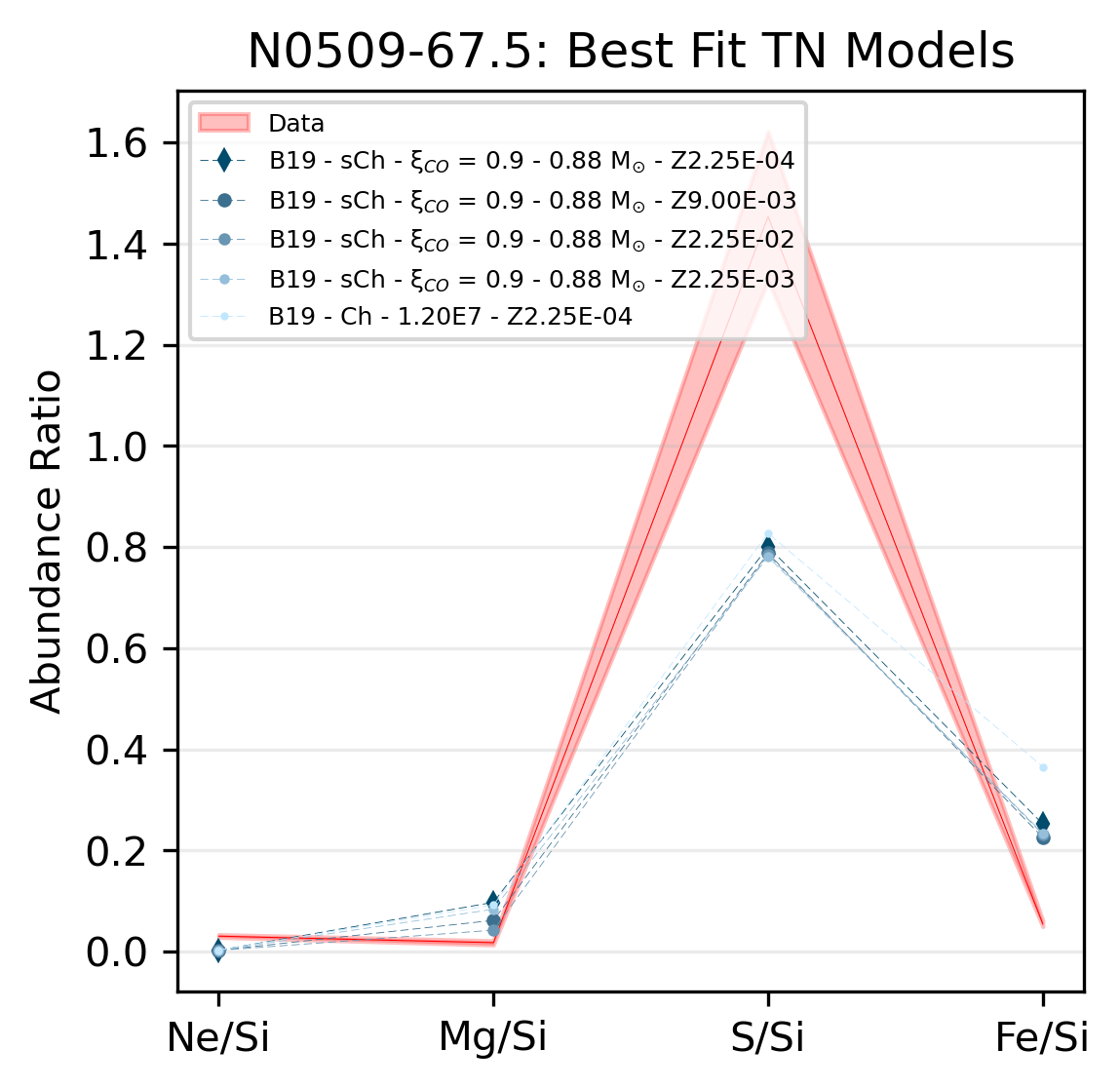}}
	\end{center}
    \caption{Continued from above.}
\end{figure*}

\begin{figure*}
    \ContinuedFloat
	\begin{center}
		\subfloat[0509--68.7]{\includegraphics[angle=0,height=0.48\textwidth]{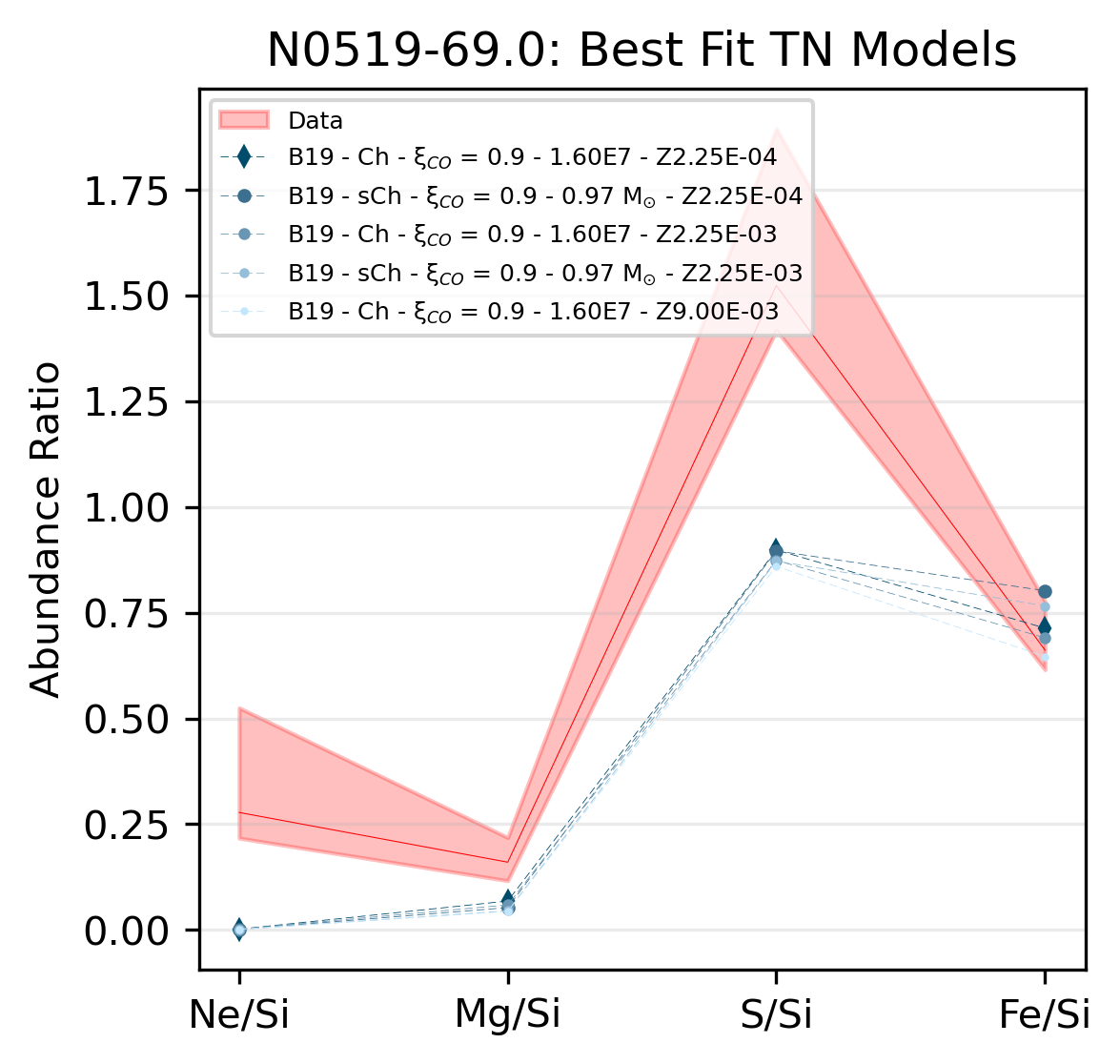}}
		\subfloat[0519--69.0]{\includegraphics[angle=0,height=0.48\textwidth,scale=0.5]{N0519_bestFitIaModels.png}} \\
		\subfloat[0534--69.9]{\includegraphics[angle=0,height=0.48\textwidth]{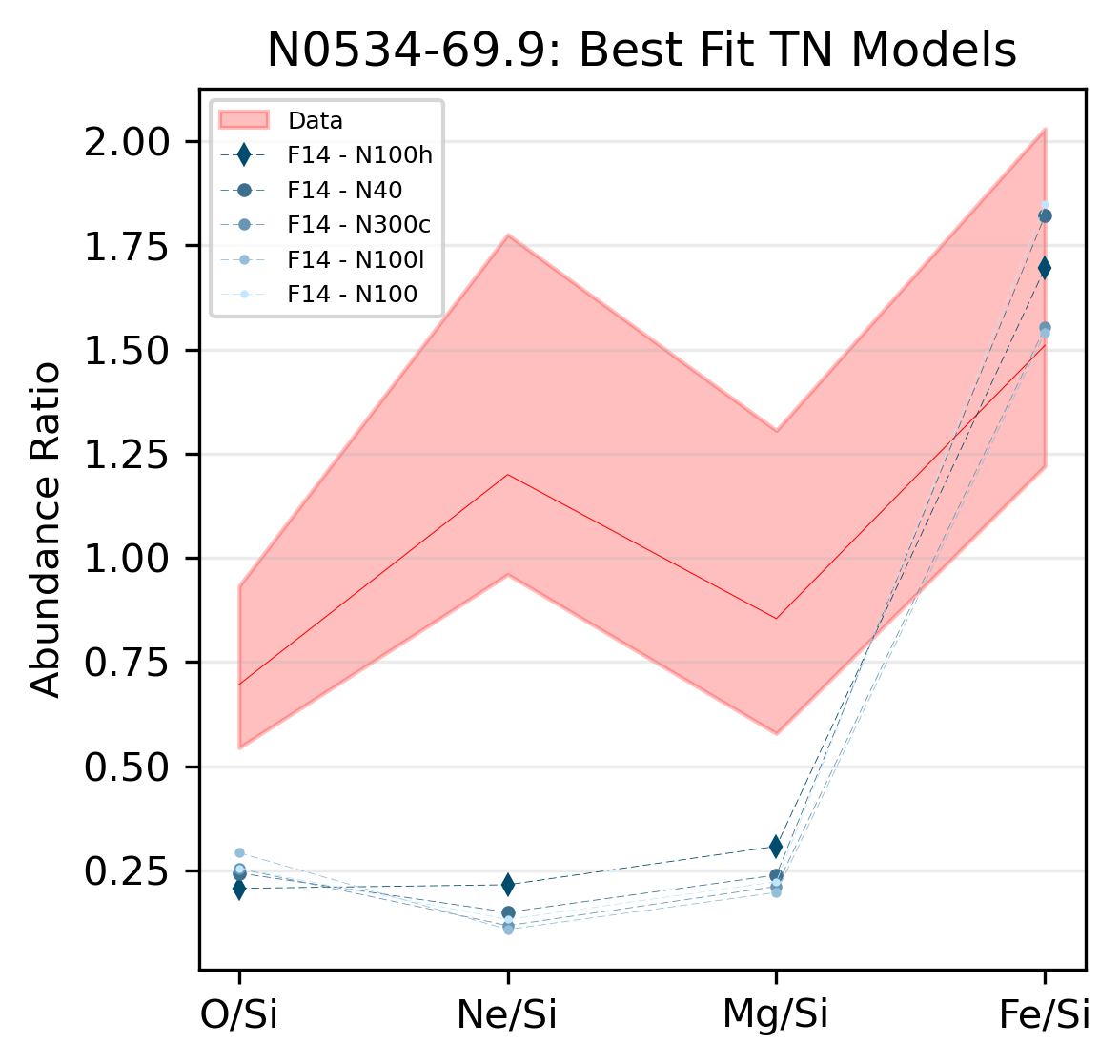}}
		\subfloat[0548--70.4]{\includegraphics[angle=0,height=0.48\textwidth,scale=0.5]{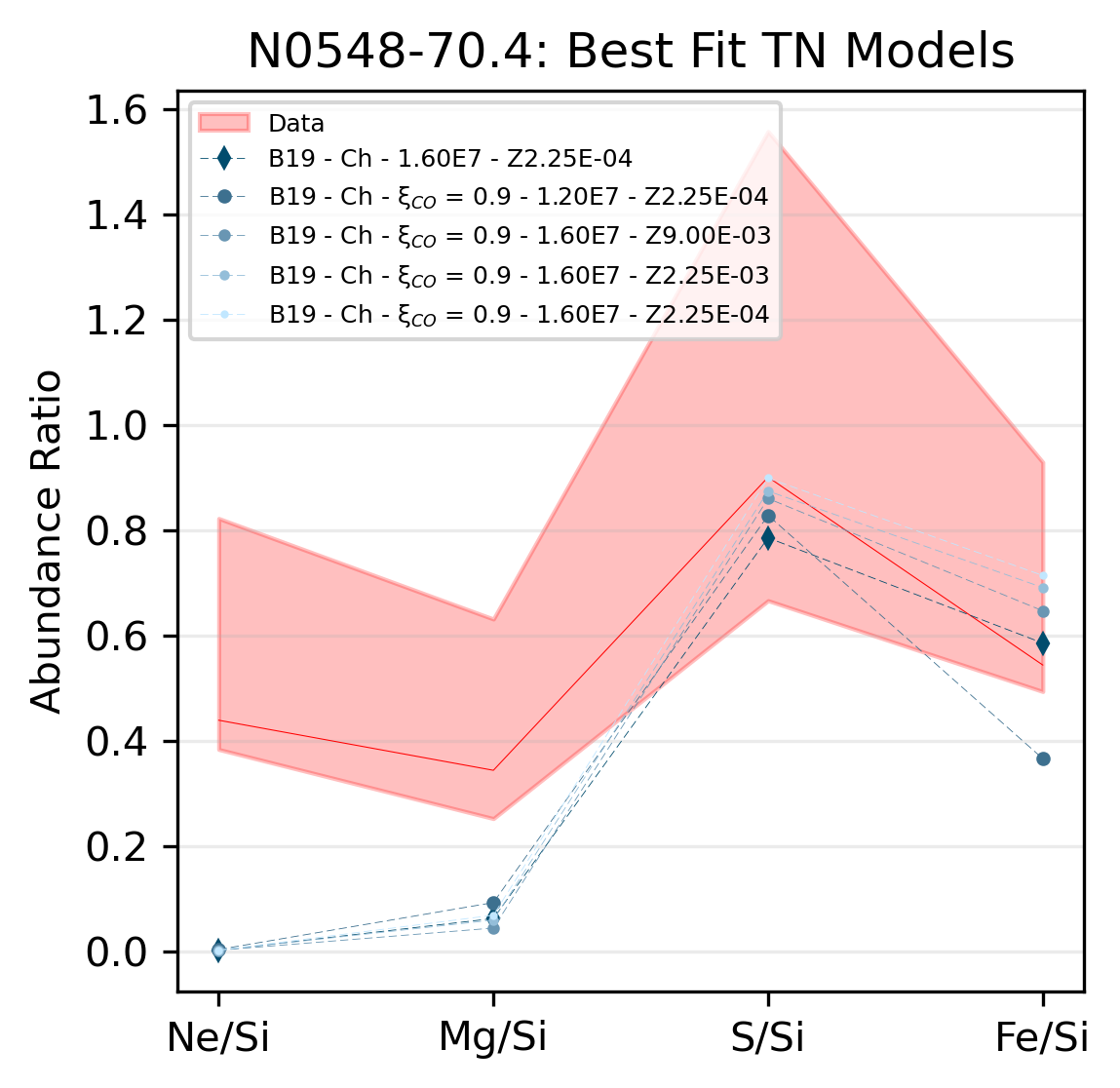}}
	\end{center}
    \caption{Continued from above.}
\end{figure*}

\begin{figure*}
	\begin{center}
		\subfloat[G1.9+0.3]{\includegraphics[angle=0,height=0.48\textwidth,scale=0.5]{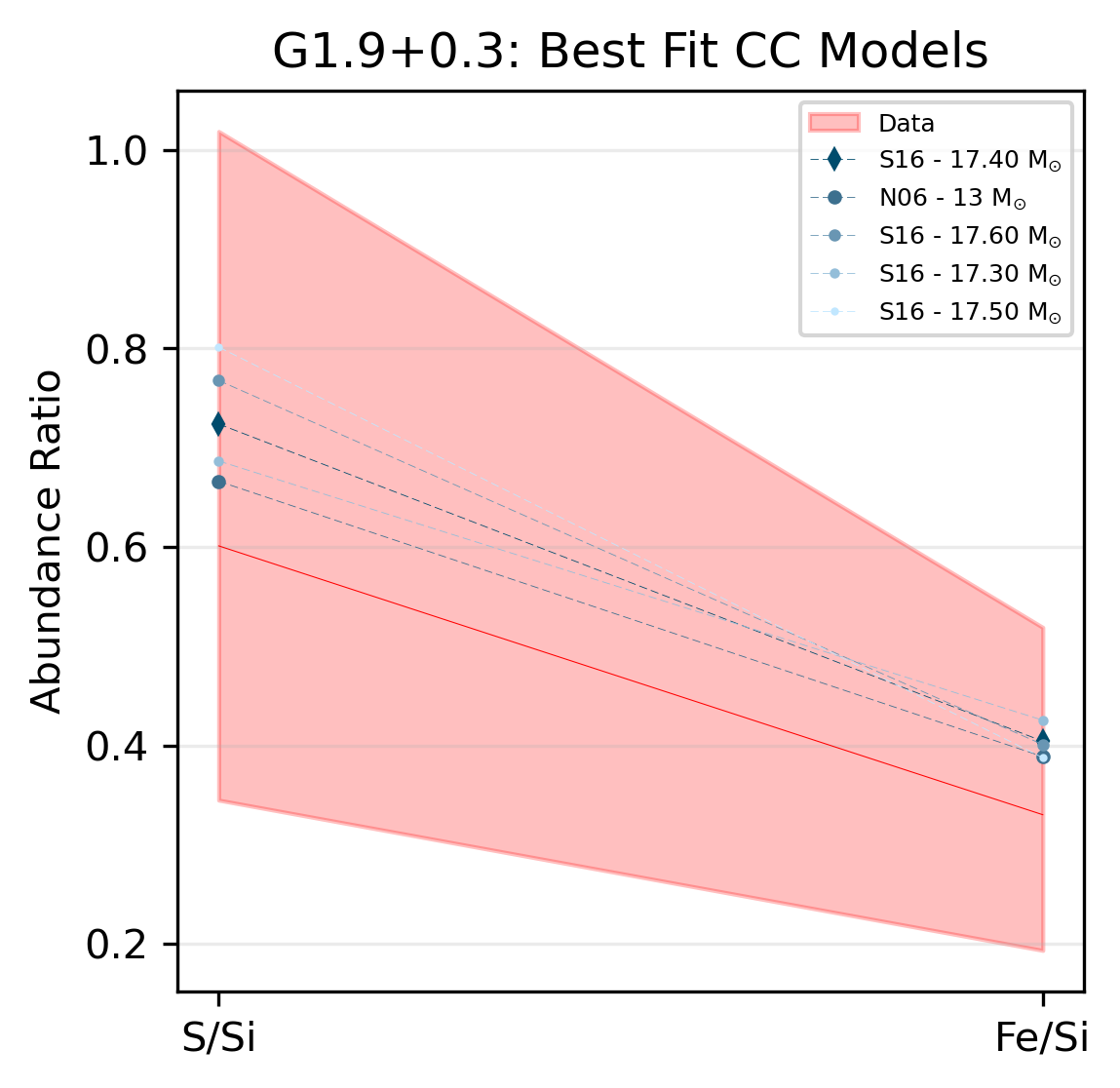}}
		\subfloat[G4.5+6.8]{\includegraphics[angle=0,height=0.48\textwidth,scale=0.5]{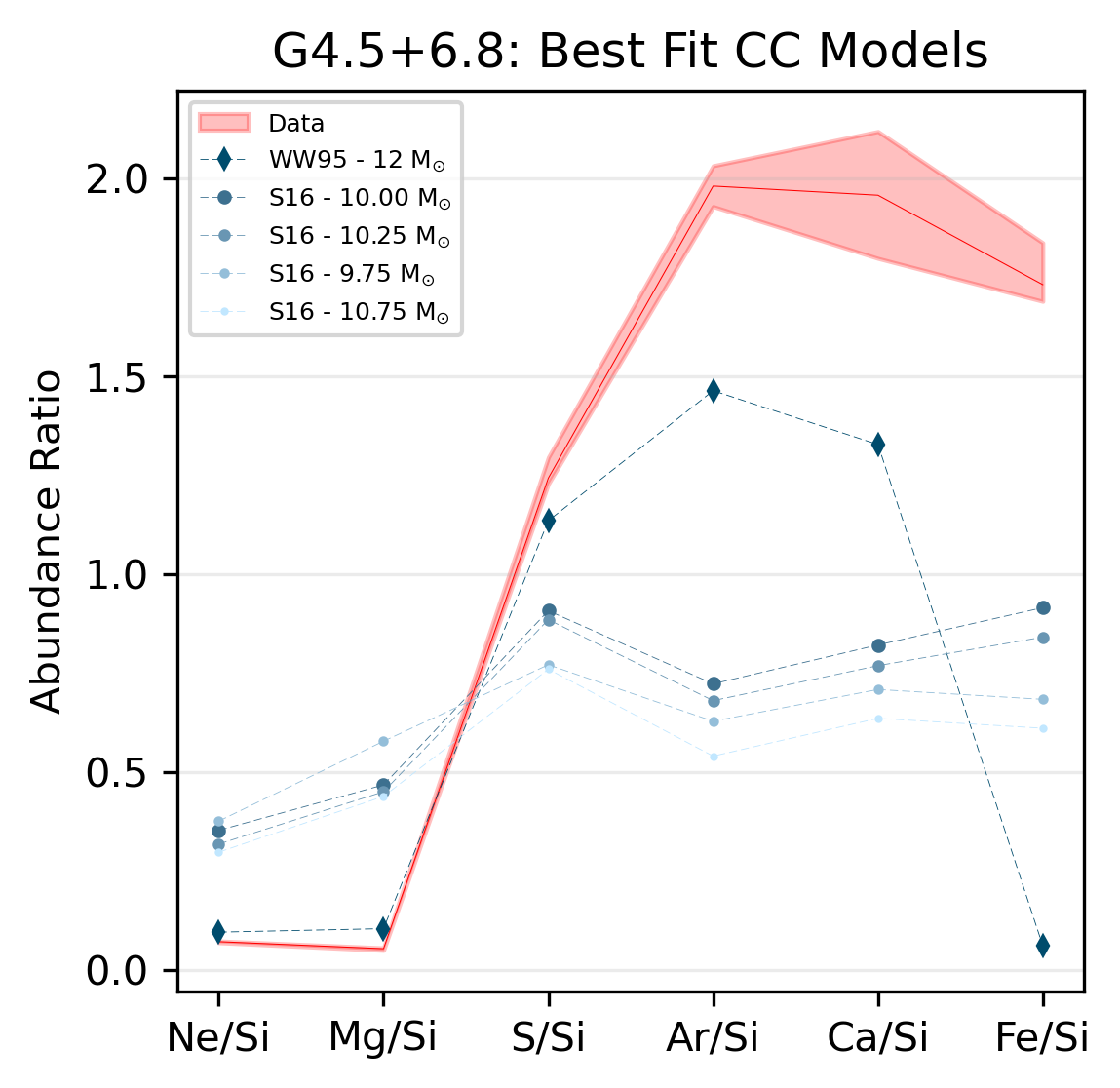}}\\
		\subfloat[G272.2--3.2]{\includegraphics[angle=0,height=0.48\textwidth]{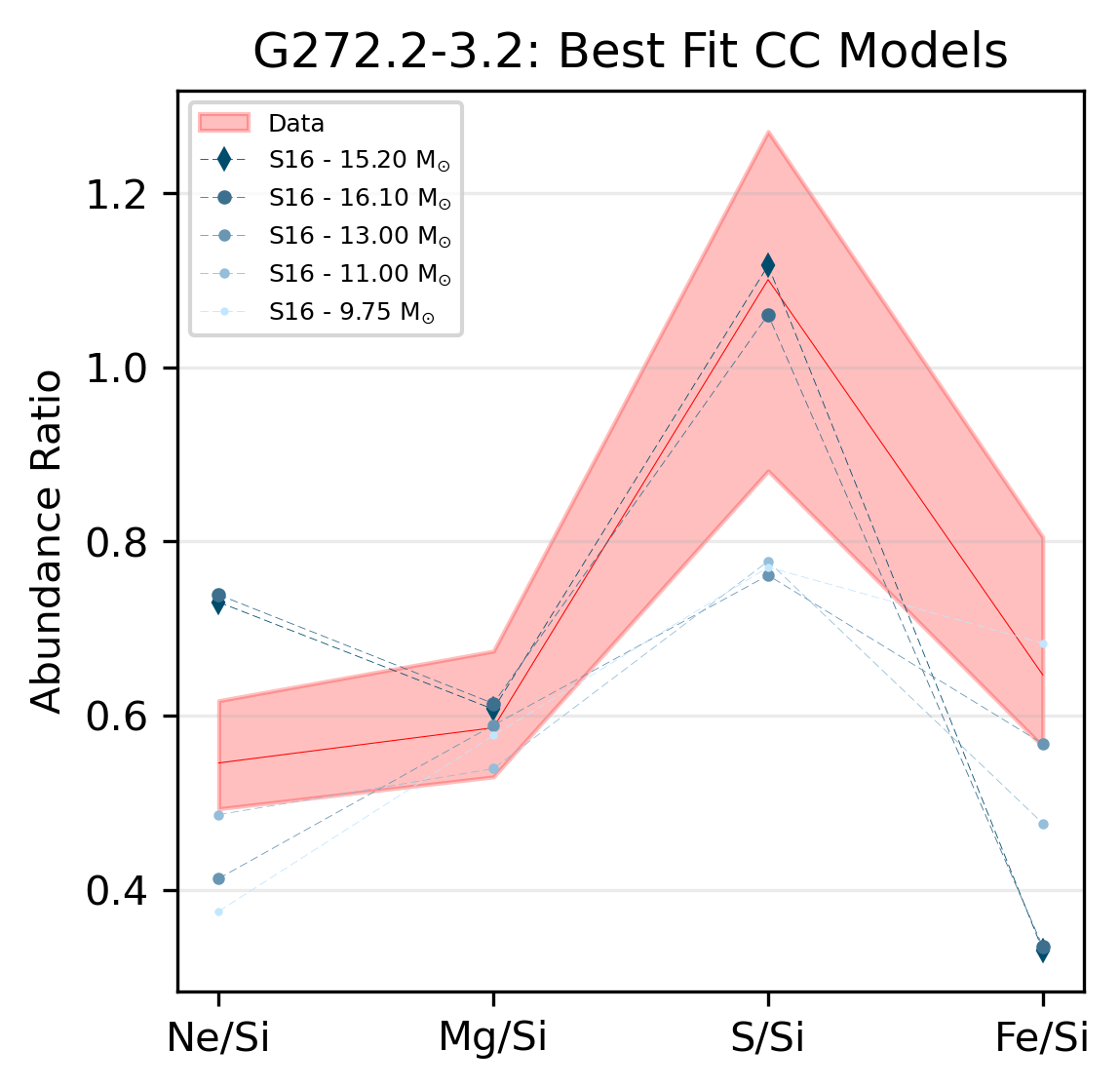}}
		\subfloat[G337.2--0.7]{\includegraphics[angle=0,height=0.48\textwidth,scale=0.5]{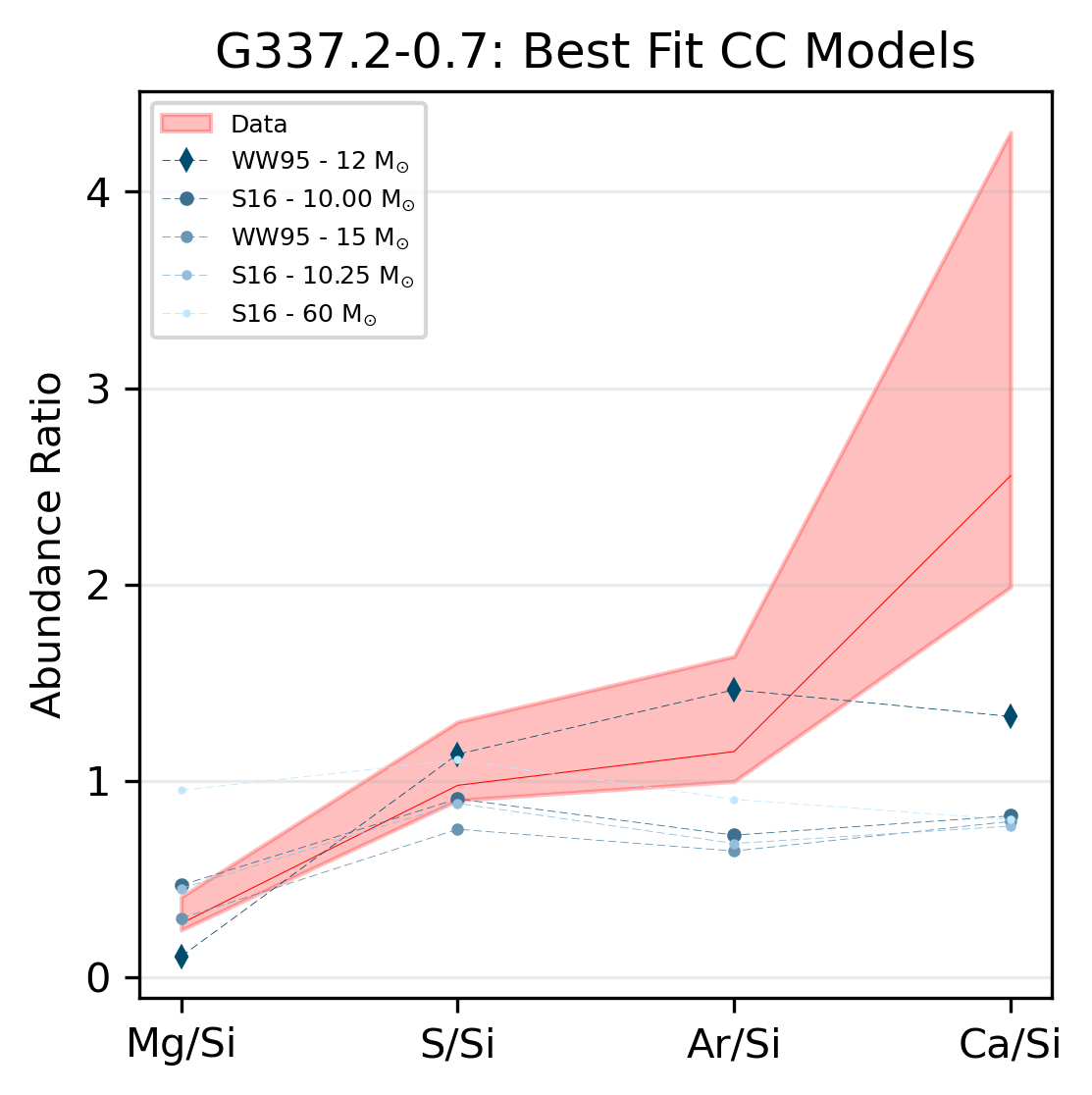}}
	\end{center}
    \caption{Comparisons between observational data and the best-fit models of nucleosynthesis in core-collapse supernovae for all objects. Shaded red areas represent observational data, while colored points represent abundance ratios from a given model and are listed in order of decreasing quality-of-fit. See \S\ref{ssec:core-collapse} and Table~\ref{tab:cc_model_summary} details of the models.}
    \label{fig:cc_nucsyn_plots}
\end{figure*}

\begin{figure*}
    \ContinuedFloat
	\begin{center}
		\subfloat[G344.7--0.1]{\includegraphics[angle=0,height=0.48\textwidth]{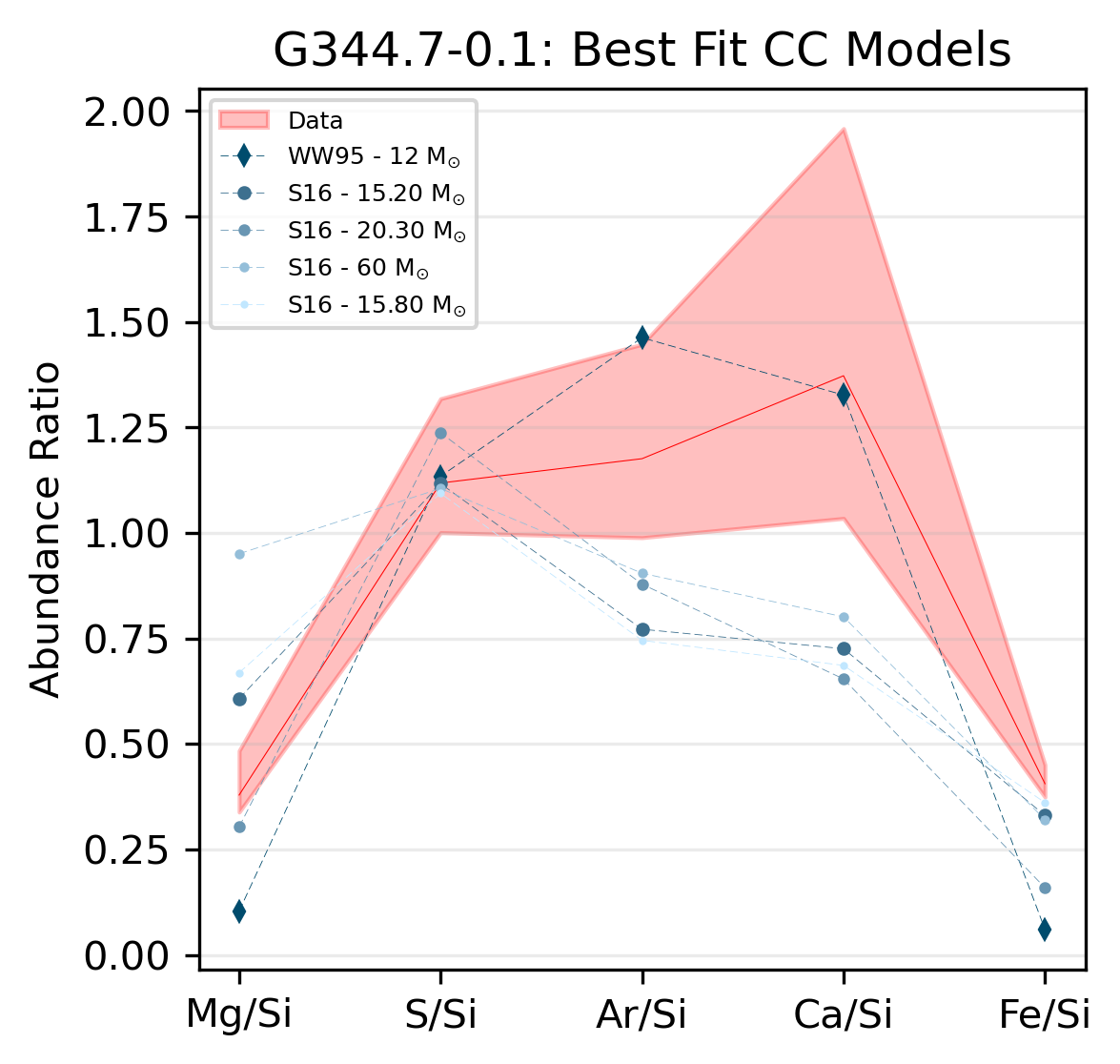}}
		\subfloat[G352.7--0.1]{\includegraphics[angle=0,height=0.48\textwidth,scale=0.5]{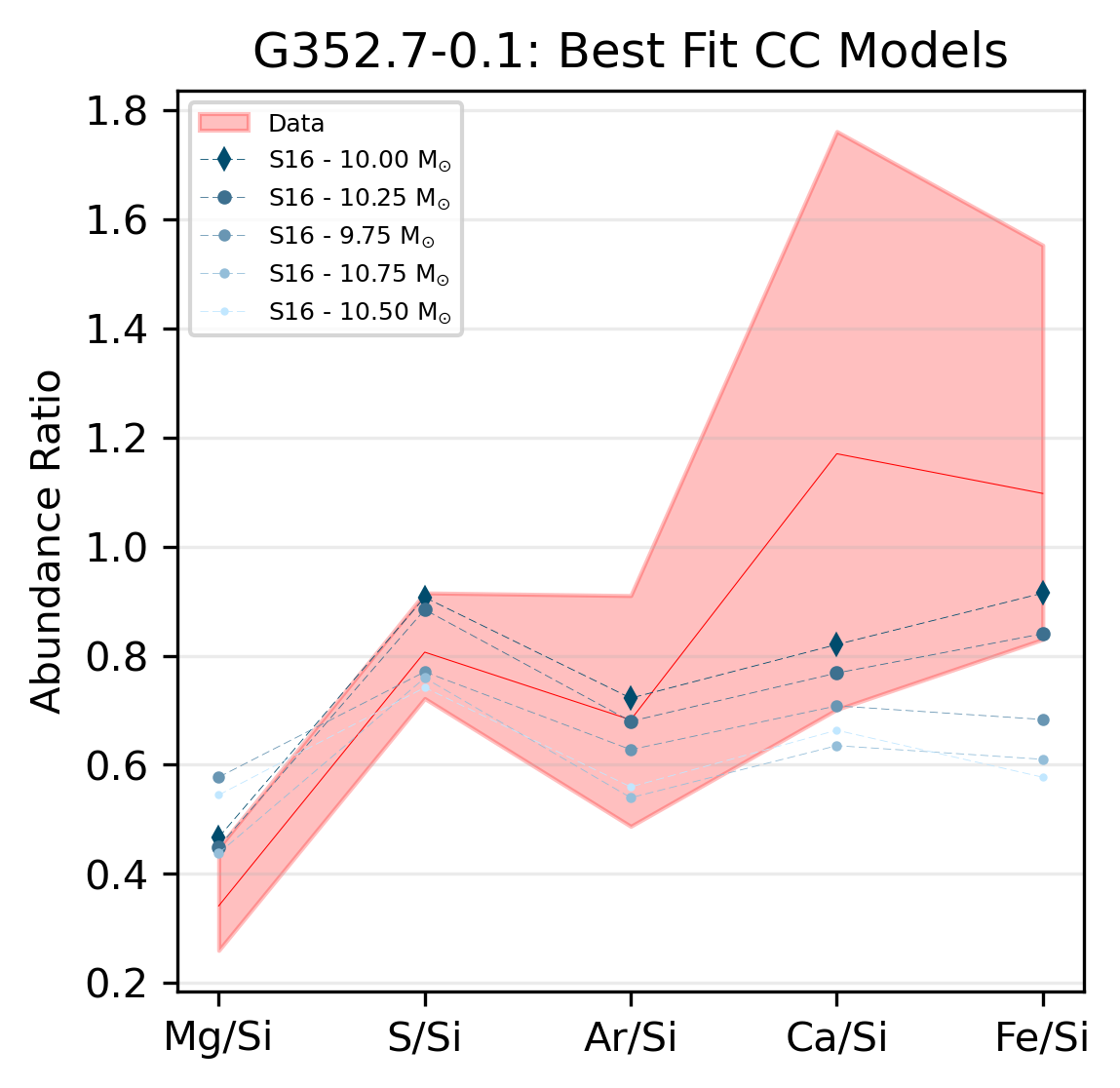}} \\
		\subfloat[0505--67.9]{\includegraphics[angle=0,height=0.48\textwidth,scale=0.5]{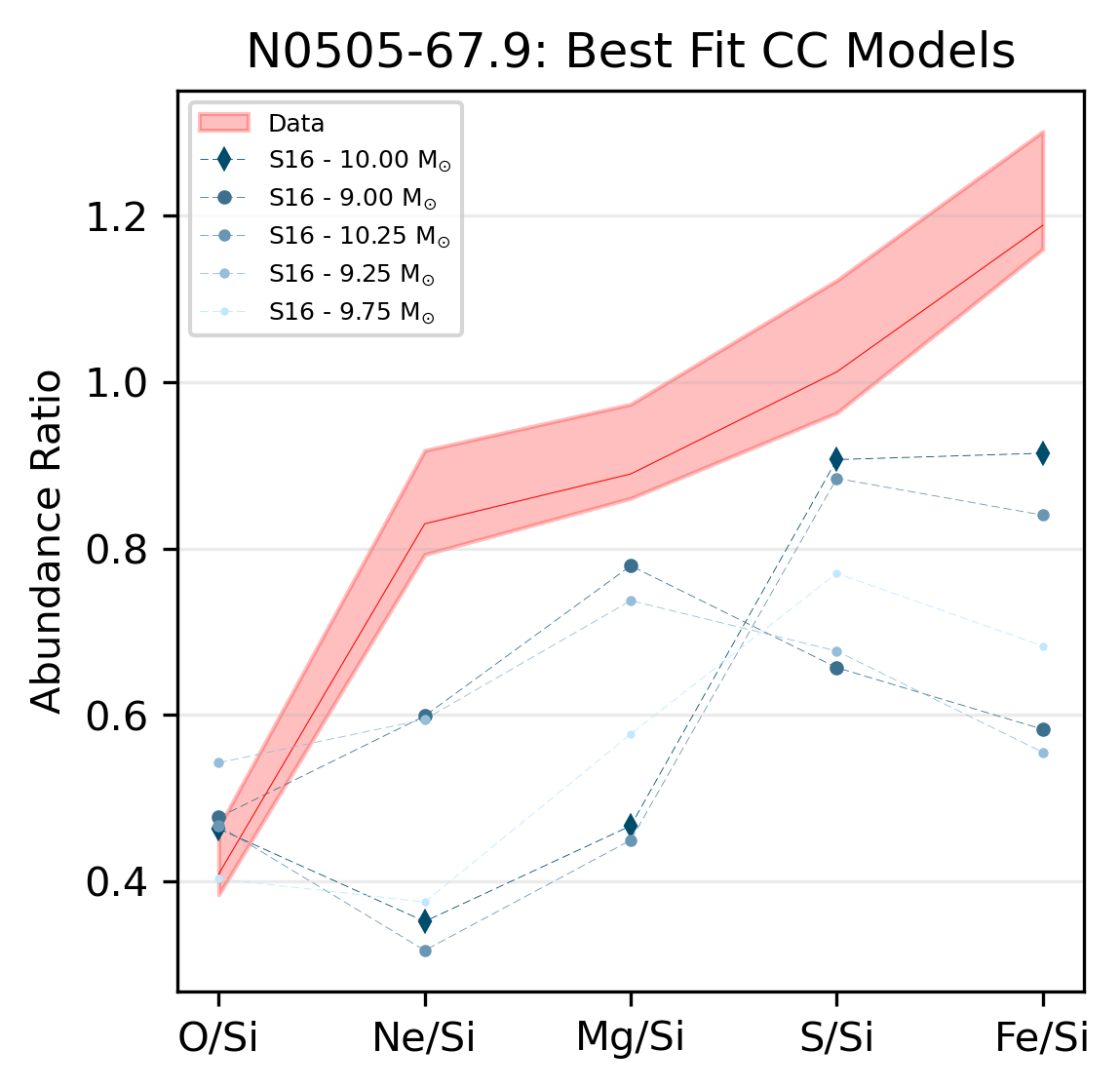}}
		\subfloat[0509--67.5]{\includegraphics[angle=0,height=0.48\textwidth]{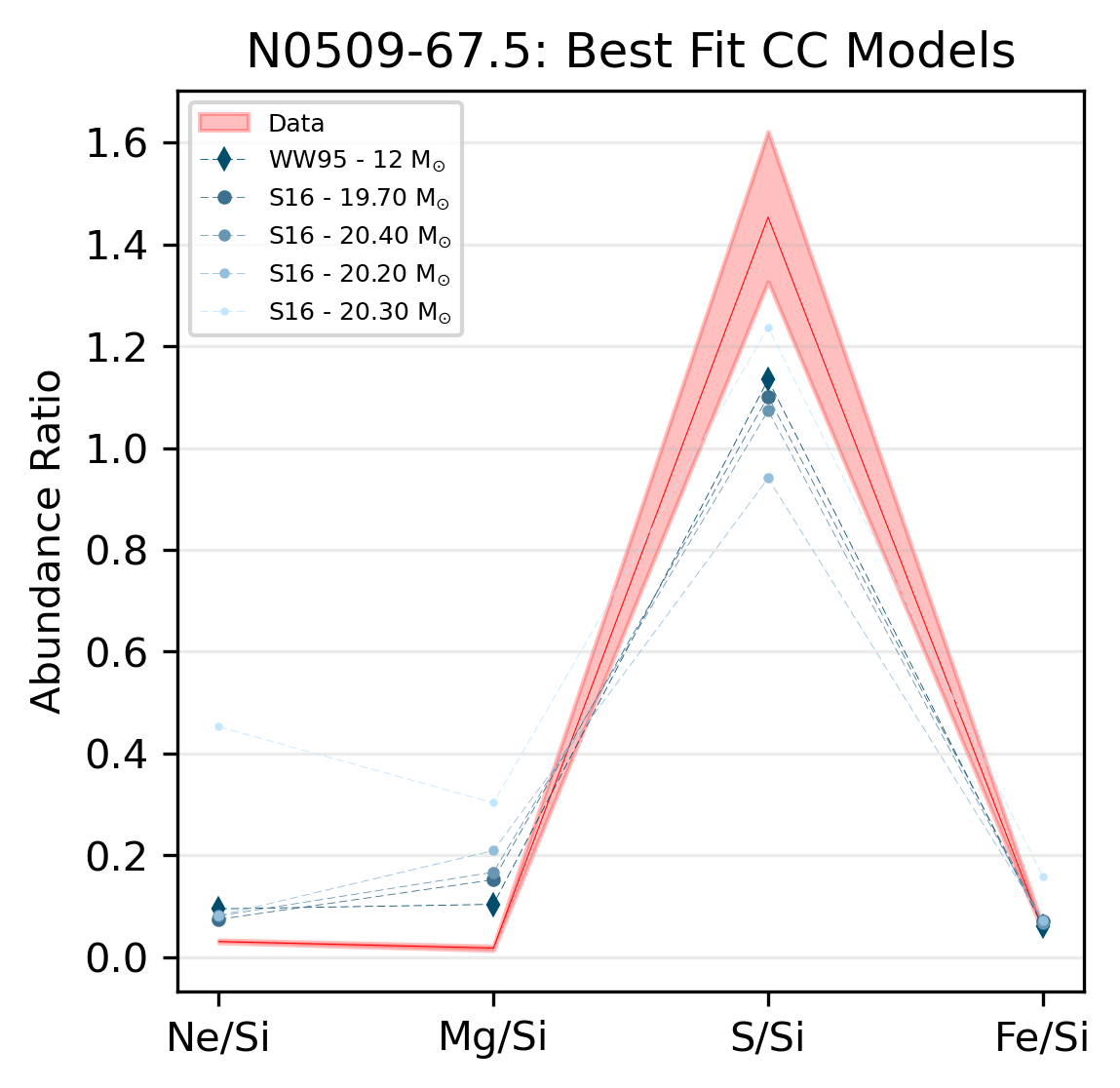}}
	\end{center}
    \caption{Continued from above.}
\end{figure*}

\begin{figure*}
    \ContinuedFloat
	\begin{center}
		\subfloat[0509--68.7]{\includegraphics[angle=0,height=0.48\textwidth]{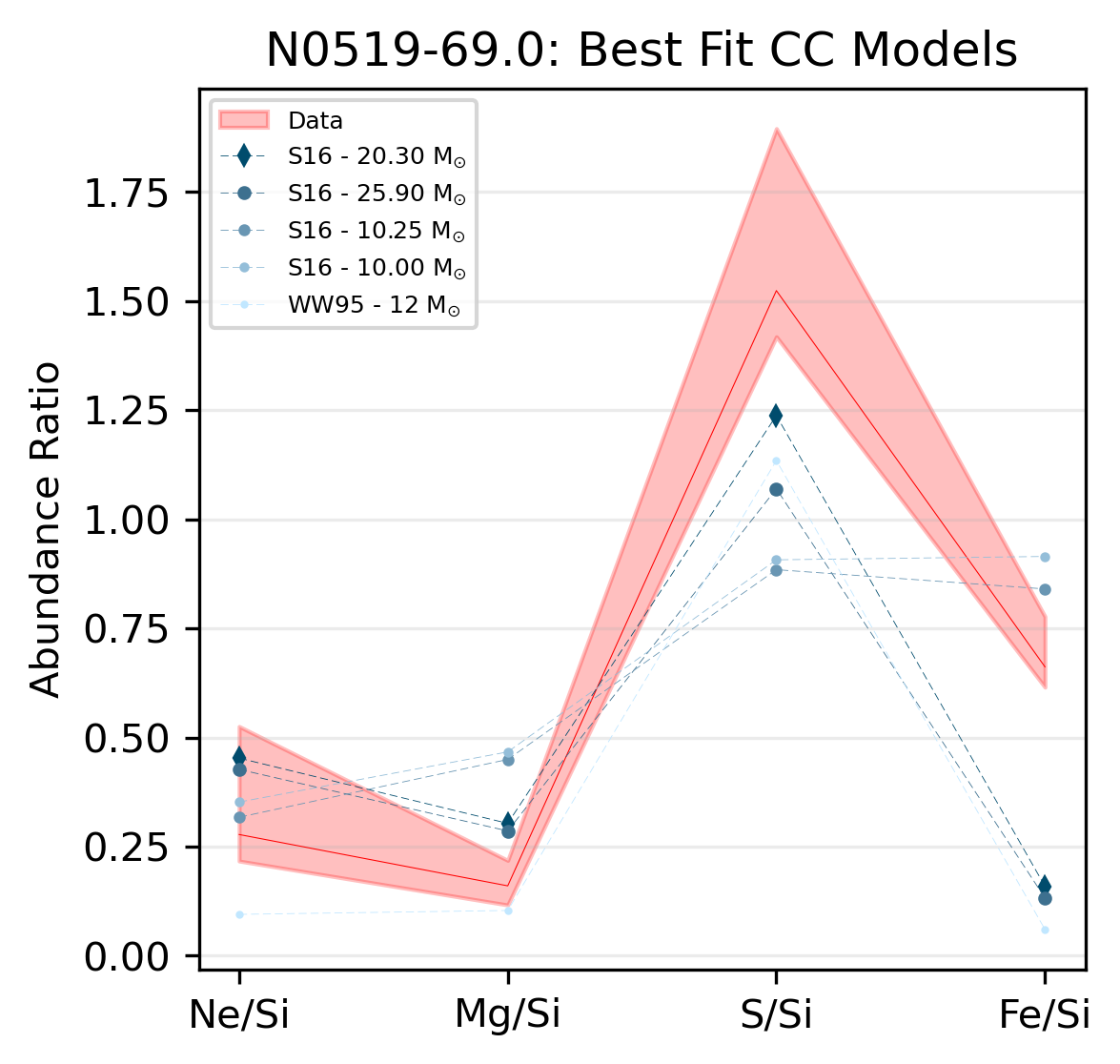}}
		\subfloat[0519--69.0]{\includegraphics[angle=0,height=0.48\textwidth,scale=0.5]{N0519_bestFitCCModels.png}} \\
		\subfloat[0534--69.9]{\includegraphics[angle=0,height=0.48\textwidth]{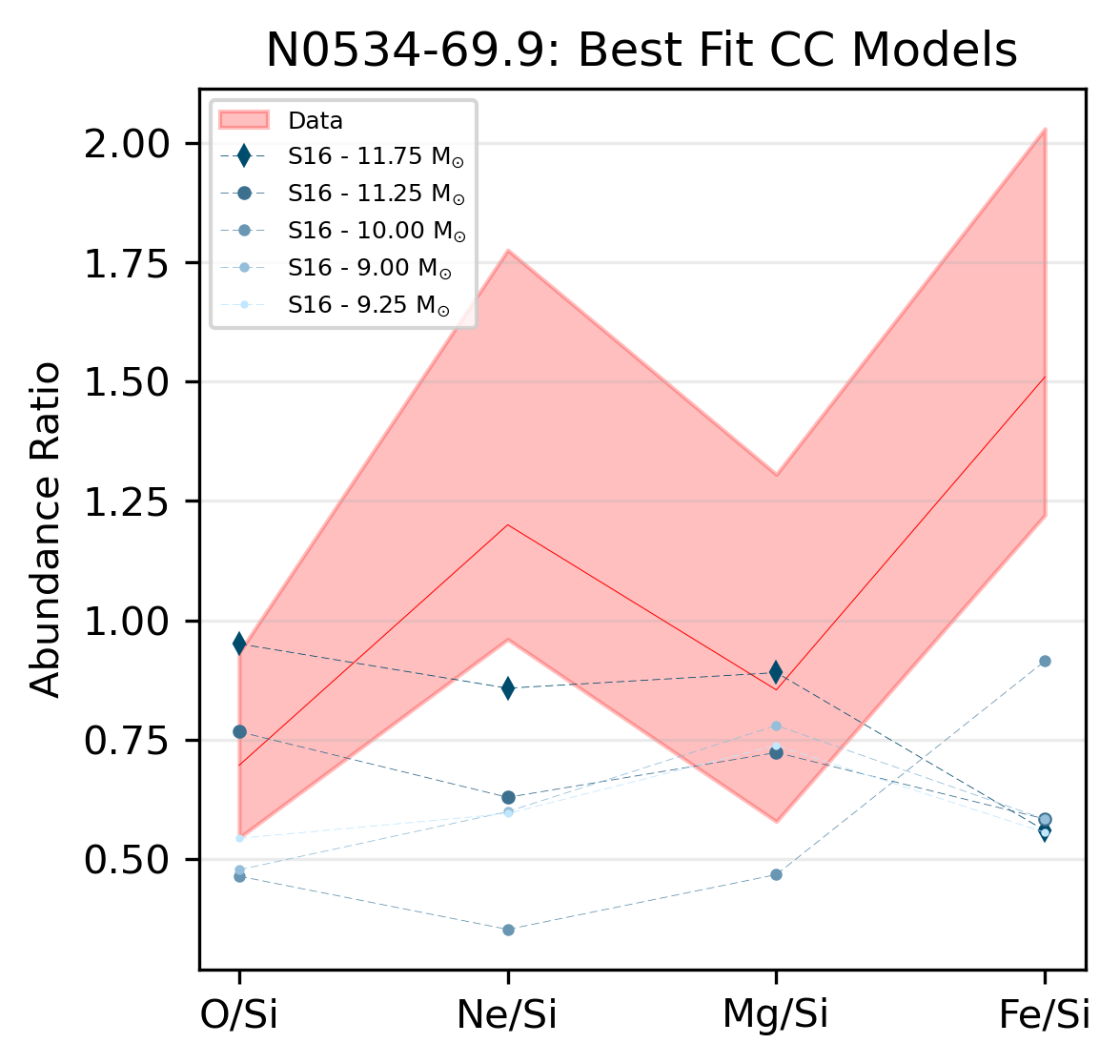}}
		\subfloat[0548--70.4]{\includegraphics[angle=0,height=0.48\textwidth,scale=0.5]{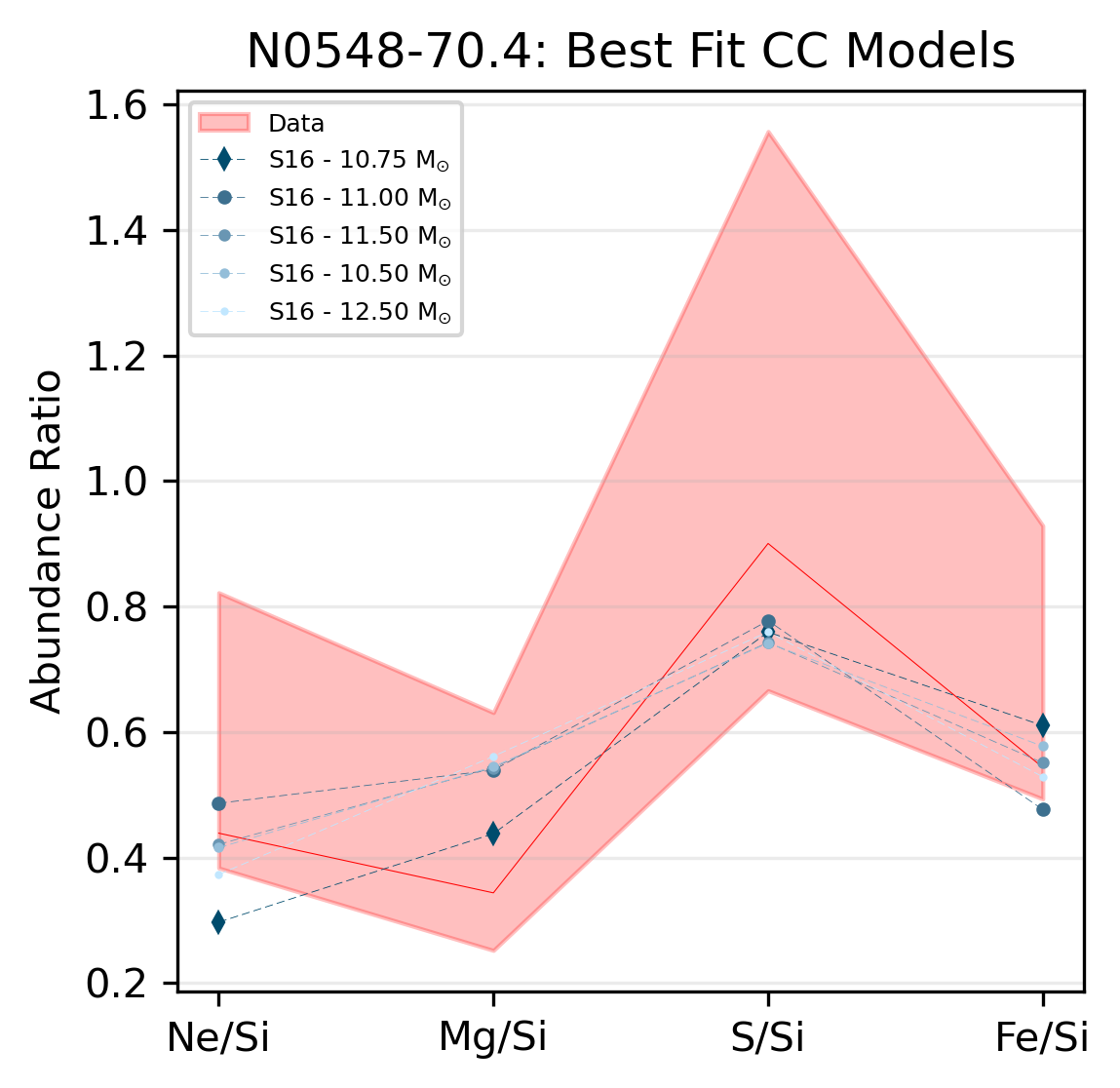}}
	\end{center}
    \caption{Continued from above.}
\end{figure*}

\begin{figure*}
	\begin{center}
		\subfloat[G120.1+1.4 - Ia - Cold Ca]{\includegraphics[angle=0,height=0.48\textwidth,scale=0.5]{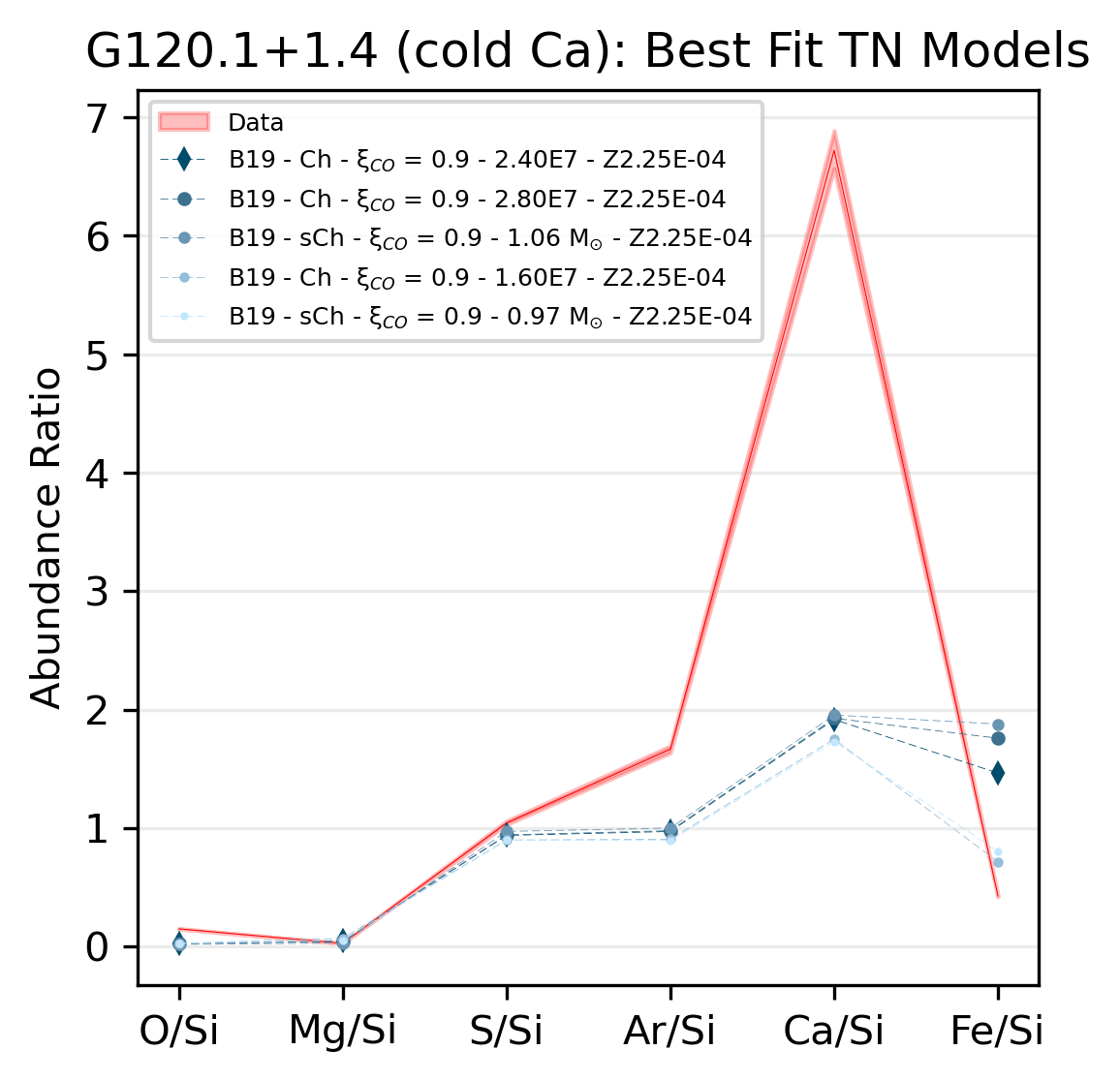}}
		\subfloat[G120.1+1.4 - CC - Cold Ca]{\includegraphics[angle=0,height=0.48\textwidth,scale=0.5]{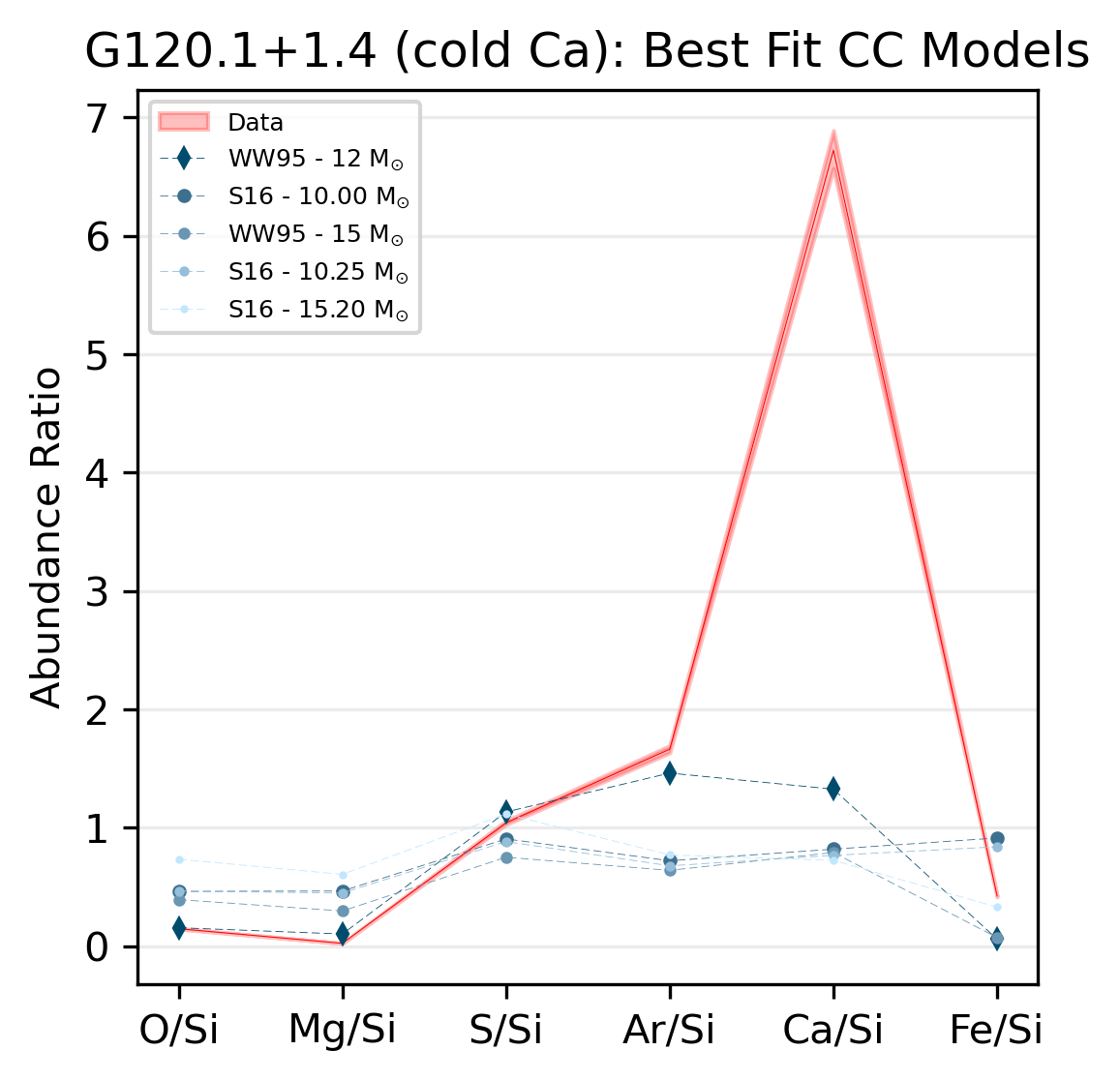}}\\
		\subfloat[G120.1+1.4 - Ia - Hot Ca]{\includegraphics[angle=0,height=0.48\textwidth]{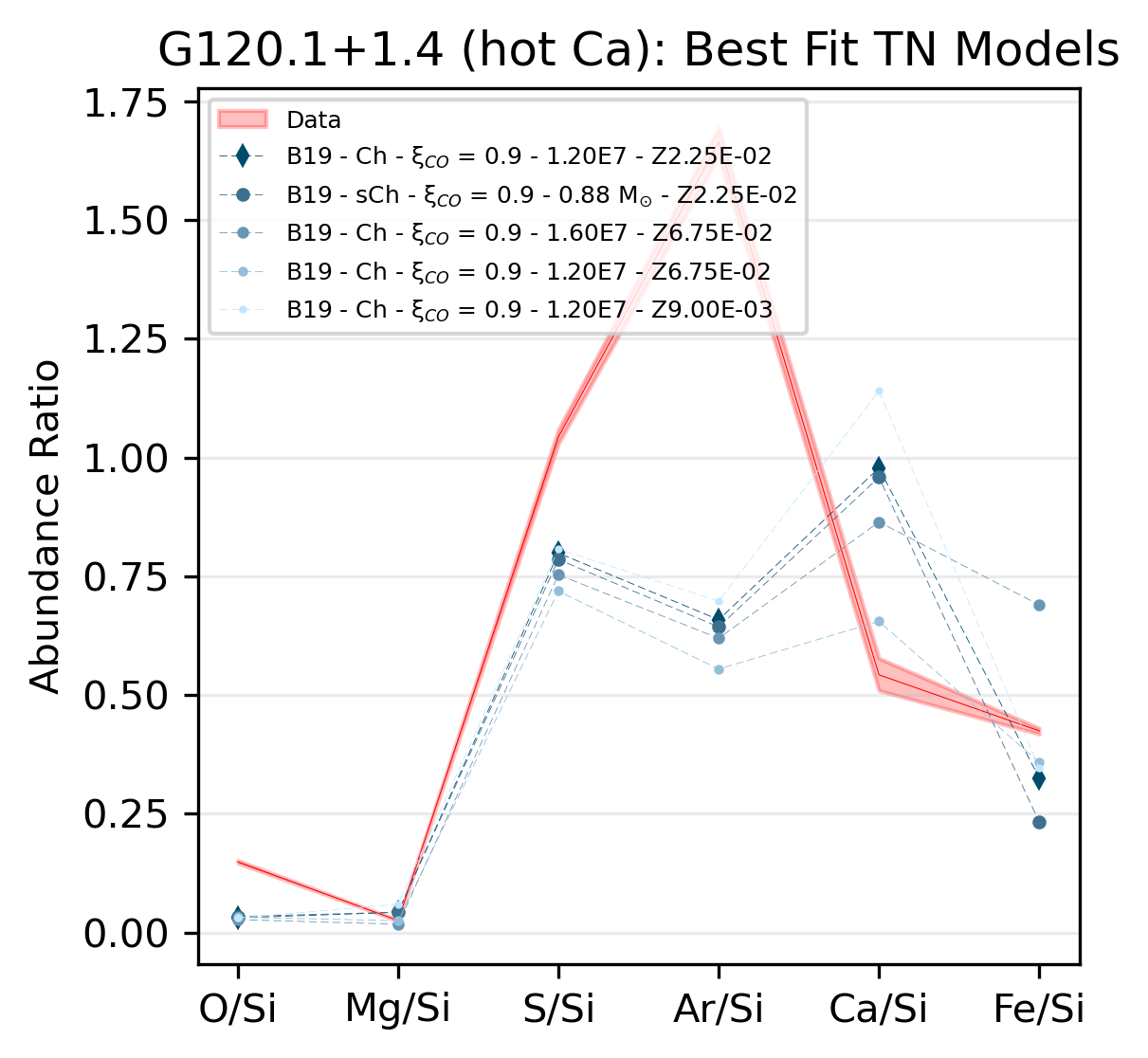}}
		\subfloat[G120.1+1.4 - CC - Hot Ca]{\includegraphics[angle=0,height=0.48\textwidth,scale=0.5]{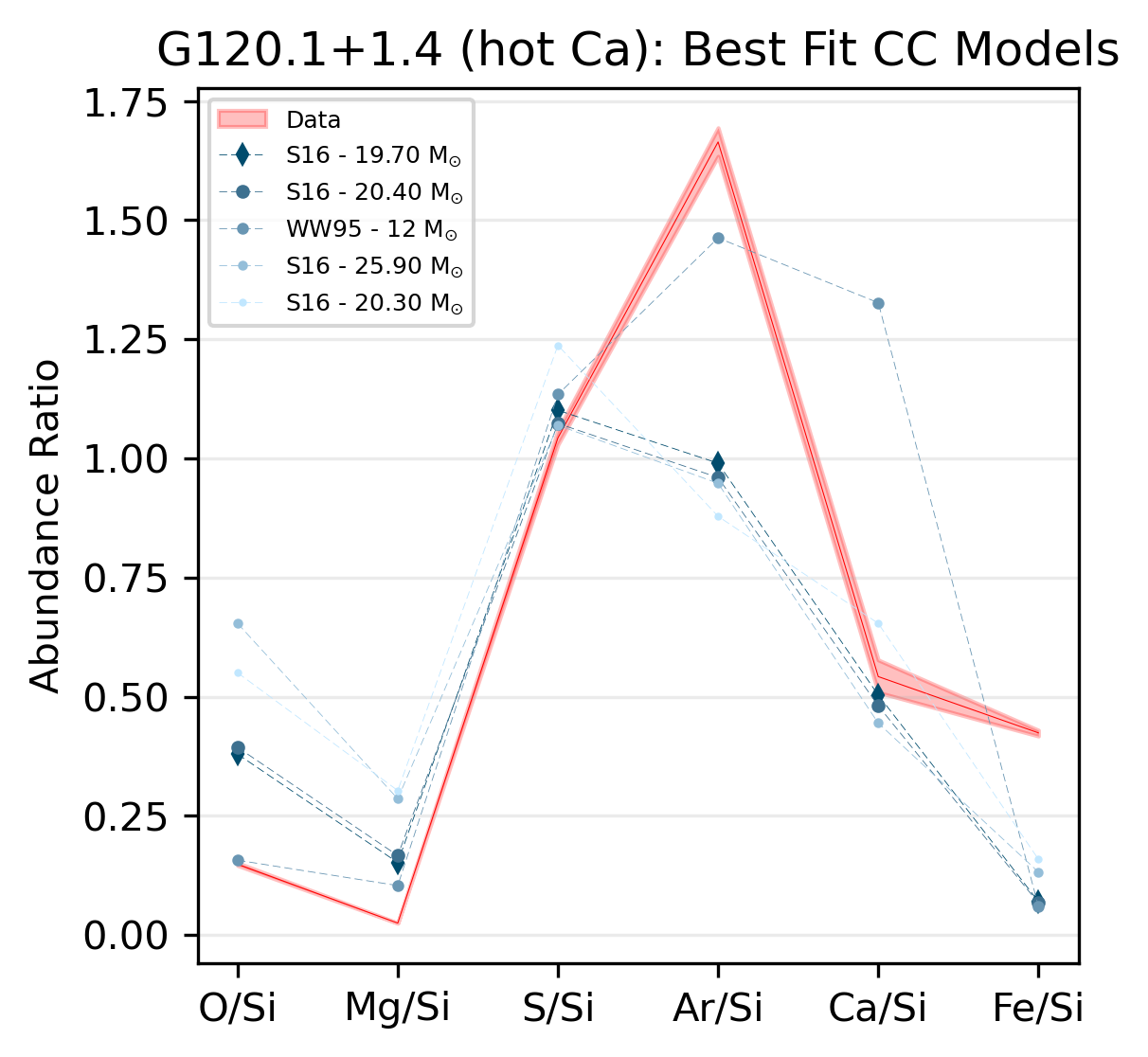}}
	\end{center}
    \caption{Comparisons between observational data for Tycho's SNR (G120.1+1.4) and the best-fit models of nucleosynthesis for both Type Ia and core-collapse supernovae. Shaded red areas represent observational data, while colored points represent abundance ratios from a given model and are listed in order of decreasing quality-of-fit. See \S\ref{sec:nucsyn_models} and Tables~\ref{tab:model_summary} and \ref{tab:cc_model_summary} for details of the models.}
    \label{fig:tycho_nucsyn_plots}
\end{figure*}

\begin{figure*}
    \ContinuedFloat
	\begin{center}
		\subfloat[G120.1+1.4 - Ia - No Ca]{\includegraphics[angle=0,height=0.48\textwidth]{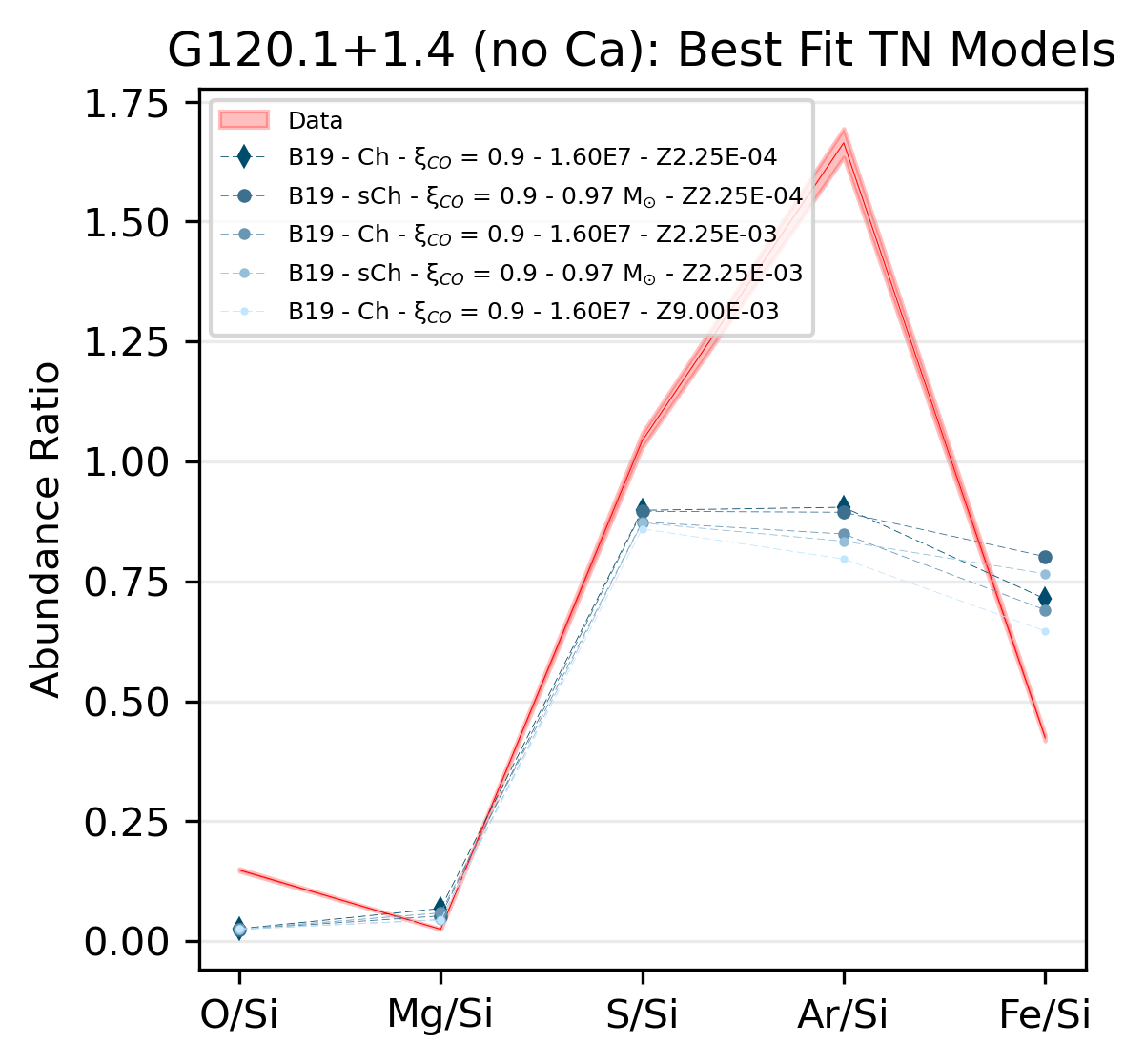}}
		\subfloat[G120.1+1.4 - CC - No Ca]{\includegraphics[angle=0,height=0.48\textwidth,scale=0.5]{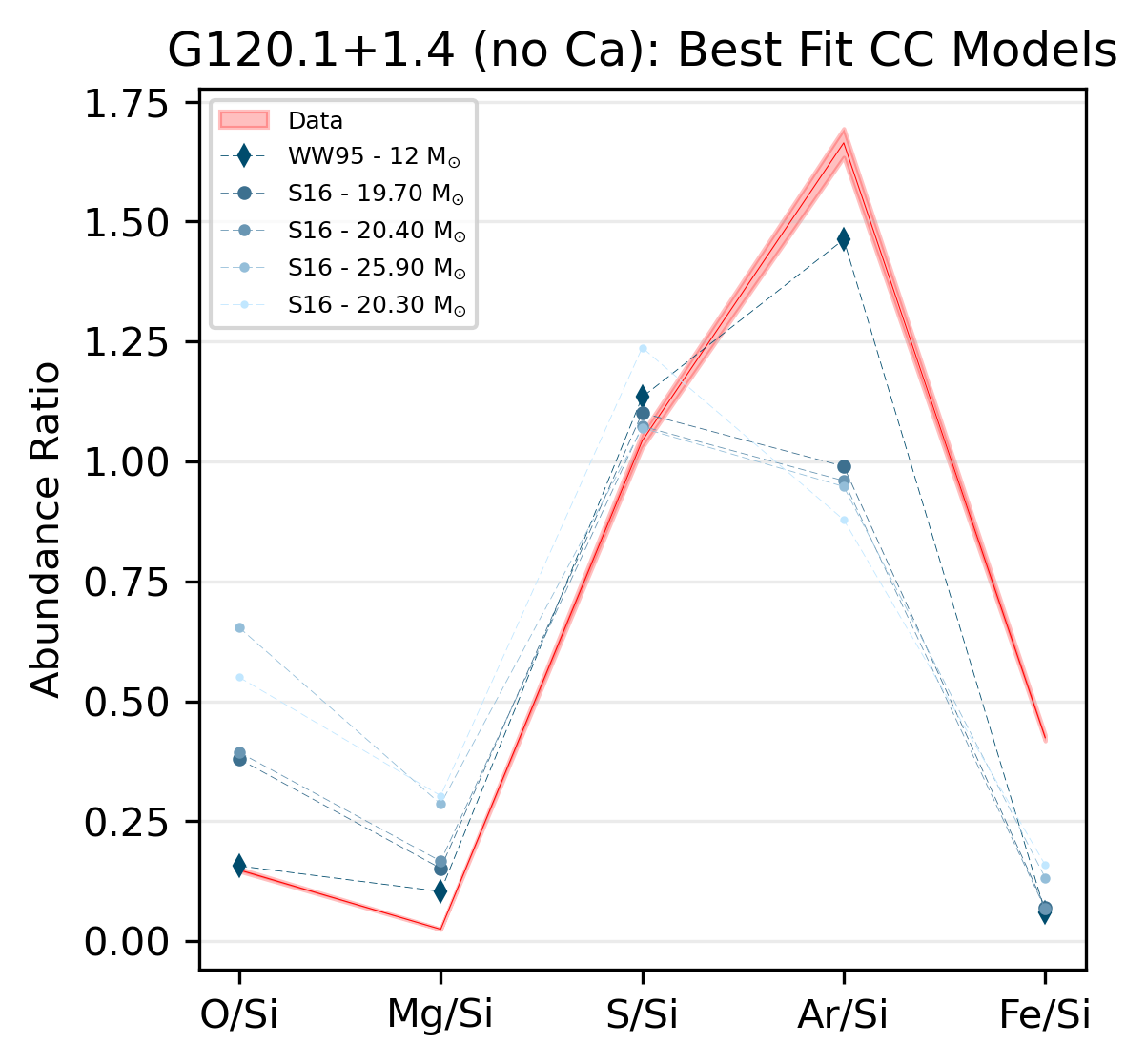}}
	\end{center}
    \caption{Continued from above.}
\end{figure*}

\section{The Suite of Nucleosynthesis Models}
\label{sec:nucsyn_models}
While supernova explosions can be broadly classified in terms of the two main types (core-collapse vs Type Ia), there is a significant amount of variation within these categories, and no single model can account for all of the sub-classifications that exist. In the case of Type Ia explosions, the effects of mass (near-$M_{Ch}$ vs. sub-$M_{Ch}$), degeneracy (SD vs. DD), and even explosion mechanism (deflagration vs detonation / delayed-detonation) have a significant impact on how the supernova unfolds. Even in the case of core-collapse explosions, the progenitor mass will have a significant impact on how the supernova proceeds, and on the end-products of the resulting supernova nucleosynthesis.

When modeling a supernova, it is necessary to take into account this diversity. Modelers do this by performing hydrodynamics simulations of supernova explosions, which attempt to capture the complexity involved in these events. However, due to the details and scales involved in such phenomena, and the typically limited amount of computational power available at a given time, it is necessary to make simplifications. One way by which this is done is through the use of the tracer particle method, in which discrete ``tracer" particles are placed throughout the exploding star. In the tracer particle method, the hydrodynamic simulation proceeds as usual, but it is not directly coupled to a comprehensive reaction network (RN). Instead, a simplified network, typically involving only a few isotopes, is used at this stage to approximate the energy release from nuclear burning during the simulation. At the beginning of the simulation, a set of tracer particles is introduced, typically distributed evenly across mass coordinates, with each particle representing an equal mass. However, particles may also be concentrated in specific regions of interest. These particles are then evolved timestep by timestep according to the local velocity field, with each particle recording the temperature and density at each step. This creates a detailed record of the evolving thermodynamic properties. After the simulation is completed, a larger reaction network can be applied in post-processing to these recorded data, rather than being integrated into the simulation itself. By assigning an appropriate weight to each tracer particle -- usually based on its mass -- detailed nuclear yield information can be derived from the explosion's nuclear reactions.

A powerful tool for the study of supernova remnants, the comparison of the results of spectral fitting to the yields generated by various models of supernova nucleosynthesis allows us to narrow down the possible progenitor class for a given supernova remnant and potentially determine several progenitor parameters such as mass, metallicity, and (for Type Ia supernovae) degeneracy. To do so, we compare the nucleosynthesis yields from multiple models -- each of which features a different description of the supernova progenitor -- to the abundance values of the ejecta component obtained from our fitted X-ray data. Here, the abundance ratios used are with respect to Si, and are given by ($X$/Si)/($X$/Si)$_{\odot}$, with $X$ being the measured ejecta mass of the relevant element with respect to the solar values of \cite{wilms_2000}. For this work, we made use of 11 sets of nucleosynthesis models altogether: seven sets of models of Type Ia supernova nucleosynthesis, and four sets of models of core-collapse supernova nucleosynthesis. The basic properties of the thermonuclear models are summarized in Table~\ref{tab:model_summary}, while those of the core-collapse models can be found in Table~\ref{tab:cc_model_summary}.

\subsection{Type Ia Supernovae}
\label{ssec:thermonuclear}

\begin{deluxetable*}{cccccccc}
\centering
\tablecaption{Summary and comparison of the basic properties of the seven sets of models of Type Ia supernova nucleosynthesis. \label{tab:model_summary}}
\tablehead{
\colhead{Paper} & 
\colhead{\makecell{No.\\ Dim.}} & 
\colhead{\makecell{Progenitor\\ Type}} & 
\colhead{\makecell{Explosion\\ Mechanism}} & 
\colhead{\makecell{No.\\ Sims.}} & 
\colhead{\makecell{Init.\\ RN}} & 
\colhead{\makecell{Final\\ RN}} & 
\colhead{Tracers}
}
\startdata
M10 & 2 & near-$M_{Ch}$ & DDT, Def. & 3 & 5 & 384 & 6400 \\ 
S13 & 3 & near-$M_{Ch}$ & DDT & 17 & 5 & 384 & 1000000 \\
F14 & 3 & near-$M_{Ch}$ & Def. & 14 & 5 & 384 & 1000000 \\
LN18 & 2 & near-$M_{Ch}$ & DDT & 21 & 7 & 495 & 25600 \\
T18 & 3 & sub-$M_{Ch}$ & D$^{6}$ & 1 & 13 & - & -  \\
B19 & 1 & \makecell{near-$M_{Ch}$\\sub-$M_{Ch}$} & \makecell{DDT\\Det.} & 100 & 722 & - & - \\
LN20 & 2 & \makecell{near-$M_{Ch}$\\sub-$M_{Ch}$} & \makecell{Def.\\Dbl. Det.} & 57 & 7 & 495 & 25600
\enddata
\end{deluxetable*}

\begin{deluxetable*}{cccccc}
\centering
\tablecaption{Summary and comparison of the basic properties of the four sets of models of core-collapse supernova nucleosynthesis. \label{tab:cc_model_summary}}
\tablehead{
\colhead{Paper} &
\colhead{\makecell{No.\\ Dim.}} &
\colhead{\makecell{Mass\\Range}} &
\colhead{\makecell{Explosion\\Type}} &
\colhead{\makecell{No.\\Sims.}}
}
\startdata
WW95 & 1 & 11--40 & Spherical & 12 \\
MN03 & 2 & 25, 40 & Bipolar & 4 \\
N06 & 1 & 13--40 & Hypernova & 11 \\
S16 & 1 & 9--120 & Spherical & 95
\enddata
\end{deluxetable*}

The seven sets of models of Type Ia supernova nucleosynthesis are as follows:  in \cite{maeda_2010}, referred hereafter as M10, the authors examine a trio of two-dimensional, asymmetric, models in near-$M_{Ch}$ progenitors: two delayed-detonation, and one pure deflagration. In \cite{seitenzahl_2013}, now referred to as S13, 17 three-dimensional, high-resolution simulations of delayed-detonation models in near-$M_{Ch}$ progenitors were used to explore variations in initial deflagration strength; the deflagration to detonation transition (DDT) was parametrized in terms of density at which the transition occurs, and the resulting ignition is parametrized by the number of ignition kernels. In \cite{fink_2014}, hereafter labeled F14, a set of 14 three-dimensional, asymmetric models similar to those from S13 were used, though this time making use of pure deflagration models rather than DDT models. In \cite{leung_nomoto_2018}, now labeled LN18, the authors performed a parameter survey, exploring the effects of metallicity, C/O mass ratio, flame shape, DDT criteria, and turbulent flame formula have on 21 two-dimensional models of near-$M_{Ch}$ DDT explosions. In \cite{tanikawa_2018}, now referred to as T18, a single unique three-dimensional smoothed particle hydrodynamics simulation was performed to examine the dynamically-driven double-degenerate double-detonation (D$^{6}$) model. In \cite{bravo_2019}, now referred to as B19, a wide set of 100 one-dimensional models explored the effects of mass, metallicity, and the density at which the transition from deflagration to detonation occurs, alongside the reaction rate of C+O using the standard C+O reaction rate, and the rate scaled down by a factor ($1-\xi_{CO}$) with $\xi_{CO}=0.9$. Lastly, in \cite{leung_nomoto_2020a, leung_nomoto_2020b}, hereafter referred to as LN20, 57 two-dimensional models were used to examine both double-detonation sub-$M_{Ch}$ and pure deflagration near-$M_{Ch}$ explosions.

\subsection{Core-Collapse Supernovae}
\label{ssec:core-collapse}
The remaining four of these sets of models are representative of a core-collapse origin. One of the first developed nucleosynthesis models, \cite{woosley_weaver_1995}, hereafter labeled WW95, explored spherical explosions for several stars of varying masses. In \cite{Maeda_2003}, which we now refer to as MN03, the authors examined bipolar supernova explosions driven by accretion-powered jets for progenitors of 25 M$_{\odot}$ and 40 M$_{\odot}$. In \cite{Nomoto_2006}, now referred to as N06, the authors used varying stellar masses and explosion energies to investigate nucleosynthesis yields for both supernovae and hypernovae. Finally, in \cite{sukhbold_2016}, now referred to as S16, a set of spherical explosions for a wide variety of progenitor masses -- ranging from 9 M$_{\odot}$ to 120 M$_{\odot}$ -- were explored, using mass steps as small as 0.1 M$_{\odot}$.

\section{Results}
\label{sec:results}
We move now to a description of our spatially resolved spectroscopy. This was performed on the regions generated via the \textit{contbin} algorithm, as described in Section \ref{ssec:regions} and seen in Figure \ref{fig:region_maps}. For the spectral fitting, the datasets were fit using the XSPEC software version 12.14.1, using the solar abundance tables of \cite{wilms_2000}. Errors for individual parameters are reported to a significance of 90\%, and were determined using the Markov Chain Monte Carlo method, using the Goodman-Weare algorithm as recommended by the XSPEC manual. The error analysis was performed with 16 walkers, a chain length of 200,000 steps, and a burn length of 100,000 steps.

The objects chosen as targets for this study were selected partially due to the presence of notable thermal emission, as noted in Section \ref{sec:targets}. As a result, we attempted to fit the extracted spectra with a variety of thermal models, using both collisional ionization equilibrium (CIE) models as well as non-equilibrium ionization (NEI) models. The models tested included:

\begin{enumerate}
    \item \texttt{APEC/VAPEC/VVAPEC}, a CIE model for diffuse gas, characterized by a constant electron temperature.
    \item \texttt{PSHOCK/VPSHOCK/VVPSHOCK}, an NEI model for a plane-parallel shocked plasma with a constant temperature.
    \item \texttt{NEI/VNEI/VVNEI}, an NEI collisional plasma model, which assumes a constant temperature and a single ionization parameter.
    \item \texttt{RNEI/VRNEI/VVRNEI}, an NEI collisional plasma model for recombining plasmas. This model assumes that the plasma started in a CIE state, and includes an initial temperature parameter.
\end{enumerate}

The general approach to the spectral fitting process was as such: for the initial step of the fitting process, once a model was chosen, the hydrogen column density (for Galactic SNRs), plasma temperature, normalization, and ionization timescale were freed and fit. The abundances in this component were then freed and fit one at a time: if a given abundance tended towards zero, it was frozen at 0; if it instead remained near to solar and did not notably impact the fit parameter, it was held frozen at solar abundance. If the resultant model was insufficient to produce a good fit to the data, a second model was added, typically to account for emission from ISM/CSM. This model was generally fit in the same way as the first, though the abundances were held to solar in all cases save for those of G4.5+6.8 and G120.1+1.4, for which the second model plasma model was taken to represent additional shocked ejecta. In several objects, a discrepancy between our observations and our fitted models was noticed around 1.2 keV, likely due to excess emission from the Fe L complex not accounted for in the used atomic models \citep{yamaguchi_2011}. In order to improve the fits and account for this emission, a Gaussian line was added to the model when relevant. The other parameters were frozen while fitting these Gaussians, so as to not significantly impact the abundance values.

When comparing our observations to models of supernova nucleosynthesis, we take the ratio of abundances with respect to that of Si. This is done to minimize systematic uncertainties in the absolute value of a given abundance, due to effects such as uncertainty in emission measures, instrumental calibrations, or possible degeneracies in the models. Si was chosen as it is typically a prominent line in supernova plasmas, and can thus be usually well constrained. These ratios are given in the form of $(\text{X}/\text{Si})/(\text{X}/\text{Si})_{\odot}$, where X is the abundance value of O, Ne, Mg, S, Ar, Ca, or Fe with respect to the solar values of \cite{wilms_2000}. These ratios were then averaged across a given SNR, with all regions being weighted equally. This was done because, while spatially-resolved spectroscopy affords us a potentially more precise method of determining abundance ratios than would be gained by observing the SNR as a whole, the models to which we are comparing our results display information pertaining to the entire SNR, and do not generally provide spatially-resolved yield data.

To compare our results to the simulations of supernova nucleosynthesis, we performed a least squares analysis between these average ratios and the same ratios for each model; the uncertainties presented in these comparisons are based on the uncertainties obtained from our spectral fits, and so include only statistical errors. The ``best fit" models are those with the lowest fit parameter from this analysis, though they are not necessarily a globally good fit to the data. A~summary of our results can be seen in Table~\ref{tab:result_summary}, though readers should consider that the ``Type Conclusion" and ``Subtype Conclusion" columns are not solely based both upon the results of this work, but rather upon the results found in past studies as well. The detailed results of our spectral fitting for specific objects can be found in Tables~\ref{tab:g1.9_results} to \ref{tab:0548_results}. The nucleosynthesis models that most closely reproduce our findings are shown in Figure~\ref{fig:ia_nucsyn_plots} (for the models of thermonuclear supernova nucleosynthesis) and Figure~\ref{fig:cc_nucsyn_plots} (for the models of core-collapse supernova nucleosynthesis). The sole exception for this are the results for Tycho's SNR (G120.1+1.4), which is shown separately in Figure~\ref{fig:tycho_nucsyn_plots} due to our peculiar results.

\begin{deluxetable*}{lcccccc}
\centering
\tablecaption{Summary of nucleosynthesis results. The ``Best Single Model" and ``Best Set of Models" columns indicate the model families that produced the best individual fit to the data and the best average fits to the data, respectively. The ``Best Match Among Model Families" column is based on general trends and prior results, not solely on the results of this work. Conclusions indicated with a ``?" were those for which there was no clear result. \label{tab:result_summary}}
\renewcommand{\arraystretch}{2}
\tablehead{
\colhead{SNR} &
\colhead{\makecell{Assumed\\Nature}} &
\colhead{Best TN Model} &
\colhead{Best CC Model} &
\colhead{\makecell{Best Single\\Model (TN/CC)}} &
\colhead{\makecell{Best Set\\of Models}} &
\colhead{\makecell{Best Match Among\\Model Families}}
}
\startdata
G1.9+0.3 & 
TN? & 
\makecell{B19 $M_{Ch}$, $\xi_{\rm CO} = 0.9$\\ $\rho_{\rm DDT} = 1.20$E+$07$, $Z= 6.75$E-$02$} & 
S16 $17.4$ M$_{\odot}$ & 
TN & 
TN & 
$M_{Ch}$ WD, low DDT density, high metallicity \\

G4.5+6.8 &
TN &
T18 D$^{6}$&
WW95 $12$ M$_{\odot}$ &
TN &
TN &
D$^{6}$ \\

G120.1+1.4 &
TN & 
\makecell{B19 $M_{Ch}$, $\xi_{\rm CO} = 0.9$\\ $\rho_{\rm DDT} =2.40$E+$07$, $Z = 2.25$E-$04$} &
WW95 $12$M$_{\odot}$ &
TN &
TN &
? \\

G272.2--3.2 &
TN? & 
\makecell{B19 $M_{Ch}$, $\xi_{\rm CO} = 0.9$\\ $\rho_{\rm DDT} = 1.60$E+$07$, $Z = 2.25$E-$04$} &
S16 $15.2$M$_{\odot}$ &
CC &
TN &
$M_{Ch}$ WD, low DDT density, low metallicity \\

G337.2--0.7 &
TN? & 
\makecell{B19 sub-$M_{Ch}$, $\xi_{\rm CO} = 0.9$\\ M$_{\rm WD}$ = $1.15$ M$_{\odot}$, $Z = 2.25$E-$04$} &
WW95 $12$M$_{\odot}$ &
TN &
TN &
sub-$M_{Ch}$ WD ($1.1-1.2$ M$_{\odot}$), low metallicity \\

G344.7--0.1 &
TN/CC & 
\makecell{B19 $M_{Ch}$, $\xi_{\rm CO} = 0.9$\\ $\rho_{\rm DDT} = 1.20$E+$07$, $Z = 2.25$E-$04$} &
WW95 $12$M$_{\odot}$ &
TN &
TN &
$M_{Ch}$ WD, low DDT density, low metallicity \\

G352.7--0.1 &
TN/CC & 
\makecell{B19 $M_{Ch}$\\ $\rho_{\rm DDT} = 2.40$E+$07$, $Z = 2.25$E-$03$} &
S16 $10.0$ M$_{\odot}$ &
TN &
TN &
$M_{Ch}$ WD, med DDT density, low-med metallicity \\

0505--67.9 &
TN? &
F14 N100Hdef &
S16 $10.0$ M$_{\odot}$ &
CC &
? &
? \\

0509--67.5 &
TN & 
\makecell{B19 sub-$M_{Ch}$, $\xi_{\rm CO} = 0.9$\\ M$_{\rm WD}$ = $0.88$ M$_{\odot}$, $Z = 2.25$E-$04$} &
WW95 $12$M$_{\odot}$ &
CC &
TN &
sub-$M_{Ch}$ WD ($<1.0$ M$_{\odot}$), low metallicity \\

0509--68.7 &
TN/CC & 
\makecell{B19 $M_{Ch}$, $\xi_{\rm CO} = 0.9$\\ $\rho_{\rm DDT} = 1.20$E+$07$, $Z = 2.25$E-$04$} &
WW95 $12$ M$_{\odot}$ &
CC &
TN &
$M_{Ch}$ WD, low DDT density, low metallicity \\

0519--69.0 &
TN? & 
\makecell{B19 $M_{Ch}$, $\xi_{\rm CO} = 0.9$\\ $\rho_{\rm DDT} = 1.60$E+$07$, $Z = 2.25$E-$04$} &
S16 $20.3$ M$_{\odot}$ &
TN &
TN &
Low-mass sCh WD / low DDT density $M_{Ch}$ WD\\

0534--69.9 &
TN/CC &
F14 N100Hdef &
S16 $11.75$ M$_{\odot}$ &
CC &
TN &
? \\

0548--70.4 &
TN/CC & 
\makecell{B19 $M_{Ch}$\\ $\rho_{\rm DDT} = 1.60$E+$07$, $Z = 2.25$E-$04$} &
S16 $10.75$ M$_{\odot}$ &
CC &
CC &
$10-12$ M$_{\odot}$ CC
\enddata
\end{deluxetable*}

\subsection{Galactic SNRs}
\label{ssec:results_galactic_snrs}

\subsubsection{G1.9+0.3}
\label{sssec:results_g1.9}
G1.9+0.3 was observed by \textit{Chandra} in nine observations between May and June 2011, totaling nearly one~megasecond of exposure. Since the remnant’s emission is dominated by non-thermal processes, the \textit{contbin} algorithm was ineffective for region mapping. Instead, we manually selected four regions based on the study by \cite{borkowski_2013}, as described in Section \ref{ssec:regions}. Due to poor photon statistics, regions were grouped with a minimum of 10 counts per bin, as opposed to the 20 used for all other SNRs.

Among the four models tested (VPSHOCK, VNEI, VRNEI, and VAPEC), VPSHOCK provided the best fit, and two-component models offered no significant improvement, so we proceeded with the simpler one-component fits. Emission lines were hard to detect due to low photon statistics and high interstellar absorption, leading us to focus on three elements: Si, S, and Fe. Si has prominent lines in all regions, S is seen in three regions (also save for the NW), and Fe appears in three regions (all save for the W). Our results can be found in Table \ref{tab:g1.9_results}.

Spectral fits reveal a consistently high hydrogen column density ($7.80 - 9.23 \times 10^{22}$ \si{cm^{-2}}), likely due to the remnant's proximity to the Galactic center. Plasma temperatures ranged from $3.54 - 5.36$ keV, and ionization timescales were low ($4.46 - 16.1 \times 10^{8}$ \si{cm^{-3}.s}), indicating the remnant's young age. Si is consistently super-solar, S is enhanced in all three regions it appeared in, and Fe shows varying enhancements: enhanced in N and NW, and sub-solar in NE. These results are consistent with those of \cite{borkowski_2013}.

With only three elements for abundance determination, our nucleosynthesis analysis (summarized in Figures~\ref{fig:ia_nucsyn_plots} and \ref{fig:cc_nucsyn_plots}) yielded two data points, making it difficult to draw strong conclusions about the progenitor. However, thermonuclear models with high-mass white dwarfs, low DDT density, and high metallicity provide the best fits. Specifically, the near-$M_{Ch}$, $1.20 \times 10^{7}$ DDT density model from B19, with a metallicity of $Z = 6.75 \times 10^{-2}$ and an attenuated CO reaction rate, produce the best fit statistically.

Core-collapse models also provide acceptable fits, with the best matches among them arising from progenitors around $17.5$ M$\odot$. The best core-collapse fit is from the $17.4$ M$\odot$ model, followed closely by $17.3$ to $17.6$ M$\odot$ models. The $13.0$ M$\odot$ model from S16 also provided a strong fit, along with similar models from $13.2$ to $13.5$ M$_\odot$. All best-fit core-collapse models are from S16.

\subsubsection{Kepler's SNR}
\label{sssec:results_kepler}
For Kepler's SNR, we used a single 140 ks \textit{XMM-Newton} observation. The \textit{contbin} algorithm generated an 18-region map from combined MOS1+MOS2 data for spectral extraction. We noticed that the pn data appeared to be slightly broadened with respect to the MOS data, and we were unable to find a spectral model capable of fitting simultaneously both the MOS and pn data while producing sensible results. This led us to exclude it from the analysis and focus on MOS1 and MOS2. We also noticed slight shift in the line centroids between our observations and the fitted models, which was likely due to bulk motion of the ejecta, so we allowed the redshift parameters to vary, tying the hotter component’s redshift to the cooler one.

Initially, we tested single-component models, but these couldn’t reproduce the spectra. We then used two-component combinations of VPSHOCK, VNEI, VRNEI, and VAPEC models, and found that the VNEI+VNEI model provided the best fit. Although the VNEI+VAPEC model also worked, we favored the former given G4.5+6.8's low age, which suggests the remnant should still be in NEI. We also noticed excess emission around 1.2 keV, likely from Fe L emission, which was not accounted for by the models. We added a Gaussian at this energy to fit the excess. All regions except R11 were well fit ($\chi_{\nu}^{2} < 2.0$), with R11 excluded from nucleosynthesis analysis due to unresolved line broadening, particularly around the Fe line. The final results of these fits can be seen in Tables \ref{tab:g4.5_results_1} and \ref{tab:g4.5_results_2}.

Spectral fitting results showed a relativity consistent hydrogen column density, which ranged between $0.44-0.76 \times 10^{22}$ \si{cm^{-2}}. The cold component had plasma temperatures from $0.41 - 0.72$ keV and ionization timescales between $1.71 - 53.5 \times 10^{11}$ \si{cm^{-3}.s}; the ionization timescale is slightly skewed by R14 and R16, which have notably higher cold component ionization timescales than other regions. The hot component showed plasma temperatures ranging from $3.64 - 17.5$ keV, with ionization timescales between $3.14 - 19.9 \times 10^{9}$ \si{cm^{-3}.s}. Abundances were fairly consistent, with Ne and Mg being sub-solar, and Si, S, Ar, Ca, and Fe being enhanced.

Our best-fit nucleosynthesis models can be seen in Figures~\ref{fig:ia_nucsyn_plots} and \ref{fig:cc_nucsyn_plots}. When comparing to nucleosynthesis models, we found the best fit with the D$^{6}$ model of T18, a double-degenerate thermonuclear model, which could help constrain the formation channel of G4.5+6.8. While the D$^{6}$ model underproduced the Ar/Si and Ca/Si ratios slightly, higher-density near-$M_{Ch}$ progenitors in B19 provided a better fit for the Ca/Si ratio. However, these models fell short for S/Si and Ar/Si ratios. A core-collapse scenario with a $12$ M$_{\odot}$ progenitor from WW95 was able to reproduce the Ne/Si, Mg/Si, and S/Si ratios, but was less successful for heavier elements. No core-collapse model matched all the ratios, though some reproduced Ne/Si reasonably well.

\subsubsection{Tycho's SNR}
\label{sssec:results_tycho}
For Tycho's SNR, we made use of four \textit{XMM-Newton} observations, which combine for a total exposure time of just under 150 ks. The \textit{contbin} algorithm was applied to a merged MOS1+MOS2 image, and generated a region map consisting of 20 regions. During initial fitting, we noticed an apparent discrepancy between the MOS and pn observations, where the pn spectra exhibited a noticeably higher count rate than the MOS spectra between roughly $1.5 - 1.8$ keV. While we attempted to correct for this discrepancy through the fitting process, our attempts were unsuccessful, leading to a notable reduction in the quality of our fits. As a result, we opted to not make use of the pn observations for this object, instead using the MOS1 and MOS2 observations.

For our spectral fits, we found that one-component models were incapable of reproducing the complexity of the observed spectra, and so we instead proceeded to fit using two- and three-component models. While both of these proved capable of producing acceptable fits, the three-component models proved unwieldy, as the spectral fitting process was more prone to becoming stuck in local minima and thus producing unrealistic or unphysical results for one or more of the involved components. The three-component models also produced only a small increase in fit quality at the expense of a significant increase in computation time; as such, we opted to use a two-component model, finding that an absorbed VPSHOCK+VPSHOCK model yielded the best results.

O, Mg, Si, S, Ar, Ca, and Fe were freed and allowed to fit. While we attempted to also fit for the abundance of Ne, we found that in doing so, the reported abundance would consistently collapse to a near-zero value. Because of this, we opted to freeze the Ne abundance at zero for all regions. Additionally, there were two energy ranges in the spectra that exhibited excess emission unaccounted for in our thermal models, likely representing lines in the Fe L complex that the used atomic models do not currently incorporate \citep{yamaguchi_2011}. To account for this excess emission, we added two Gaussians to our spectral fits: one at $0.72$ keV, and a second at $1.2$ keV. As a result, we were able to obtain acceptable fits ($\chi_{\nu}^{2} < 2$) to 18 of our 20 regions. The remaining two regions -- R05 and R19 -- both appeared to exhibit notable line broadening that we were unable to account for with our models. While the values and abundances within these regions were comparable to those of our other regions, we opted to exclude them from our nucleosynthesis calculations as a result. Our final results can be found in Tables \ref{tab:g120_results_1} and \ref{tab:g120_results_2}.

The abundance of Ca in Tycho's SNR proved unique in our study. While it is clear that this abundance is enhanced to super-solar values, it was present at two distinct temperatures, which led to our allowing the Ca abundance to vary in both the hot and cold components. While this led to a notable improvement in the quality of the fit over simply allowing the Ca to be present in either one of the two components, it also led to the Ca abundance -- particularly, that determined within the cold component -- to become poorly constrained when compared to the other abundances within our fit, finding itself enhanced to values well above what would normally be expected. We thus note that the Ca abundance in these results may best be served as an upper limit, and urge the reader to treat it as such.

The comparison of our results to models of supernova nucleosynthesis can be found in Figure~\ref{fig:tycho_nucsyn_plots}, where we display the best-fit models for both the thermonuclear (Figure~\ref{fig:ia_nucsyn_plots}) and core-collapse scenarios (Figure~\ref{fig:cc_nucsyn_plots}). we find good agreement between our observations and the models for the O/Si, Mg/Si, and S/Si ratios. Both thermonuclear and core-collapse models are able to reproduce these ratios, though the thermonuclear models fare better, as the core-collapse models tend to overproduce O and Mg relative to Si. The observed Fe/Si ratio aligns closer to models of core-collapse nucleosynthesis; however, thermonuclear models featuring low-mass (for the sub-$M_{Ch}$ case) or low DDT-density (for the near-$M_{Ch}$ case) WDs are also able to reproduce our findings, as these explosions tend towards producing smaller amounts of Fe. The Ar/Si ratio is high, and is only reproduced by two models: the D$^{6}$ model of T18, and the 12 M$_{\odot}$ model of WW95. The large abundance of Ca makes it difficult to come to a conclusion, as no tested model is able to reproduce such a high ratio of Ca/Si. As a result, we opt to perform this analysis in three parts: once utilizing the Ca abundance from the cold component, once using the Ca abundance from the hot component, and once omitting the Ca abundance altogether. We find a general agreement between the comparisons using the cold Ca, and those omitting Ca altogether, with a medium-DDT density near-$M_{Ch}$ WD producing the best results for both models. The comparison to the hot Ca, however, is suggestive instead of a slightly lower DDT density. Further, the comparisons differ in that the hot Ca more strongly aligns with a high-metallicity progenitor, while the cold Ca and no Ca comparisons suggest a low-metallicity progenitor instead. The core-collapse models, for their part, are somewhat inconsistent: estimates for progenitor mass among the best fits ranges from approximately $10$ M$_{\odot}$ to $26$ M$_{\odot}$. The $12$ M$_{\odot}$ model of WW95 fares particularly well, largely due to its ability to reproduce the Ar/Si ratio. Given the uncertainty that arises from our Ca abundance, we cannot determine with certainty a progenitor for this SNR. However, as it has been previously established that G120.1+1.4 arose as a result of a thermonuclear supernova, we suggest a near-$M_{Ch}$ progenitor with a low-to-medium DDT density as a possible source.

\subsubsection{G272.2--3.2}
\label{sssec:results_g272}
There is only one \textit{XMM-Newton} observation of G272.2–3.2, a 38 ks exposure from 2001. Due to the object's large size relative to the \textit{XMM-Newton} EPIC field of view, we used the ESAS pipeline to create a merged, background-subtracted, and vignetting-corrected image. The \textit{contbin} algorithm generated seven regions for spectral extraction. In the \textit{XMM-Newton} observation used for this study, G272.2--3.2 has the peculiar property of being largely undetected above $3.0$ keV: when viewed in such a regime, it is indistinguishable from the background, an issue also noted by \cite{sanchez-ayaso_2013}. As a result, we analyzed the 0.5–3.0 keV range instead of the full 0.5–8.0 keV range used for other remnants.

When fitting the spectra for G272.2--3.2, we initially fit each spectra with an absorbed single-component thermal model. Out of the four models used in this study -- VPSHOCK, VNEI, VRNEI, and VAPEC -- we found that the VNEI model was best able to reproduce the observations. Attempts to add a second thermal component yielded little improvement, and so we proceeded with a single absorbed VNEI component. We freed the column density, temperature, ionization timescale, and normalizations first, followed by freeing and fitting the individual abundances (Si, S, Fe, Mg, Ne) one at a time. Abundances of Ne and Mg were generally near solar, while Si, S, and Fe tended towards super-solar values. Four regions showed an excess around 1.2 keV, likely related to the Fe L shell, which was modeled with a Gaussian. All regions were well fit ($\chi_{\nu}^{2} < 1.25$), and our final results can be seen in Table \ref{tab:g272_results}.

The results showed little variation in column density, plasma temperature, and ionization timescale across the remnant: $1.09-1.37 \times 10^{22}$ \si{cm^{-2}}, $0.70 - 0.98$ keV, and $3.85 - 6.78 \times 10^{10}$ \si{cm^{-3}.s}, respectively, with the low ionization timescale suggesting the remnant is still in NEI. Abundances were generally consistent and typically supersolar throughout the remnant, except in the brightest region (R00), which showed particularly high values for Si, S, and Fe.

Comparing our results to supernova nucleosynthesis models (see Figures~\ref{fig:ia_nucsyn_plots} and \ref{fig:cc_nucsyn_plots}), we find disagreement with thermonuclear models, particularly in the Ne/Si and Mg/Si ratios. The S/Si and Fe/Si ratios align more closely with these models, with good fits found using low-metallicity models, such as those from B19 and LN18. The best-fit thermonuclear models suggest a low-density, possibly near-$M_{Ch}$ progenitor, with both near-$M_{Ch}$ and sub-$M_{Ch}$ models providing good fits.

Core-collapse models show slightly better agreement, particularly for lower progenitor masses. The S/Si ratio can be reproduced by models across a range of progenitor masses, but the Fe/Si ratio is best matched by models with progenitor masses between 9 and 13 M${\odot}$, with a peak around 10 M${\odot}$. The best overall fit is found with the $15.2$ M$_{\odot}$ model from S16, which reproduces the Mg/Si and S/Si ratios well. However, the consistency of the thermonuclear models and their progenitor mass range suggests a thermonuclear origin, likely from a near-$M_{Ch}$ progenitor.

\subsubsection{G337.2--0.7}
\label{sssec:results_g337}
For G337.2--0.7, a single 40 ks observation was used, and a \textit{contbin} region map of seven regions was generated. A single absorbed VNEI model was applied across the $0.5 - 8.0$ keV energy range, with column density, plasma temperature, ionization timescale, and normalisation constant allowed to vary. The abundances were freed one at a time and fit; the fits proved insensitive to Fe, and as a result, that abundance was frozen at solar values. All regions had good fits ($\chi_{\nu}^{2} < 1.25$), and our results can be found in Table \ref{tab:g337_results}.

Our results show hydrogen column densities ranging from $3.44 - 4.83 \times 10^{22}$ \si{cm^{-2}}, with plasma temperatures between $0.91 - 1.56$ keV. The ionization timescale ranges from $3.89 - 14.7 \times 10^{10}$ \si{cm^{-3}.s}, highest in the central regions, suggesting the remnant is still in a state of NEI. Elemental abundances are generally consistent, though Ca is enhanced in the outer regions.

Comparison with supernova nucleosynthesis models shows strong agreement, particularly with sub-$M_{Ch}$ progenitors from B19 that have a reduced CO reaction rate and low density (see Figure~\ref{fig:ia_nucsyn_plots}). These models match most abundance ratios, except Mg/Si, which is underproduced by the models in comparison to our findings. Four out of five best-fitting models are from sub-$M_{Ch}$ WDs above $1$ M${\odot}$, with one near-$M_{Ch}$ WD model (high DDT density) also fitting well. Core-collapse models (see Figure~\ref{fig:cc_nucsyn_plots}) fail to reproduce the results for heavier elements; as a result, a thermonuclear origin is favored, with a likely progenitor being a WD with a mass between $1.00 - 1.20$ M${\odot}$ and low metallicity.

\subsubsection{G344.7--0.1}
\label{sssec:results_g344}
Our analysis of G344.7--0.1 made use of two observations for a total of $22.2$ ks of observation time. While Obs. Id. 0111210401 was made using the EPIC MOS cameras in Full Frame mode and EPIC pn in Extended Full Frame mode, Obs. Id. 0111210101 only made use of EPIC pn in Extended Full Frame mode; this observation did not contain data from the EPIC MOS cameras. A combined MOS image from Obs. Id. 0111210401 was processed using \textit{contbin}, generating a map with eight regions. The column density, plasma temperature, ionization timescale, and normalisation constant were allowed to vary for the initial fit, while the abundances for the elements of Si, S, Ar, Ca, and Mg were freed and allowed to fit one at a time for subsequent fits. All regions had good fits, with $\chi_{\nu}^{2} < 1.35$. The final results are shown in Table \ref{tab:g344_results}.

Results  show consistent absorption coefficients ($4.84 - 5.97 \times 10^{22}$ \si{cm^{-2}}), plasma temperatures ($1.16 - 2.18$ keV), and ionization timescales ($2.99 - 8.61 \times 10^{10}$ \si{cm^{-3}.s}). R04, located on the northeastern edge of the remnant, showed higher temperature and lower column density and ionization timescale, suggesting it contains recently shock-heated material. Abundances of Si, S, Ar, and Ca are enhanced, with Ca frozen to solar values in R04 and R05.

As shown in Figure~\ref{fig:ia_nucsyn_plots}, thermonuclear models from B19, particularly those of near-$M_{Ch}$ WDs, fit our findings best, although Mg/Si, S/Si, and Ar/Si ratios are underproduced in most models. The Ca/Si ratio is well-explained by models with low metallicity, and the Fe/Si ratio is overestimated by most models, with B19's low DDT density near-$M_{Ch}$ WD models offering the best match. Core-collapse models fail to reproduce the Ar/Si and Ca/Si ratios, as shown in Figure~\ref{fig:cc_nucsyn_plots}, though a $12$ M$_{\odot}$ model from WW95 provides a good fit for some ratios, despite underproducing Mg/Si and Fe/Si. Overall, the abundance pattern supports a thermonuclear origin, with a near-$M_{Ch}$ WD progenitor with low metallicity and low DDT density being the most likely source.

\subsubsection{G352.7--0.1}
\label{sssec:results_g352}
G352.7--0.1 was observed by \textit{XMM-Newton} for $30$ ks. Using these data, we generated a four-region map with \textit{contbin} and fit the spectra in the $0.5-8.0$ keV range using an absorbed, single-component VNEI model. Initial tests with other models (VPSHOCK, VRNEI, and VAPEC) showed no improvement, and a two-component model did not significantly enhance the fit. The initial fitting allowed column density, plasma temperature, ionization timescale, and normalisation constant to vary freely, while the abundances of Si, S, Fe, Ar, Ca, and Mg were subsequently freed one by one. All regions had a good fit with $\chi_{\nu}^{2} < 1.5$, and the results can be seen in Table \ref{tab:g352_results}.

The spectral fits show column densities ranging from $3.36 - 5.08 \times 10^{22}$ \si{cm^{-2}}, high plasma temperatures ($1.79 - 3.89$ keV), and low ionization timescales ($2.52 - 3.92 \times 10^{10}$ \si{cm^{-3}.s}), indicating the remnant is in a state of non-equilibrium ionization (NEI). Elemental abundances are enhanced to super-solar values, though region R01 shows lower abundances, possibly due to interaction with a nearby molecular cloud \citep{zhang_2023}.

Comparisons to supernova nucleosynthesis models can be seen in Figures~\ref{fig:ia_nucsyn_plots} and \ref{fig:cc_nucsyn_plots}. These comparisons suggest a near-$M_{Ch}$ progenitor from B19. All best-fit models indicate a near-$M_{Ch}$ white dwarf (WD) with medium-high density of delayed detonation (DDT), low metallicity, and standard CO reaction rates, producing good agreement for S/Si, Ar/Si, Ca/Si, and Fe/Si ratios. However, the Mg/Si ratio is lower than observed. Core-collapse models, especially from a $\sim 10$ M$_{\odot}$ progenitor, also fit the data but fail to reproduce the observed Fe/Si ratio. Overall, thermonuclear models from a low-metallicity, medium DDT-density, near-$M_{Ch}$ WD progenitor are the most consistent with the observed abundances.

\subsection{LMC SNRs}
\label{ssec:results_lmc_snrs}
For the LMC objects, a slightly different approach to the spectral fitting process was needed to account for the differences in composition between the LMC and our own galaxy. To that end, two steps were taken: firstly, rather than use the TBABS model to account for interstellar absorption, we opted to use the TBVARABS model instead. In addition to accounting for hydrogen column density, the TBVARABS model allows one to vary the elemental abundances of the absorbing material. We chose to set these abundances to the average metal abundances of the LMC as determined by \cite{dopita_2019}. As well, given that the LMC objects are located outside of our own galaxy, we also opted to use fixed values of $\text{N}_\text{H}$ for each object. These absorption values were determined by using the HEASARC tool $\text{N}_\text{H}$,\footnote{\url{https://heasarc.gsfc.nasa.gov/cgi-bin/Tools/w3nh/w3nh.pl}} using data provided by the HI4PI full-sky HI survey \citep{HI4PI}. Secondly, whenever our fit seemed to require a second component, we opted to set the abundances in said component to the same average metal abundances of the LMC as were used for the absorption parameters.

\subsubsection{0505--67.9}
\label{sssec:results_0505}
0505--67.9 was observed by \textit{XMM-Newton} for nearly 130 ks, with regions generated using the \textit{contbin} algorithm. Initial fitting with single-component thermal models (VPSHOCK, VNEI, VRNEI, VAPEC) failed to produce good fits, so a two-component VPSHOCK model was used instead, with LMC average abundances and a column density fixed at $N_\text{H} = 0.170 \times 10^{22}$ \si{cm^{-2}}. Plasma temperatures, ionization timescales, and normalisation constants were allowed to vary, and abundances for Si, S, Fe, Mg, Ne, and O were fitted individually. Six regions had good fits ($\chi_{\nu}^2 < 1.5$), with one region slightly worse ($\chi_{\nu}^2 = 1.52$); our results can be found in Table \ref{tab:0505_results}.

Plasma temperatures for the cold component were consistent ($0.20 - 0.22$ keV), and ionization timescales were low in the two innermost regions (R00 and R01). The hot component had plasma temperatures ranging from $0.78 - 0.84$ keV and ionization timescales from $2.61 - 6.17 \times 10^{11}$ \si{cm^{-3}.s}. Elemental abundances were mostly consistent, with O subsolar in all regions, Ne and Mg slightly enhanced or subsolar, and all other abundances slightly above solar. R03, the outermost region, had the lowest abundances and ionization timescale, suggesting recent shock-heating.

Regarding nucleosynthesis, models had poor agreement with the data. Thermonuclear models (Figure~\ref{fig:ia_nucsyn_plots}) struggled with the O/Si, Ne/Si, and Mg/Si ratios, which were much higher in the observations than in any thermonuclear model. The S/Si and Fe/Si ratios were better matched by models, especially those from high-density near-$M_{Ch}$ or high-mass sub-$M_{Ch}$ progenitors (B19), with LN18 models performing the best. Core-collapse models (Figure~\ref{fig:cc_nucsyn_plots}) showed better agreement for O/Si, Ne/Si, and Mg/Si, but none could simultaneously reproduce these ratios. The S/Si ratio was generally well-matched by S16 models for progenitors with masses above $15$ M$_{\odot}$, though the Fe/Si ratio was underproduced by all models. While core-collapse models overall fit the abundance pattern better, no model could fully explain the data.

\subsubsection{0509--67.5}
\label{sssec:results_0509}
A 44.3 ks \textit{XMM-Newton} observation of 0509--67.5 was analyzed, resulting in a map of four regions after applying the \textit{contbin} algorithm. An absorbed two-component VPSHOCK+VAPEC model provided the best fits, with the VAPEC abundances set to LMC averages. The column density was fixed at $N_\text{H} = 0.166 \times 10^{22}$ \si{cm^{-2}}, and plasma temperatures, ionization timescales, and normalisation constants were allowed to vary. Abundances for Si, S, Fe, Mg, and Ne in the VPSHOCK component were varied individually. Excess emission around 1.2 keV was modeled with a Gaussian, resulting in a good fit ($\chi_\nu^2 < 1.5$) for all regions. The final results can be seen in Table \ref{tab:0509_results}.

The plasma temperatures for the cold component were consistent ($0.37-0.41$ keV), while the hot component ranged from $3.41$ to $4.86$ keV. Ionization timescales were also consistent, ranging from $3.81 \times 10^{11}$ to $5.31 \times 10^{11}$ \si{cm^{-3}.s}. Elemental abundances varied across regions: Ne and Fe were slightly enhanced, Mg was below solar, and S and Si were highly enhanced (around $50 \times$ solar).

For nucleosynthesis, both thermonuclear and core-collapse models (Figures~\ref{fig:ia_nucsyn_plots} and \ref{fig:cc_nucsyn_plots}, respectively) showed moderate agreement. Thermonuclear models matched the Ne/Si and Mg/Si ratios well but struggled with the S/Si and Fe/Si ratios. The S/Si ratio was high and only matched by a few models, while the Fe/Si ratio was low, with the lowest-mass models (LN20b) producing the closest match. The absence of higher-energy Fe K-shell emission in our data could suggest a lower limit on the Fe/Si ratio, potentially improving agreement with thermonuclear models, such as those from LN18 and T18, which produce better matches for Ne/Si, Mg/Si, and S/Si but with a larger Fe/Si ratio.

Core-collapse models, particularly from S16 and WW95, fit the S/Si and Fe/Si ratios better but overproduce Ne and Mg, resulting in worse Ne/Si and Mg/Si ratios. Given the missing Fe K-shell emission and better match to the lighter elements, we propose a low-mass, low-metallicity, sub-$M_{Ch}$ WD progenitor for this SNR. These results also support the findings of \cite{2025NatAs...9.1356D}, who propose a double-degenerate, sub-$M_{Ch}$ progenitor scenario for this SNR.

\subsubsection{0509--68.7}
\label{sssec:results_n103b}
0509--68.7 was observed by \textit{XMM-Newton} for 26.5 ks in 2000 using various filters for MOS1, MOS2, and pn. After processing, \textit{contbin} generated a map of four regions. A single-component model failed to fit the data, but an absorbed two-component VAPEC+VPSHOCK model provided the best fit. The VAPEC component's abundances were fixed to LMC averages, and the column density was set to $N_\text{H} = 0.255 \times 10^{22}$ \si{cm^{-2}}. Plasma temperatures, ionization timescales, and normalisation constants were allowed to vary. Abundances for Si, S, Fe, Ar, Ca, Mg, Ne, and O were varied individually. A~Gaussian was added at 1.2 keV to account for underfitting of Fe L-shell emission. The final model fit each region well ($\chi_\nu^2 < 1.25$), and the results can be seen in Table \ref{tab:n103b_results}.

Plasma temperatures for the cold component were consistent at $0.74-0.76$ keV, while the hot component varied from $2.89$ to $3.44$ keV. Abundances in the hot component were enhanced above solar values, particularly in outer regions, though R03 showed similar abundances to the innermost region, R00. Ionization timescales ranged from $1.09 - 1.34 \times 10^{11}$ \si{cm^{-3}.s}, showing little variation across the remnant.

The nucleosynthesis comparisons can be seen in Figures~\ref{fig:ia_nucsyn_plots} and \ref{fig:cc_nucsyn_plots}. For the thermonuclear models, a low-density, low-metallicity near-Ch-mass WD with a reduced CO reaction rate (B19) was favored. These models reproduce the general abundance pattern, particularly the Ca/Si ratio, though they slightly underproduce Mg/Si, S/Si, O/Si, Ne/Si, and Ar/Si. The Fe/Si ratio can be matched by low DDT-density models, but at the expense of Ar/Si. Curiously, the best fit comes from the 12 M$_{\odot}$ model of WW95, which accurately matches O/Si, Mg/Si, S/Si, Ar/Si, and Ca/Si ratios. However, while core-collapse models (S16) reproduce some ratios, they are less consistent across elements. Overall, low DDT-density, low-metallicity near-Ch-mass models yield the most consistent results for this remnant.

\subsubsection{0519--69.0}
\label{sssec:results_0519}
We analyzed a 48.4 ks \textit{XMM-Newton} observation of 0519--69.0. For this observation, MOS1 made use of both the Full Frame, Medium Filter mode as well as the Full Frame, Thin Filter mode. For both the MOS2 and pn observations, only the Full Frame, Medium Filter mode was used. Each of these different observation modes was processed individually, and after doing so, a map of five regions was generated through the use of \textit{contbin}. Initial fitting with a single thermal model was inadequate, but adding a second component resulted in the best fit using a VPSHOCK+VAPEC model, with the VAPEC abundances fixed to LMC averages. We set the column density to $N_\text{H} = 0.217 \times 10^{22}$ \si{cm^{-2}} and allowed plasma temperatures, ionization timescales, and normalisation constants to vary. Abundances for Si, S, Fe, Mg, and Ne were fit individually. Excess Fe L emission around 1.2 keV was modeled with a Gaussian, resulting in good fits ($\chi_{\nu}^2 < 1.5$) across all regions. The final results can be found in Table \ref{tab:0519_results}.

Plasma temperatures ranged from $0.36$ to $0.41$ keV, and ionization timescales from $1.66 \times 10^{12}$ to $4.80 \times 10^{12}$ \si{cm^{-3}.s}, with elemental abundances showing consistency across regions. Mg was subsolar, Ne was near solar or subsolar, and Si, S, and Fe were enhanced. The outermost region, R04, had higher plasma temperatures, lower ionization timescales, and higher abundances. While the plasma temperature and ionization timescale are suggestive of a possible state of NEI and a blast wave origin, the enhanced metal abundances make this scenario less likely.

From a nucleosynthesis perspective, both thermonuclear and core-collapse models (Figures~\ref{fig:ia_nucsyn_plots} and \ref{fig:cc_nucsyn_plots}, respectively) reproduced the abundances in 0519--69.0. Core-collapse models matched the Ne/Si and Mg/Si ratios, though they typically overproduced these elements. Thermonuclear models, on the other hand, better matched the S/Si and Fe/Si ratios, particularly the C-DEF model from M10 and the low-density $Z = 0.1$ model from LN18 for S/Si. Thermonuclear models with low-density explosions, producing lower amounts of Fe, gave the best fits. Given the overall abundance patterns, a thermonuclear origin, particularly from a low-mass, low-density WD with low metallicity, seems most likely.

\subsubsection{0534--69.9}
\label{sssec:results_0534}
A single 61.9 ks \textit{XMM-Newton} observation of 0534--69.9 was analyzed, and spectra were extracted from six regions generated using the \textit{contbin} algorithm. Spectral fitting with VPSHOCK, VNEI, VRNEI, and VAPEC models multiplied by a TBVARABS absorption model showed that the absorbed VPSHOCK model provided the best fit across the remnant. The column density was set to $N_\text{H} = 0.255 \times 10^{22}$ \si{cm^{-2}}, and plasma temperature, ionization timescale, and normalisation constants were allowed to vary. Abundances for Si, Fe, Mg, Ne, and O were freed individually. Excess emission around 1.2 keV in regions R01 to R05 was modeled with a Gaussian, resulting in a good fit for all regions ($\chi_{\nu}^2 < 1.5$). The final results can be found in Table \ref{tab:0534_results}.

Plasma temperatures ranged from $0.54$ to $0.88$ keV, generally increasing toward the edges of the remnant. O, Ne, Mg, Si, and Fe showed a similar pattern, with O and Mg being subsolar, and Ne and Fe supersolar. Ionization timescales decreased toward the remnant's edge, ranging from $2.90 \times 10^{10}$ to $6.95 \times 10^{10}$ \si{cm^{-3}.s}, except in the bright central region (R00), which had a higher, though poorly constrained, ionization timescale of $10.9 \times 10^{11}$ \si{cm^{-3}.s}. These ionization timescales suggest that the remnant is still in NEI.

For nucleosynthesis, comparing the results to thermonuclear and core-collapse models (Figures~\ref{fig:ia_nucsyn_plots} and \ref{fig:cc_nucsyn_plots}, respectively), we found the Fe/Si ratio favored a thermonuclear origin, while O/Si, Ne/Si, and Mg/Si ratios leaned towards core-collapse. Among thermonuclear models, the F14 N100H model gave the closest match but underproduced O, Ne, and Mg. Core-collapse models from S16 for progenitors with masses between $9-12$ M${\odot}$ fit best, with the 11.75 M${\odot}$ model closely matching O and Mg abundances, though it underproduced Fe. The $30$ M$_{\odot}$ model from N06 also matched the O/Si, Ne/Si, and Mg/Si ratios but produced a low Fe/Si ratio. Despite some matches from core-collapse models, the overall low O abundances suggest a thermonuclear origin is more likely.

\subsubsection{0548--70.4}
\label{sssec:results_0548}
We analyzed a 60 ks \textit{XMM-Newton} observation of 0548--70.4, extracting spectra from the four \textit{contbin} regions. Initial fitting with a single thermal component model failed to reproduce the observed spectrum, leading us to test combinations of VPSHOCK, VNEI, VRNEI, and VAPEC models. The best fit was achieved with an absorbed VPSHOCK+VPSHOCK model. We fixed the column density at $N_\text{H} = 0.377 \times 10^{22}$ \si{cm^{-2}}, and set the second VPSHOCK component's abundances to LMC average values. While the plasma temperatures and normalisation constants were allowed to vary independently, tying the ionization timescale of the cold component to that of the hot component yielded an improvement to the fit. Abundances of Si, S, Fe, Mg, and Ne were varied, with O only notably improving the fit in region R03. As in other objects, there was excess emission around the $1.2$ keV mark that the models used could not account for; to compensate, we added a Gaussian at this energy. The regions were well fit, with $\chi_\nu^2 < 1.25$ for all regions, and the final results can be found in Table \ref{tab:0548_results}.

The plasma temperatures were consistent across the remnant, with the cold component at $0.29-0.31$ keV and the hot component at $0.68-0.69$ keV. Abundances of Ne, Mg, and Fe were near-solar, peaking at the core and decreasing outward, while Si and S showed the opposite trend. The outermost region (R03) had the lowest average abundances, with subsolar values for O, Ne, Mg, and Fe. Ionization timescales ranged from $1.24 \times 10^{11}$ to $6.18 \times 10^{11}$ \si{cm^{-3}.s}, with the lowest value in R03.

For nucleosynthesis, comparing our results to supernova models, we found that thermonuclear models (Figure~\ref{fig:ia_nucsyn_plots}) suggest a low-metallicity near-Ch-mass WD, based on the S/Si and Fe/Si ratios. However, the Mg/Si and Ne/Si ratios favor core-collapse models (Figure~\ref{fig:cc_nucsyn_plots}), particularly those from S16 for a progenitor star mass of $10-12$ M${\odot}$, with the best match from the 10.75 M${\odot}$ model.

\section{Discussion}
\label{sec:discussion}

\subsection{Caveats}
\label{ssec:caveats}
Above, we have shown that there is a widely varying range of agreement between our data and the suite of nucleosynthesis models used in this study. Here, we will list some possible caveats in our analysis that may account for these discrepancies.
 
First, the ejecta components used to infer our metal abundances are quite complex, with the level of complexity likely varying across different regions. Although we have attempted to account for this through the use of two-component models in our spectral fits where appropriate, these models may still oversimplify the system, and additional components could be necessary for a more accurate representation of the data. Second, while this study is spatially resolved, the comparisons presented here involve averaging the ejecta yields across the entirety of a given SNR. Certain regions within these objects may more accurately reflect the expected nucleosynthetic yields, particularly given the likelihood of asymmetries in the progenitor explosions. Third, systematic uncertainties related to telescope calibrations and plasma models used in this study (in this case, ATOMDB) could also affect the accuracy of our results. Fourth, while our analysis focuses on the shock-heated ejecta within these SNRs, it is likely that some ejecta have not undergone shock heating, or are otherwise not detectable within the X-ray band--this will be the subject of future work. Finally, the spectral resolution of the \textit{XMM-Newton} and \textit{Chandra} telescopes are not able to fully resolve every emission line within the observed energy range, which may result in the introduction of some degeneracy in the model parameters and result in less precise measurements of plasma properties. These last two points will be addressed in future work, e.g. \cite{2026ApJ...998..297M, 2026ApJ...999..151F}.

\subsection{The \textsuperscript{12}C +\textsuperscript{16}\text{O} Reaction Rate}
\label{ssec:co_reaction_rate}
The Ca/S mass ratio in thermonuclear supernova remnants offers a valuable probe of ejecta neutronization levels. However, \cite{martinez-rodriguez_2017} found that nucleosynthesis models predicted Ca/S ratios approximately 50\% lower than observations, tracing this discrepancy to the \textsuperscript{12}C +\textsuperscript{16}\text{O} reaction rate. Since Ca/S ratios scale with the square of $\alpha$-particle abundance, and the \textsuperscript{12}C +\textsuperscript{16}\text{O} reaction depletes $\alpha$-particles, the observations suggested the reaction rate in nature must be reduced by up to 90\% relative to standard values—well beyond the ~50\% uncertainty supported by experimental measurements.

Despite theoretical difficulties with such large attenuation, \cite{shen_2018} found that 90\% reduction improved model agreement, prompting \cite{bravo_2019} to conduct systematic investigations using both near-Ch-mass and sub-Ch scenarios with standard and attenuated (90\%) \textsuperscript{12}C +\textsuperscript{16}\text{O} rates. Their ``SNR-calibrated" models with reduced rates produced substantially higher Ca/S and Ar/S ratios, better matching observed abundances.

Our analysis confirms these SNR-calibrated models perform exceptionally well. Among 213 thermonuclear models examined, the \cite{bravo_2019} set (100 models) provided best fits to 10 of our 13 objects, with 8 of these 10 using the reduced \textsuperscript{12}C +\textsuperscript{16}\text{O} rate. The primary improvement stems from increased Ar/Si and Ca/Si ratios, along with enhanced S/Si and Fe/Si ratios. While lighter elements (O, Ne, Mg) are also affected, their typically low yields in thermonuclear supernovae make them less influential for fit quality.

Although one might expect other reaction rates to be similarly uncertain, sensitivity studies demonstrate that most yields remain relatively insensitive to rate variations—even tenfold changes produce only modest yield differences—except for specific reactions involving \textsuperscript{12}C and \textsuperscript{16}\text{O}. Consequently, uncertainties from explosion physics generally dominate over reaction rate uncertainties, though the latter merit consideration in future work.

\subsection{The Discrepancy Between Observations and Simulations}
\label{ssec:obs_vs_sim}
As has been shown in Section \ref{sec:results}, for most of the objects in this study, our results cannot be fully reproduced by the existing models of supernova nucleosynthesis.

First, let us consider what did not work. In general, the abundance ratios for the lightest elements considered -- those of O/Si, Ne/Si, and Mg/Si -- tended towards being underproduced by the models when compared to our observations. This is more noticeably the case when considering the Type Ia models, particularly so for the models of B19, which otherwise tended to most closely reproduce our results. This may be due to the 1D-nature of these simulations: as these elements are typically found in the outer layers of a progenitor where conditions are less likely to reach the required values for explosive burning, they are more likely to be unburned or only partially burned during the supernova. While the amount of burning that these elements experience can be adjusted through the turbulent mixing of these outer layers with the inner layers, this is a process that 1D simulations are less capable of reproducing than simulations with higher dimensionality. 

The abundance ratios for the intermediate-mass elements -- such as those of S/Si, Ar/Si, and Ca/Si -- proved more reliable than those of the lighter elements when it came to determining between a Type Ia or a core-collapse origin, but not consistently so. They did, however, often prove to be strong determinants of sub-models, owing to the frequent strong variation in these abundances between different models. The production of these elements was notably impacted by the \textsuperscript{12}C +\textsuperscript{16}\text{O} reaction rate as mentioned in Section \ref{ssec:co_reaction_rate}, suggesting that an attenuated reaction rate should be considered in future simulations. Furthermore, the production of these elements is strongly influenced by the metallicity of the progenitor, as a lower electron fraction $Y_{e}$ alters which isotopes are favored during nuclear burning. Variations in $Y_{e}$ during the burning phase can substantially alter the resulting yields of these intermediate-mass elements, e.g. a lower $Y_{e}$ will result in the production of more neutron-rich isotopes. As such, precise modeling of the neutronization rate and the effects of metallicity are of great importance when one aims to accurately reproduce the observed abundance yields for these elements.

In comparison to the above abundance ratios, which worked either seldom or sometimes, the ratio of Fe/Si tended to work much more frequently, with the abundance ratios predicted by Type Ia models tending to be higher than those predicted by core-collapse models. This is largely expected as $^{56}$Ni -- which decays via $^{56}$Co into $^{56}$Fe -- is the most prominent isotope produced by a Type Ia supernova, and is considered one of the main sources of power behind the optical light curve for these events. Since there is a relation between the peak luminosity of a Type Ia event and the amount of $^{56}$Ni produced in the associated supernova, $^{56}$Ni is often used as a calibration marker for simulations of Type Ia supernova explosions. It stands to reason, then, that the abundance of $^{56}$Fe -- as the main decay product of $^{56}$Ni -- would serve as a fairly reliable marker when comparing between observations and simulations. While we do find this to be a general trend across our results, it is worth noting that this was not always the case. In particular, the Fe/Si ratios predicted by low-mass ($\sim 10-15$ M$_{\odot}$) core-collapse models were comparable to the lowest Fe/Si ratios predicted by some Type Ia models, leaving room for ambiguity. As such, while the Fe/Si ratio provides important information and is useful in distinguishing between sub-models, it is difficult to recommend its use as a sole discriminator between SN classifications.

One aspect that remains underexplored in nucleosynthesis models is that of progenitor metallicity. Metallicity is believed to be a primary source of neutron excess (defined as $\eta = 1 - 2 Y_{e}$) in the progenitors of Type Ia supernovae, though in this regard, it may be degenerate with other mechanisms, such as the amount of carbon simmering prior to the explosion, or the state of neutron-rich nuclear statistical equilibrium \citep{martinez-rodriguez_2017}. While it is unclear whether this effect is due solely to the different values of metallicity or to other mechanisms with which the metallicity may be degenerate, and while only two of the tested model sets -- LN18 and B19 -- clearly incorporated progenitor metallicity as a variable, lower-metallicity models consistently produced higher abundance ratios for the intermediate-mass elements, yielding better agreement with the majority of our observations. This preference for low metallicity could be physically motivated by delay-time considerations: Type Ia supernovae originate from white dwarfs formed by lower-mass stars, which have longer lifetimes (inversely proportional to mass) and thus belong to older, more metal-poor stellar populations. Our results support lower-metallicity progenitors for these events, as the production of IMEs appears to be notably influenced by this parameter. While earlier studies typically assumed solar metallicity, recent work has increasingly included metallicity as a parameter, which our findings suggest is crucial for future simulations.

Finally, while it is not something that was directly addressed in this study, we consider explosion energy. The canonical supernova explosion energy is $10^{51}$ erg, which many models adopt as a single fixed value. However, observational studies (e.g., \cite{foley_2009, lovegrove_2013, stritzinger_2014}) have measured explosion energies that deviate significantly—sometimes by orders of magnitude—from this standard value. While some simulations have explored nominally lower-energy explosion mechanisms (e.g., deflagrations versus detonations), only \cite{fryer_2018} have parametrized explosion energy itself, doing so for core-collapse supernovae based on injection energy (see also \citealt{2023MNRAS.525.6257B}). To our knowledge, no comparable study exists for thermonuclear supernovae, despite observational evidence for energy variations. A single canonical explosion energy clearly cannot represent the diversity of supernova events, making this parameter essential for future simulations to explore.

\section{Conclusions and Future Work}
\label{sec:conclusion}
We have presented a systematic X-ray study of 13 supernova remnants --  G1.9+0.3, G4.5+6.8 (Kepler's SNR), G120.1+1.4 (Tycho's SNR), G272.2–3.2, G337.2–0.7, G344.7–0.1, G352.7–0.1, N0505–67.9 (DEM L71), N0509–67.5, N0509–68.7 (N103B), N0519–69.0, N0534–69.9, and N0548–70.4 -- which were selected as a result of their young ages, the presence of thermal emission and enhanced ejecta, and their presumed Type Ia origins. As a spatially-resolved spectroscopic study, regions were selected using a nearest-neighbor algorithm that follows surface brightness variations, given the established correlation between surface brightness and physical properties including density, temperature, and elemental abundances \citep{sanders_2006}. The exception was SNR G1.9+0.3, where regions were chosen based on a previous study due to the remnant's dominant non-thermal emission. X-ray spectra extracted from each region were fitted with appropriate plasma models to derive temperatures and elemental abundances. These measured abundances were then compared against 335 individual models spanning 11 commonly-used supernova nucleosynthesis simulation sets from the literature: WW95, MN03, N06, M10, S13, F14, S16, LN18, T18, B19, and LN20.

Through these comparisons, we find that, while for some objects there is strong agreement between the models and observations, in general, the observed abundances do not match the entirety of the predicted yields for any of the models. As a result, while we present conclusions for the classification of each involved object, these conclusions are largely based upon the ratios of only a select few elements, rather than all of those which were examined for a given object. We identify some limitations in the current nucleosynthesis models that might lead to these disagreements, and explore them further. We examine what aspects of the models tend to work well, as well as those that fare more poorly, identifying some possible physical considerations that might result in the various levels of agreement. We provide feedback for the modeling community to help inform future simulations by identifying some parameters that would lend themselves well to further exploration. Specifically, we identify four areas that future simulations should focus on that might help account for our findings: a refinement of the reaction rates for certain nuclear reactions, such as that of \textsuperscript{12}C +\textsuperscript{16}\text{O}; an increase in dimensionality to ensure simulations can accurately capture the effects of turbulence, mixing, and convection; exploring the parameter space of progenitor metallicity or neutronization due to its effects on certain reaction rates; and investigating further the possibility of explosion energies that vary from the canonical value of $10^{51}$ erg due to observations and studies that suggest that such a value may not always be the case. Further, while an exploration of the possible connections between abundance ratios and physical parameters may be worth pursuing, it is difficult to do so with the currently existing models owing to their limited exploration of their parameter spaces, and the inherent differences between any given sets of simulations (e.g. dimensionality, how particular parameters are treated, etc.). Should future models explore more thoroughly these parameter spaces, it may be possible to gain further insight into how a given physical parameter affects the production of particular abundance ratios. Finally, we must advocate for the inclusion of spatially-resolved yield data in the models of supernova nucleosynthesis yields: while modern X-ray telescopes provide us with the necessary angular resolution to perform spatially-resolved studies and determine the abundance yields present in arcsecond-scale regions across an SNR, modern models of supernova nucleosynthesis present information as it pertains to the entire SNR. While not all models will be able to do so (owing to differences in the simulations, such as dimensionality), including such information could allow for a significantly more thorough analysis by making use of one of the greatest strengths of modern X-ray telescopes.

In addition to necessary improvements in the models of supernova nucleosynthesis, these results highlight the need for improvements in the observational data itself. As noted in \S\ref{ssec:caveats}, any future study will benefit immensely from the high-resolution X-ray spectroscopy available to the recently-launched \textit{XRISM} mission, and from future missions such as \textit{NewAthena} \citep{2025hsa..conf..148C, 2025NatAs...9...36C}. The improvements -- specifically in areas such as spectral resolution -- would help reduce degeneracies in the parameter space, and allow for a more accurate measurement of the chemical properties of these objects. Sensitive, high-resolution imaging instruments -- such as those aboard the proposed \textit{AXIS} Probe mission \citep{2023SPIE12678E..1ER, 2023arXiv231107673S} -- would also strongly benefit future studies, allowing for ejecta to be more accurately localized and isolated from swept-up material.

\begin{acknowledgments}

This research made use of HEASARC, the High-Energy Astrophysics Archive Research Centre maintained at the Goddard Space Flight Center, and of SNRcat, the high-energy catalogue of Supernova Remnants maintained at the University of Manitoba. We acknowledge support from the Natural Sciences and Research Council of Canada (NSERC) through the Discovery Grants and the Canada Research Chairs programs, from the University of Manitoba's GETS (Graduate Enhancement for Tri-Agency Stipends) program, and from the Canadian Space Agency. We also thank the referee for their thorough reading of the manuscript, and for their detailed comments that helped improve the clarity and quality of the paper.

This paper employs a list of Chandra datasets, obtained by the Chandra X-ray Observatory, contained in the Chandra Data Collection ~\dataset[DOI:~10.25574/cdc.618]{https://doi.org/10.25574/cdc.618}

\end{acknowledgments}

\facilities{XMM, CXO}
\software{HEAsoft \citep{heasoft}, XMM-SAS \citep{xmm-sas}, CIAO \citep{ciao}, SAOImage~DS9 \citep{sao_ds9}, XSPEC v12.11.1 \citep{xspec}}

\begin{deluxetable*}{ccccc}
\centering
\tablecaption{\textbf{G1.9+0.3}: Best fit parameters per region \label{tab:g1.9_results}}
\tablehead{
\colhead{Parameter} & 
\colhead{N} & 
\colhead{NE} & 
\colhead{NW} & 
\colhead{W} 
}
\startdata
\makecell{N$_{\rm H}$\\($10^{22}$ cm$^{-2}$)} &
  $8.27_{-0.32}^{+0.52}$ &
  $7.80_{-0.82}^{+1.73}$ &
  $9.23_{-0.45}^{+0.92}$ &
  $8.50_{-0.35}^{+0.93}$ \\
\makecell{kT\\(keV)} &
  $3.54_{-0.01}^{+0.00}$ &
  $5.36_{-0.60}^{+0.43}$ &
  $3.61_{-0.16}^{+0.10}$ &
  $4.43_{-0.68}^{+0.10}$ \\
Si                     & $4.50_{-1.54}^{+2.77}$ & $4.35_{-3.64}^{+8.83}$ & $3.88_{-3.38}^{+4.86}$ & $3.65_{-2.92}^{+2.16}$ \\
S                      & $4.63_{-1.40}^{+1.36}$ & $2.35_{-2.15}^{+3.81}$ & -                      & $2.10_{-1.24}^{+1.22}$ \\
Fe                     & $2.64_{-0.76}^{+0.84}$ & $0.45_{-0.39}^{+0.99}$ & $1.39_{-1.02}^{+1.11}$ & -                      \\
\makecell{$\tau$\\($10^{8}$ cm$^{-3}$ s )} &
  $16.1_{-9.8}^{+8.5}$ &
  $8.99_{-7.24}^{+28.6}$ &
  $4.46_{-2.70}^{+40.4}$ &
  $9.69_{-8.14}^{+69.1}$ \\
\makecell{Redshift\\($10^{-3}$)} &
  $-6.75_{-6.15}^{+3.78}$ &
  $6.91_{-28}^{+57}$ &
  $-5.57_{-13.9}^{+12.9}$ &
  $3.69_{-8.17}^{+14.4}$ \\
$\chi_{\nu}^{2}$ (DoF) & 0.93 (553)             & 0.92 (277)             & 0.92 (249)             & 0.91 (497)   
\enddata
\end{deluxetable*}

\begin{deluxetable*}{ccccccccccc}
\centering
\tablecaption{\textbf{G4.5+6.8}: Best fit parameters per region for regions R00 through R09 \label{tab:g4.5_results_1}}
\tablehead{
\colhead{Parameter} & 
\colhead{R00} & 
\colhead{R01} & 
\colhead{R02} & 
\colhead{R03} &
\colhead{R04} & 
\colhead{R05} & 
\colhead{R06} & 
\colhead{R07} & 
\colhead{R08} & 
\colhead{R09}
}
\startdata
\makecell{N$_{\rm H}$\\($\times 10^{22}$ cm$^{-2}$)} &
  $0.51_{-0.02}^{+0.02}$ &
  $0.48_{-0.01}^{+0.01}$ &
  $0.50_{-0.01}^{+0.01}$ &
  $0.48_{-0.01}^{+0.02}$ &
  $0.50_{-0.01}^{+0.01}$ &
  $0.56_{-0.01}^{+0.01}$ &
  $0.48_{-0.01}^{+0.01}$ &
  $0.62_{-0.02}^{+0.01}$ &
  $0.52_{-0.01}^{+0.01}$ &
  $0.45_{-0.01}^{+0.02}$ \\
\makecell{kT$_{\rm c}$\\(keV)} &
  $0.72_{-0.01}^{+0.01}$ &
  $0.71_{-0.01}^{+0.01}$ &
  $0.71_{-0.01}^{+0.01}$ &
  $0.71_{-0.01}^{+0.01}$ &
  $0.71_{-0.01}^{+0.01}$ &
  $0.64_{-0.01}^{+0.01}$ &
  $0.66_{-0.01}^{+0.01}$ &
  $0.56_{-0.01}^{+0.01}$ &
  $0.70_{-0.01}^{+0.01}$ &
  $0.66_{-0.01}^{+0.01}$ \\
Ne &
  $0.74_{-0.03}^{+0.05}$ &
  $0.37_{-0.05}^{+0.05}$ &
  $0.37_{-0.05}^{+0.03}$ &
  $0.55_{-0.05}^{+0.03}$ &
  $0.28_{-0.03}^{+0.01}$ &
  $0.29_{-0.03}^{+0.05}$ &
  $0.60_{-0.04}^{+0.04}$ &
  $0.22_{-0.01}^{+0.03}$ &
  $0.27_{-0.01}^{+0.01}$ &
  $0.49_{-0.03}^{+0.05}$ \\
Mg &
  $0.66_{-0.07}^{+0.08}$ &
  $0$ &
  $0.12_{-0.05}^{+0.06}$ &
  $0.46_{-0.06}^{+0.05}$ &
  $0.04_{-0.04}^{+0.02}$ &
  $0.01_{-0.01}^{+0.07}$ &
  $0.32_{-0.06}^{+0.05}$ &
  $0.01_{-0.01}^{+0.02}$ &
  $0.09_{-0.05}^{+0.06}$ &
  $0.51_{-0.08}^{+0.02}$ \\
Si &
  $4.10_{-0.15}^{+0.13}$ &
  $5.53_{-0.17}^{+0.25}$ &
  $5.12_{-0.21}^{+0.15}$ &
  $4.45_{-0.16}^{+0.14}$ &
  $5.08_{-0.18}^{+0.18}$ &
  $6.73_{-0.13}^{+0.23}$ &
  $4.99_{-0.16}^{+0.16}$ &
  $5.60_{-0.19}^{+0.22}$ &
  $6.52_{-0.13}^{+0.30}$ &
  $6.18_{-0.26}^{+0.07}$ \\
S &
  $5.02_{-0.20}^{+0.23}$ &
  $7.57_{-0.30}^{+0.30}$ &
  $6.71_{-0.28}^{+0.21}$ &
  $5.15_{-0.24}^{+0.18}$ &
  $7.42_{-0.35}^{+0.24}$ &
  $8.63_{-0.24}^{+0.37}$ &
  $5.55_{-0.24}^{+0.19}$ &
  $7.40_{-0.21}^{+0.35}$ &
  $8.74_{-0.18}^{+0.42}$ &
  $7.83_{-0.33}^{+0.26}$ \\
Ar &
  $7.64_{-0.81}^{+0.68}$ &
  $11.5_{-1.0}^{+0.9}$ &
  $9.66_{-0.85}^{+0.85}$ &
  $7.41_{-0.69}^{+0.50}$ &
  $10.9_{-0.9}^{+0.7}$ &
  $12.5_{-1.0}^{+1.0}$ &
  $6.94_{-0.98}^{+0.75}$ &
  $12.8_{-1.8}^{+0.3}$ &
  $15.0_{-1.0}^{+1.1}$ &
  $11.2_{-1.4}^{+1.3}$ \\
\makecell{$\tau_{\rm c}$\\($\times 10^{11}$ cm$^{-3}$ s )} &
  $2.08_{-0.13}^{+0.16}$ &
  $2.69_{-0.25}^{+0.12}$ &
  $2.39_{-0.14}^{+0.23}$ &
  $1.96_{-0.13}^{+0.10}$ &
  $3.09_{-0.34}^{+0.15}$ &
  $2.93_{-0.22}^{+0.24}$ &
  $2.45_{-0.13}^{+0.19}$ &
  $2.83_{-0.14}^{+0.93}$ &
  $2.30_{-0.10}^{+0.24}$ &
  $2.02_{-0.14}^{+0.13}$ \\
\makecell{kT$_{\rm h}$\\(keV)} &
  $5.36_{-0.54}^{+0.82}$ &
  $10.8_{-1.0}^{+0.9}$ &
  $9.33_{-0.54}^{+0.57}$ &
  $6.18_{-0.78}^{+0.33}$ &
  $10.2_{-1.0}^{+0.4}$ &
  $5.34_{-0.31}^{+0.26}$ &
  $6.70_{-0.78}^{+0.75}$ &
  $4.46_{-0.31}^{+0.30}$ &
  $10.9_{-0.7}^{+2.4}$ &
  $4.88_{-0.42}^{+0.17}$ \\
Ca &
  $9.05_{-3.06}^{+1.55}$ &
  $17.4_{-5.0}^{+3.2}$ &
  $17.3_{-3.7}^{+3.2}$ &
  $12.3_{-3.3}^{+3.6}$ &
  $21.8_{-4.8}^{+4.1}$ &
  $10.8_{-4.0}^{+4.4}$ &
  $8.51_{-3.46}^{+4.26}$ &
  $9.51_{-2.72}^{+3.62}$ &
  $15.5_{-3.4}^{+2.4}$ &
  $10.6_{-7.5}^{+5.3}$ \\
Fe &
  $7.35_{-0.99}^{+0.85}$ &
  $8.71_{-0.52}^{+0.74}$ &
  $7.05_{-0.36}^{+0.58}$ &
  $10.2_{-0.6}^{+1.9}$ &
  $9.61_{-0.31}^{+0.69}$ &
  $13.2_{-0.8}^{+1.2}$ &
  $7.22_{-0.73}^{+1.02}$ &
  $15.9_{-1.3}^{+1.6}$ &
  $4.45_{-0.51}^{+0.22}$ &
  $25.1_{-2.2}^{+2.1}$ \\
\makecell{$\tau_{\rm h}$\\($\times 10^{9}$ cm$^{-3}$ s )} &
  $4.65_{-0.32}^{+0.39}$ &
  $6.99_{-0.29}^{+0.22}$ &
  $6.50_{-0.31}^{+0.29}$ &
  $4.60_{-0.34}^{+0.16}$ &
  $6.69_{-0.33}^{+0.11}$ &
  $4.37_{-0.15}^{+0.11}$ &
  $4.41_{-0.27}^{+0.28}$ &
  $3.74_{-0.22}^{+0.06}$ &
  $6.01_{-0.19}^{+0.58}$ &
  $3.58_{-0.18}^{+0.08}$ \\
\makecell{Redshift\\($10^{-3}$)} &
  $-3.96_{-0.06}^{+0.03}$ &
  $-1.77_{-0.03}^{+0.04}$ &
  $-4.03_{-0.01}^{+0.05}$ &
  $-2.98_{-0.02}^{+0.01}$ &
  $-2.97_{-0.02}^{+0.07}$ &
  $-1.77_{-0.02}^{+0.04}$ &
  $-1.77_{-0.02}^{+0.07}$ &
  $-2.97_{-0.02}^{+0.06}$ &
  $-1.77_{-0.01}^{+0.15}$ &
  $2.45_{-0.01}^{+0.13}$ \\
\makecell{Line Centroid\\(keV)} &
  $1.21_{-0.01}^{+0.01}$ &
  $1.21_{-0.01}^{+0.01}$ &
  $1.22_{-0.01}^{+0.01}$ &
  $1.20_{-0.01}^{+0.01}$ &
  $1.21_{-0.01}^{+0.01}$ &
  $1.18_{-0.01}^{+0.00}$ &
  $1.20_{-0.01}^{+0.01}$ &
  $1.20_{-0.01}^{+0.01}$ &
  $1.20_{-0.01}^{+0.01}$ &
  $1.14_{-0.01}^{+0.01}$ \\
$\chi_{\nu}^{2}$ (DoF) &
  $1.44$ ($1126$) &
  $1.89$ ($1172$) &
  $1.72$ ($1172$) &
  $1.46$ ($1113$) &
  $1.83$ ($1188$) &
  $1.89$ ($1181$) &
  $1.59$ ($1151$) &
  $1.88$ ($1196$) &
  $1.73$ ($1240$) &
  $1.71$ ($1102$)
\enddata
\end{deluxetable*}

\begin{deluxetable*}{ccccccccc}
\centering
\tablecaption{\textbf{G4.5+6.8}: Best fit parameters per region for regions R10 through R17\label{tab:g4.5_results_2}}
\tablehead{
\colhead{Parameter} & 
\colhead{R10} & 
\colhead{R11} & 
\colhead{R12} & 
\colhead{R13} &
\colhead{R14} & 
\colhead{R15} & 
\colhead{R16} & 
\colhead{R17}
}
\startdata
\makecell{N$_{\rm H}$\\($\times 10^{22}$ cm$^{-2}$)} &
  $0.44_{-0.01}^{+0.03}$ &
  $0.55_{-0.01}^{+0.02}$ &
  $0.53_{-0.01}^{+0.01}$ &
  $0.56_{-0.01}^{+0.01}$ &
  $0.44_{-0.01}^{+0.01}$ &
  $0.64_{-0.02}^{+0.02}$ &
  $0.76_{-0.01}^{+0.01}$ &
  $0.59_{-0.01}^{+0.01}$ \\
\makecell{kT$_{\rm c}$\\(keV)} &
  $0.68_{-0.01}^{+0.00}$ &
  $0.57_{-0.01}^{+0.01}$ &
  $0.60_{-0.01}^{+0.01}$ &
  $0.58_{-0.01}^{+0.01}$ &
  $0.51_{-0.01}^{+0.00}$ &
  $0.61_{-0.01}^{+1.27}$ &
  $0.41_{-0.01}^{+0.01}$ &
  $0.51_{-0.01}^{+0.00}$ \\
Ne &
  $0.53_{+0.01}^{+0.01}$ &
  $0.31_{-0.06}^{+0.01}$ &
  $0.54_{-0.04}^{+0.04}$ &
  $0.35_{-0.04}^{+0.05}$ &
  $0.15_{-0.01}^{+0.01}$ &
  $0.39_{-0.19}^{+0.04}$ &
  $0.06_{-0.03}^{+0.02}$ &
  $0.29_{-0.07}^{+0.04}$ \\
Mg &
  $0.62_{-0.05}^{+0.05}$ &
  $0.01_{-0.01}^{+0.03}$ &
  $0.32_{-0.06}^{+0.04}$ &
  $0.22_{-0.11}^{+0.04}$ &
  $0.39_{-0.03}^{+0.03}$ &
  $0.39_{-0.36}^{+0.05}$ &
  $0.07_{-0.04}^{+0.01}$ &
  $0.16_{-0.04}^{+0.07}$ \\
Si &
  $4.46_{-0.14}^{+0.14}$ &
  $6.18_{-0.26}^{+0.04}$ &
  $6.13_{-0.23}^{+0.15}$ &
  $6.90_{-0.22}^{+0.21}$ &
  $8.05_{-0.15}^{+0.33}$ &
  $5.53_{-0.04}^{+1.15}$ &
  $8.19_{-0.23}^{+0.22}$ &
  $7.93_{-0.29}^{+0.36}$ \\
S &
  $5.14_{-0.22}^{+0.24}$ &
  $7.12_{-0.31}^{+0.09}$ &
  $5.87_{-0.27}^{+0.27}$ &
  $8.92_{-0.31}^{+0.39}$ &
  $9.82_{-0.31}^{+0.69}$ &
  $7.16_{-0.14}^{+4.76}$ &
  $11.2_{-0.5}^{+0.6}$ &
  $9.07_{-0.25}^{+0.62}$ \\
Ar &
  $7.24_{-0.95}^{+1.03}$ &
  $8.78_{-0.51}^{+0.77}$ &
  $6.40_{-1.04}^{+0.75}$ &
  $14.2_{-1.4}^{+1.6}$ &
  $20.0_{-1.6}^{+1.7}$ &
  $12.1_{-1.6}^{+1.2}$ &
  $32.8_{-2.5}^{+2.2}$ &
  $13.1_{-2.2}^{+1.7}$ \\
\makecell{$\tau_{\rm c}$\\($\times 10^{11}$ cm$^{-3}$ s )} &
  $1.71_{-0.91}^{+0.92}$ &
  $4.05_{-0.91}^{+0.02}$ &
  $2.69_{-0.12}^{+0.25}$ &
  $3.40_{-0.33}^{+0.32}$ &
  $20.6_{-4.65}^{+4.16}$ &
  $2.66_{-2.49}^{+0.56}$ &
  $53.5_{-29.3}^{+41.7}$ &
  $5.16_{-0.64}^{+1.46}$ \\
\makecell{kT$_{\rm h}$\\(keV)} &
  $5.26_{-1.15}^{+0.20}$ &
  $5.57_{-0.67}^{+0.05}$ &
  $4.61_{-0.35}^{+0.28}$ &
  $4.32_{-0.10}^{+0.50}$ &
  $17.5_{-0.8}^{+1.2}$ &
  $5.00_{-2.51}^{+0.50}$ &
  $4.40_{-0.02}^{+0.06}$ &
  $3.64_{-0.19}^{+0.17}$ \\
Ca &
  $12.4_{-4.0}^{+7.0}$ &
  $7.29_{-3.92}^{+2.28}$ &
  $9.00_{-3.84}^{+4.60}$ &
  $8.99_{-5.13}^{+4.14}$ &
  $6.59_{-0.25}^{+0.33}$ &
  $11.03_{-2.71}^{+2.63}$ &
  $10.8_{-2.9}^{+1.1}$ &
  $6.00_{-4.20}^{+2.61}$ \\
Fe &
  $12.4_{-0.5}^{+5.3}$ &
  $9.90_{-0.56}^{+1.03}$ &
  $12.5_{-0.8}^{+1.8}$ &
  $12.6_{-2.4}^{+0.9}$ &
  $2.65_{-0.15}^{+0.18}$ &
  $4.52_{-0.13}^{+6.66}$ &
  $5.28_{-0.39}^{+0.21}$ &
  $13.2_{-1.1}^{+1.0}$ \\
\makecell{$\tau_{\rm h}$\\($\times 10^{9}$ cm$^{-3}$ s )} &
  $3.63_{-0.49}^{+0.05}$ &
  $3.88_{-0.25}^{+0.03}$ &
  $3.66_{-0.20}^{+0.08}$ &
  $3.50_{-0.08}^{+0.26}$ &
  $19.9_{-0.62}^{+0.57}$ &
  $4.17_{-0.20}^{+0.19}$ &
  $3.16_{-0.03}^{+0.10}$ &
  $3.14_{-0.08}^{+0.09}$ \\
\makecell{Redshift\\($10^{-3}$)} &
  $1.06_{-0.01}^{+0.01}$ &
  $-1.78_{-0.03}^{+0.01}$ &
  $-1.76_{-0.01}^{+0.05}$ &
  $1.15_{-0.09}^{+0.03}$ &
  $1.16_{-0.16}^{+0.23}$ &
  $-3.00_{-0.01}^{+0.02}$ &
  $-2.34_{-0.05}^{+0.01}$ &
  $-2.28_{-0.01}^{+0.21}$ \\
\makecell{Line Centroid\\(keV)} &
  $1.14_{-0.01}^{+0.01}$ &
  $1.18_{-0.01}^{+0.01}$ &
  $1.18_{-0.01}^{+0.01}$ &
  $1.16_{-0.01}^{+0.02}$ &
  $1.19_{-0.01}^{+0.00}$ &
  $1.22_{-0.01}^{+0.04}$ &
  $1.22_{-0.01}^{+0.01}$ &
  $1.15_{-0.01}^{+0.01}$ \\
$\chi_{\nu}^{2}$ (DoF) &
  $1.69$ ($1101$) &
  $2.32$ ($1249$) &
  $1.87$ ($1166$) &
  $1.70$ ($1206$) &
  $1.88$ ($1205$) &
  $1.60$ ($1306$) &
  $1.84$ ($1426$) &
  $1.81$ ($1245$)
\enddata
\end{deluxetable*}

\begin{deluxetable*}{ccccccccccc}
\centering
\tablecaption{\textbf{G120.1+1.4}: Best fit parameters per region for regions R00 through R09 \label{tab:g120_results_1}}
\tablehead{
\colhead{Parameter} & 
\colhead{R00} & 
\colhead{R01} & 
\colhead{R02} & 
\colhead{R03} &
\colhead{R04} & 
\colhead{R05} & 
\colhead{R06} & 
\colhead{R07} & 
\colhead{R08} & 
\colhead{R09}
}
\startdata
\makecell{N$_{\rm H}$\\($\times 10^{22}$ cm$^{-2}$)} &
  $0.98_{-0.01}^{+0.01}$ &
  $0.98_{-0.02}^{+0.01}$ &
  $1.09_{-0.01}^{+0.01}$ &
  $1.01_{-0.01}^{+0.01}$ &
  $1.10_{-0.01}^{+0.01}$ &
  $1.21_{-0.01}^{+0.01}$ &
  $1.17_{-0.02}^{+0.01}$ &
  $1.08_{-0.01}^{+0.01}$ &
  $1.28_{-0.01}^{+0.01}$ &
  $1.05_{-0.01}^{+0.01}$ \\
\makecell{kT$_{\rm c}$\\(keV)} &
  $0.56_{-0.02}^{+0.06}$ &
  $0.68_{-0.01}^{+0.01}$ &
  $0.52_{-0.01}^{+0.04}$ &
  $0.55_{-0.06}^{+0.02}$ &
  $0.44_{-0.01}^{+0.02}$ &
  $0.48_{-0.04}^{+0.01}$ &
  $0.46_{-0.06}^{+0.01}$ &
  $0.61_{-0.02}^{+0.01}$ &
  $0.47_{-0.05}^{+0.01}$ &
  $0.83_{-0.03}^{+0.03}$ \\
O &
  $1.78_{-0.07}^{+0.20}$ &
  $1.96_{-0.09}^{+0.09}$ &
  $2.31_{-0.01}^{+0.34}$ &
  $2.53_{-0.12}^{+0.16}$ &
  $1.13_{-0.09}^{+0.08}$ &
  $2.13_{-0.13}^{+0.03}$ &
  $3.75_{-0.27}^{+0.28}$ &
  $2.90_{-0.09}^{+0.06}$ &
  $7.32_{-0.24}^{+0.49}$ &
  $3.78_{-0.21}^{+0.11}$ \\
Mg &
  $0.56_{-0.02}^{+0.06}$ &
  $0.68_{-0.01}^{+0.01}$ &
  $0.52_{-0.01}^{+0.04}$ &
  $0.55_{-0.06}^{+0.02}$ &
  $0.44_{-0.01}^{+0.02}$ &
  $0.48_{-0.04}^{+0.01}$ &
  $0.46_{-0.06}^{+0.01}$ &
  $0.61_{-0.02}^{+0.01}$ &
  $0.47_{-0.05}^{+0.01}$ &
  $0.83_{-0.03}^{+0.03}$ \\
Si &
  $27.2_{-0.6}^{+1.3}$ &
  $28.9_{-0.1}^{+1.0}$ &
  $31.6_{-0.3}^{+1.5}$ &
  $23.8_{-1.2}^{+0.6}$ &
  $11.6_{-0.2}^{+0.5}$ &
  $24.8_{-1.5}^{+0.1}$ &
  $30.0_{-0.9}^{+1.9}$ &
  $15.4_{-0.4}^{+0.2}$ &
  $36.0_{-0.6}^{+1.9}$ &
  $25.7_{-0.7}^{+0.6}$ \\
S &
  $29.0_{-0.6}^{+1.6}$ &
  $29.7_{-0.1}^{+1.0}$ &
  $33.7_{-0.3}^{+1.8}$ &
  $25.3_{-1.4}^{+0.6}$ &
  $12.3_{-0.2}^{+0.6}$ &
  $30.6_{-1.7}^{+0.6}$ &
  $31.7_{-1.2}^{+1.7}$ &
  $16.1_{-0.6}^{+0.3}$ &
  $36.7_{-0.9}^{+2.0}$ &
  $26.1_{-0.6}^{+1.0}$ \\
Ar &
  $42.9_{-1.2}^{+2.3}$ &
  $42.9_{-0.5}^{+2.4}$ &
  $55.0_{-1.3}^{+3.0}$ &
  $38.4_{-0.9}^{+2.2}$ &
  $18.5_{-0.9}^{+0.7}$ &
  $65.3_{-2.2}^{+4.9}$ &
  $53.3_{-2.7}^{+1.3}$ &
  $24.7_{-0.5}^{+0.7}$ &
  $63.9_{-1.7}^{+3.9}$ &
  $32.3_{-1.1}^{+2.8}$ \\
Ca &
  $112_{-5}^{+7}$ &
  $117_{-9}^{+5}$ &
  $162_{-14}^{+14}$ &
  $113_{-4}^{+7}$ &
  $79.6_{-4.1}^{+6.6}$ &
  $254_{-12}^{+16}$ &
  $256_{-30}^{+9}$ &
  $68.9_{-2.1}^{+1.9}$ &
  $326_{-17}^{+33}$ &
  $102_{-5}^{+10}$ \\
\makecell{$\tau_{\rm c}$\\($\times 10^{11}$ cm$^{-3}$ s )} &
  $2.22_{-0.11}^{+0.22}$ &
  $2.47_{-0.13}^{+0.05}$ &
  $2.18_{-0.15}^{+0.12}$ &
  $2.09_{-0.24}^{+0.12}$ &
  $1.41_{-0.10}^{+0.14}$ &
  $0.53_{-0.02}^{+0.03}$ &
  $2.27_{-0.12}^{+0.36}$ &
  $3.82_{-0.29}^{+0.83}$ &
  $2.84_{-0.12}^{+0.05}$ &
  $1.89_{-0.16}^{+0.07}$ \\
\makecell{kT$_{\rm h}$\\(keV)} &
  $3.19_{-0.07}^{+0.09}$ &
  $3.25_{-0.01}^{+0.15}$ &
  $3.60_{-0.01}^{+0.01}$ &
  $3.13_{-0.13}^{+0.10}$ &
  $4.11_{-0.30}^{+0.24}$ &
  $2.14_{-0.08}^{+0.02}$ &
  $2.99_{-0.06}^{+0.17}$ &
  $6.21_{-0.39}^{+0.60}$ &
  $3.01_{-0.07}^{+0.09}$ &
  $3.54_{-0.09}^{+0.14}$ \\
Ca &
  $25.2_{-4.1}^{+4.0}$ &
  $26.3_{-4.8}^{+2.4}$ &
  $39.4_{-5.7}^{+7.5}$ &
  $7.46_{-4.95}^{+2.89}$ &
  $7.46_{-2.24}^{+1.79}$ &
  - &
  $12.0_{-2.6}^{+1.9}$ &
  $4.25_{-1.39}^{+0.73}$ &
  $9.28_{-1.92}^{+1.87}$ &
  $26.0_{-5.5}^{+4.2}$ \\
Fe &
  $14.1_{-0.7}^{+0.3}$ &
  $15.4_{-1.5}^{+0.1}$ &
  $13.1_{-0.4}^{+0.2}$ &
  $11.2_{-0.2}^{+0.5}$ &
  $4.48_{-0.40}^{+0.32}$ &
  $21.3_{-0.9}^{+3.2}$ &
  $10.5_{-1.0}^{+0.2}$ &
  $1.20_{-0.30}^{+0.12}$ &
  $8.09_{-0.42}^{+0.28}$ &
  $14.5_{-0.6}^{+0.5}$ \\
\makecell{$\tau_{\rm h}$\\($\times 10^{9}$ cm$^{-3}$ s )} &
  $7.36_{-0.06}^{+0.29}$ &
  $7.14_{-0.04}^{+0.20}$ &
  $7.09_{-0.02}^{+0.27}$ &
  $6.68_{-0.15}^{+0.11}$ &
  $6.11_{-0.25}^{+0.35}$ &
  $4.05_{-0.22}^{+0.10}$ &
  $6.54_{-0.03}^{+0.15}$ &
  $0.10_{-0.02}^{+0.20}$ &
  $6.05_{-0.09}^{+0.18}$ &
  $6.75_{-0.13}^{+0.26}$ \\
\makecell{Redshift\\($10^{-3}$)} &
  $-0.58_{-0.09}^{+0.03}$ &
  $-2.38_{-0.06}^{+0.01}$ &
  $-0.55_{-0.06}^{+0.16}$ &
  $-0.51_{-0.05}^{+0.01}$ &
  $-2.96_{-0.03}^{+0.10}$ &
  $-0.64_{-0.01}^{+0.09}$ &
  $-1.58_{-0.20}^{+0.01}$ &
  $-2.10_{-0.01}^{+0.01}$ &
  $-1.78_{-0.01}^{+0.01}$ &
  $-0.30_{-0.01}^{+0.01}$ \\
\makecell{Line Centroid\\(keV)} &
  $0.74_{-0.01}^{+0.01}$ &
  $0.74_{-0.01}^{+0.01}$ &
  $0.73_{-0.01}^{+0.01}$ &
  $0.73_{-0.01}^{+0.01}$ &
  $0.73_{-0.01}^{+0.01}$ &
  $0.73_{-0.01}^{+0.01}$ &
  $0.73_{-0.01}^{+0.01}$ &
  $0.72_{-0.01}^{+0.01}$ &
  $0.73_{-0.01}^{+0.01}$ &
  $0.73_{-0.01}^{+0.01}$ \\
\makecell{Line Centroid\\(keV)} &
  $1.23_{-0.01}^{+0.01}$ &
  $1.23_{-0.01}^{+0.01}$ &
  $1.23_{-0.01}^{+0.01}$ &
  $1.23_{-0.01}^{+0.01}$ &
  $1.24_{-0.01}^{+0.01}$ &
  $1.23_{-0.01}^{+0.01}$ &
  $1.23_{-0.01}^{+0.01}$ &
  $1.23_{-0.01}^{+0.01}$ &
  $1.23_{-0.01}^{+0.01}$ &
  $1.23_{-0.01}^{+0.01}$ \\
$\chi_{\nu}^{2}$ (DoF) &
  $1.79$ ($4824$) &
  $1.93$ ($4844$) &
  $1.74$ ($4811$) &
  $1.72$ ($4961$) &
  $1.51$ ($5404$) &
  $2.31$ ($5426$) &
  $1.90$ ($5169$) &
  $1.69$ ($5645$) &
  $1.72$ ($5610$) &
  $1.79$ ($4948$)
\enddata
\end{deluxetable*}

\begin{deluxetable*}{ccccccccccc}
\centering
\tablecaption{\textbf{G120.1+1.4}: Best fit parameters per region for regions R10 through R19 \label{tab:g120_results_2}}
\tablehead{
\colhead{Parameter} & 
\colhead{R10} & 
\colhead{R11} & 
\colhead{R12} & 
\colhead{R13} &
\colhead{R14} & 
\colhead{R15} & 
\colhead{R16} & 
\colhead{R17} & 
\colhead{R18} & 
\colhead{R19}
}
\startdata
\makecell{N$_{\rm H}$\\($\times 10^{22}$ cm$^{-2}$)} &
  $1.26_{-0.01}^{+0.01}$ &
  $0.99_{-0.01}^{+0.01}$ &
  $1.14_{-0.01}^{+0.01}$ &
  $1.19_{-0.01}^{+0.01}$ &
  $1.13_{-0.01}^{+0.01}$ &
  $1.14_{-0.01}^{+0.02}$ &
  $1.14_{-0.01}^{+0.01}$ &
  $1.28_{-0.01}^{+0.01}$ &
  $1.28_{-0.01}^{+0.01}$ &
  $1.20_{-0.01}^{+0.02}$ \\
\makecell{kT$_{\rm c}$\\(keV)} &
  $0.51_{-0.02}^{+0.02}$ &
  $0.60_{-0.03}^{+0.03}$ &
  $0.95_{-0.06}^{+0.09}$ &
  $0.40_{-0.01}^{+0.02}$ &
  $0.63_{-0.01}^{+0.02}$ &
  $0.47_{-0.02}^{+0.03}$ &
  $0.81_{-0.04}^{+0.04}$ &
  $0.44_{-0.04}^{+0.01}$ &
  $0.76_{-0.05}^{+0.04}$ &
  $0.42_{-0.03}^{+0.02}$ \\
O &
  $2.42_{-0.35}^{+0.06}$ &
  $2.22_{-0.08}^{+0.08}$ &
  $4.12_{-0.18}^{+0.43}$ &
  $5.12_{-0.18}^{+0.13}$ &
  $1.41_{-0.01}^{+0.11}$ &
  $4.19_{-0.62}^{+0.35}$ &
  $2.94_{-0.23}^{+0.16}$ &
  $7.86_{-0.32}^{+0.65}$ &
  $4.40_{-0.48}^{+0.36}$ &
  $2.91_{-0.34}^{+0.31}$ \\
Mg &
  $0.51_{-0.02}^{+0.02}$ &
  $0.60_{-0.03}^{+0.03}$ &
  $0.95_{-0.06}^{+0.09}$ &
  $0.40_{-0.01}^{+0.02}$ &
  $0.63_{-0.01}^{+0.02}$ &
  $0.47_{-0.02}^{+0.03}$ &
  $0.81_{-0.04}^{+0.04}$ &
  $0.44_{-0.04}^{+0.01}$ &
  $0.76_{-0.05}^{+0.04}$ &
  $0.42_{-0.03}^{+0.02}$ \\
Si &
  $27.0_{-1.8}^{+0.6}$ &
  $27.2_{-0.4}^{+0.3}$ &
  $35.7_{-1.7}^{+3.1}$ &
  $8.79_{-0.11}^{+0.17}$ &
  $17.8_{-0.1}^{+0.8}$ &
  $31.3_{-2.6}^{+1.9}$ &
  $28.8_{-1.1}^{+1.1}$ &
  $26.5_{-0.7}^{+2.0}$ &
  $37.1_{-2.7}^{+1.7}$ &
  $26.3_{-2.5}^{+1.5}$ \\
S &
  $33.9_{-2.6}^{+0.5}$ &
  $26.5_{-0.5}^{+0.3}$ &
  $34.8_{-1.6}^{+3.3}$ &
  $8.55_{-0.10}^{+0.16}$ &
  $20.0_{-0.3}^{+0.85}$ &
  $32.5_{-2.1}^{+1.7}$ &
  $27.7_{-1.2}^{+1.2}$ &
  $26.5_{-0.6}^{+2.1}$ &
  $39.0_{-3.3}^{+1.4}$ &
  $26.6_{-1.6}^{+0.8}$ \\
Ar &
  $75.6_{-9.4}^{+1.1}$ &
  $35.9_{-0.5}^{+1.5}$ &
  $51.3_{-2.5}^{+5.6}$ &
  $9.93_{-0.20}^{+0.25}$ &
  $35.6_{-1.0}^{+1.2}$ &
  $50.7_{-0.3}^{+5.3}$ &
  $43.9_{-2.4}^{+1.0}$ &
  $42.0_{-1.4}^{+3.0}$ &
  $78.6_{-6.5}^{+7.3}$ &
  $47.0_{-3.0}^{+3.7}$ \\
Ca &
  $296_{-23}^{+17}$ &
  $104_{-7}^{+7}$ &
  $233_{-16}^{+32}$ &
  $25.5_{-1.9}^{+1.6}$ &
  $132_{-6}^{+8}$ &
  $237_{-4}^{+23}$ &
  $222_{-15}^{+20}$ &
  $193_{-5}^{+20}$ &
  $588_{-527}^{+48}$ &
  $258_{-11}^{+18}$ \\
\makecell{$\tau_{\rm c}$\\($\times 10^{11}$ cm$^{-3}$ s )} &
  $0.69_{-0.05}^{+0.01}$ &
  $1.95_{-0.08}^{+0.06}$ &
  $3.24_{-0.06}^{+0.44}$ &
  $1.69_{-0.05}^{+0.09}$ &
  $0.45_{-0.01}^{+0.01}$ &
  $3.70_{-0.62}^{+0.27}$ &
  $2.59_{-0.18}^{+0.11}$ &
  $2.74_{-0.03}^{+0.27}$ &
  $1.71_{-0.09}^{+0.10}$ &
  $1.43_{-0.31}^{+0.16}$ \\
\makecell{kT$_{\rm h}$\\(keV)} &
  $2.57_{-0.04}^{+0.13}$ &
  $3.53_{-0.01}^{+0.01}$ &
  $3.03_{-0.07}^{+0.09}$ &
  $10.3_{-2.1}^{+0.7}$ &
  $2.62_{-0.11}^{+0.02}$ &
  $3.00_{-0.05}^{+0.07}$ &
  $2.90_{-0.05}^{+0.09}$ &
  $3.47_{-0.13}^{+0.04}$ &
  $2.74_{-0.07}^{+0.04}$ &
  $2.53_{-0.05}^{+0.05}$ \\
Ca &
  $11.6_{-7.2}^{+1.2}$ &
  $29.5_{-3.3}^{+11.4}$ &
  $17.9_{-0.4}^{+3.6}$ &
  $2.76_{-0.61}^{+0.87}$ &
  - &
  $12.5_{-3.9}^{+2.1}$ &
  $15.8_{-3.5}^{+2.5}$ &
  $5.98_{-0.63}^{+2.93}$ &
  $7.65_{-1.81}^{+1.99}$ &
  $7.70_{-3.49}^{+3.65}$ \\
Fe &
  $11.5_{-0.9}^{+0.3}$ &
  $15.9_{-0.5}^{+0.7}$ &
  $10.6_{-0.2}^{+0.3}$ &
  $0.50_{-0.06}^{+0.14}$ &
  $8.62_{-0.13}^{+0.47}$ &
  $13.5_{-1.2}^{+0.4}$ &
  $11.0_{-0.9}^{+0.5}$ &
  $6.11_{-0.12}^{+0.30}$ &
  $9.62_{-0.36}^{+0.79}$ &
  $17.1_{-0.9}^{+1.5}$ \\
\makecell{$\tau_{\rm h}$\\($\times 10^{9}$ cm$^{-3}$ s )} &
  $4.96_{-0.08}^{+0.25}$ &
  $6.72_{-0.08}^{+0.09}$ &
  $5.23_{-0.09}^{+0.10}$ &
  $7.53_{-1.37}^{+0.52}$ &
  $3.87_{-0.03}^{+0.13}$ &
  $5.84_{-0.07}^{+0.29}$ &
  $4.93_{-0.08}^{+0.08}$ &
  $6.37_{-0.19}^{+0.01}$ &
  $5.01_{-0.21}^{+0.12}$ &
  $5.17_{-0.10}^{+0.13}$ \\
\makecell{Redshift\\($10^{-3}$)} &
  $-1.76_{-0.01}^{+0.03}$ &
  $-2.67_{-0.01}^{+0.01}$ &
  $-2.54_{-0.07}^{+0.05}$ &
  $-0.26_{-0.02}^{+0.05}$ &
  $-2.26_{-0.09}^{+0.05}$ &
  $-2.97_{-0.02}^{+0.38}$ &
  $-2.50_{-0.11}^{+0.02}$ &
  $-0.59_{-0.19}^{+0.04}$ &
  $-1.83_{-0.27}^{+0.06}$ &
  $-2.67_{-0.02}^{+0.01}$ \\
\makecell{Line Centroid\\(keV)} &
  $0.73_{-0.01}^{+0.01}$ &
  $0.74_{-0.01}^{+0.01}$ &
  $0.73_{-0.01}^{+0.01}$ &
  $0.72_{-0.01}^{+0.01}$ &
  $0.73_{-0.01}^{+0.01}$ &
  $0.73_{-0.01}^{+0.01}$ &
  $0.73_{-0.01}^{+0.01}$ &
  $0.73_{-0.01}^{+0.01}$ &
  $0.72_{-0.01}^{+0.01}$ &
  $0.73_{-0.01}^{+0.01}$ \\
\makecell{Line Centroid\\(keV)} &
  $1.23_{-0.01}^{+0.01}$ &
  $1.23_{-0.01}^{+0.01}$ &
  $1.23_{-0.01}^{+0.01}$ &
  $1.22_{-0.01}^{+0.01}$ &
  $1.23_{-0.01}^{+0.01}$ &
  $1.23_{-0.01}^{+0.01}$ &
  $1.23_{-0.01}^{+0.01}$ &
  $1.23_{-0.01}^{+0.01}$ &
  $1.23_{-0.01}^{+0.01}$ &
  $1.23_{-0.01}^{+0.01}$ \\
$\chi_{\nu}^{2}$ (DoF) &
  $1.76$ ($5121$) &
  $1.83$ ($4894$) &
  $1.55$ ($5071$) &
  $1.76$ ($5757$) &
  $1.95$ ($6426$) &
  $1.97$ ($5657$) &
  $1.50$ ($5257$) &
  $1.64$ ($5659$) &
  $1.71$ ($5167$) &
  $2.23$ ($5106$)
\enddata
\end{deluxetable*}

\begin{deluxetable*}{cccccccc}
\centering
\tablecaption{\textbf{G272.2--3.2}: Best fit parameters per region \label{tab:g272_results}}
\tablehead{
\colhead{Parameter} & 
\colhead{R00} & 
\colhead{R01} & 
\colhead{R02} & 
\colhead{R03} &
\colhead{R04} &
\colhead{R05} &
\colhead{R06} 
}
\startdata
\makecell{N$_{\rm H}$\\($10^{22}$ cm$^{-2}$)} &
  $1.37_{-0.04}^{+0.19}$ &
  $1.09_{-0.09}^{+0.10}$ &
  $1.28_{-0.04}^{+0.05}$ &
  $1.26_{-0.19}^{+0.04}$ &
  $1.17_{-0.03}^{+0.05}$ &
  $1.32_{-0.04}^{+0.12}$ &
  $1.21_{-0.05}^{+0.04}$ \\
\makecell{kT\\(keV)} &
  $0.73_{-0.03}^{+0.05}$ &
  $0.70_{-0.12}^{+1.08}$ &
  $0.74_{-0.05}^{+0.07}$ &
  $0.79_{-0.06}^{+1.91}$ &
  $0.85_{-0.07}^{+0.10}$ &
  $0.77_{-0.14}^{+0.09}$ &
  $0.98_{-0.10}^{+0.14}$ \\
Ne &
  $1.08_{-0.21}^{+0.71}$ &
  $1.05_{-0.10}^{+0.13}$ &
  $1.11_{-0.12}^{+0.15}$ &
  $0.99_{-0.41}^{+0.17}$ &
  $1.05_{-0.12}^{+0.59}$ &
  $1.03_{-0.18}^{+0.20}$ &
  $0.95_{-0.15}^{+0.17}$ \\
Mg &
  $1.23_{-0.13}^{+1.46}$ &
  $1.08_{-0.16}^{+0.37}$ &
  $1.05_{-0.12}^{+0.18}$ &
  $1.12_{-0.12}^{+0.21}$ &
  $1.19_{-0.26}^{+0.17}$ &
  $1.03_{-0.22}^{+0.19}$ &
  $1.14_{-0.13}^{+0.19}$ \\
Si &
  $5.12_{-1.41}^{+0.61}$ &
  $1.06_{-0.13}^{+0.29}$ &
  $1.82_{-0.17}^{+0.27}$ &
  $2.79_{-0.25}^{+0.32}$ &
  $2.27_{-1.04}^{+0.30}$ &
  $2.28_{-0.26}^{+0.38}$ &
  $1.29_{-0.15}^{+0.22}$ \\
S &
  $5.58_{-2.28}^{+0.64}$ &
  $0.72_{-0.25}^{+0.74}$ &
  $2.24_{-0.32}^{+0.47}$ &
  $3.21_{-0.59}^{+0.53}$ &
  $2.73_{-2.73}^{+0.59}$ &
  $3.05_{-0.59}^{+1.52}$ &
  $1.31_{-0.29}^{+0.38}$ \\
Fe &
  $2.89_{-0.34}^{+5.06}$ &
  $0.87_{-0.19}^{+0.21}$ &
  $1.17_{-0.18}^{+0.27}$ &
  $1.60_{-0.23}^{+0.30}$ &
  $1.06_{-0.84}^{+0.14}$ &
  $1.16_{-0.32}^{+0.35}$ &
  $1.23_{-0.18}^{+0.26}$ \\
\makecell{$\tau$\\($10^{10}$ cm$^{-3}$ s )} &
  $6.78_{-1.18}^{+27.3}$ &
  $6.51_{-5.16}^{+1.63}$ &
  $6.57_{-1.12}^{+1.23}$ &
  $5.00_{-3.88}^{+1.35}$ &
  $4.65_{-3.61}^{+1.04}$ &
  $4.84_{-1.38}^{+1.64}$ &
  $3.85_{-0.75}^{+0.94}$ \\
\makecell{Line Centroid\\(keV)} &
  $1.24_{-0.11}^{+0.01}$ &
  - &
  - &
  $1.24_{-0.01}^{+0.01}$ &
  $1.22_{-0.09}^{+0.01}$ &
  $1.24_{-0.01}^{+0.01}$ &
  - \\
$\chi_{\nu}^{2}$ (DoF) &
  $1.24$ ($855$) &
  $1.09$ ($640$) &
  $1.19$ ($792$) &
  $1.09$ ($814$) &
  $0.99$ ($899$) &
  $1.03$ ($799$) &
  $1.06$ ($825$)
\enddata
\end{deluxetable*}

\begin{deluxetable*}{cccccccc}
\centering
\tablecaption{\textbf{G337.2--0.7}: Best fit parameters per region \label{tab:g337_results}}
\tablehead{
\colhead{Parameter} & 
\colhead{R00} & 
\colhead{R01} & 
\colhead{R02} & 
\colhead{R03} &
\colhead{R04} &
\colhead{R05} &
\colhead{R06} 
}
\startdata
\makecell{N$_{\rm H}$\\($10^{22}$ cm$^{-2}$)} &
  $3.44_{-0.17}^{+0.20}$ &
  $3.68_{-0.13}^{+0.12}$ &
  $4.06_{-0.22}^{+0.29}$ &
  $4.15_{-0.23}^{+0.37}$ &
  $4.35_{-0.31}^{+0.46}$ &
  $3.95_{-0.26}^{+0.36}$ &
  $4.83_{-0.36}^{+0.87}$ \\
\makecell{kT\\(keV)} &
  $1.38_{-0.20}^{+0.15}$ &
  $1.28_{-0.10}^{+0.12}$ &
  $0.93_{-0.14}^{+0.14}$ &
  $1.16_{-0.22}^{+0.19}$ &
  $1.56_{-0.30}^{+0.34}$ &
  $0.91_{-0.15}^{+0.15}$ &
  $0.99_{-0.27}^{+0.22}$ \\
Mg &
  $0.67_{-0.26}^{+0.31}$ &
  $0.76_{-0.18}^{+0.24}$ &
  $1.28_{-0.29}^{+0.38}$ &
  $1.29_{-0.33}^{+2.64}$ &
  $1.47_{-0.42}^{+0.98}$ &
  $0.96_{-0.22}^{+0.34}$ &
  $1.34_{-0.40}^{+1.35}$ \\
Si &
  $4.48_{-0.62}^{+0.80}$ &
  $3.97_{-0.35}^{+0.50}$ &
  $3.77_{-0.61}^{+0.72}$ &
  $3.93_{-0.82}^{+6.20}$ &
  $5.41_{-0.90}^{+1.53}$ &
  $3.03_{-0.47}^{+0.71}$ &
  $4.36_{-0.71}^{+1.66}$ \\
S &
  $4.36_{-0.55}^{+0.78}$ &
  $3.74_{-0.32}^{+0.46}$ &
  $3.61_{-0.46}^{+0.64}$ &
  $3.86_{-0.56}^{+5.75}$ &
  $4.38_{-0.63}^{+1.09}$ &
  $3.32_{-0.42}^{+0.69}$ &
  $4.69_{-0.69}^{+1.25}$ \\
Ar &
  $4.60_{-1.03}^{+1.34}$ &
  $4.34_{-0.67}^{+0.80}$ &
  $4.48_{-1.12}^{+1.69}$ &
  $5.48_{-1.48}^{+7.08}$ &
  $5.37_{-1.26}^{+2.16}$ &
  $3.01_{-1.44}^{+2.34}$ &
  $5.88_{-2.36}^{+5.99}$ \\
Ca &
  $9.45_{-2.08}^{+2.86}$ &
  $7.36_{-1.33}^{+1.65}$ &
  $5.55_{-4.07}^{+6.57}$ &
  $6.64_{-4.33}^{+10.4}$ &
  $12.4_{-4.0}^{+5.7}$ &
  $13.2_{-7.6}^{+17.4}$ &
  $17.9_{-9.6}^{+42.4}$ \\
\makecell{$\tau$\\($10^{10}$ cm$^{-3}$ s )} &
  $12.7_{-2.4}^{+6.1}$ &
  $14.7_{-2.6}^{+3.5}$ &
  $8.56_{-2.39}^{+4.19}$ &
  $4.42_{-2.26}^{+2.28}$ &
  $4.23_{-1.01}^{+1.54}$ &
  $6.49_{-1.86}^{+3.62}$ &
  $3.89_{-0.94}^{+3.23}$ \\
$\chi_{\nu}^{2}$ (DoF) &
  $1.17$ ($302$) &
  $1.13$ ($558$) &
  $1.01$ ($318$) &
  $1.12$ ($305$) &
  $1.21$ ($324$) &
  $1.03$ ($309$) &
  $1.05$ ($307$)
\enddata
\end{deluxetable*}

\begin{deluxetable*}{ccccccccc}
\centering
\tablecaption{\textbf{G344.7--0.1}: Best fit parameters per region \label{tab:g344_results}}
\tablehead{
\colhead{Parameter} & 
\colhead{R00} & 
\colhead{R01} & 
\colhead{R02} & 
\colhead{R03} &
\colhead{R04} &
\colhead{R05} &
\colhead{R06} &
\colhead{R07}
}
\startdata
\makecell{N$_{\rm H}$\\($10^{22}$ cm$^{-2}$)} &
  $5.70_{-0.38}^{+1.00}$ &
  $5.49_{-0.39}^{+0.63}$ &
  $5.41_{-0.41}^{+0.48}$ &
  $5.97_{-0.32}^{+0.46}$ &
  $4.84_{-0.71}^{+0.59}$ &
  $5.72_{-0.55}^{+0.68}$ &
  $4.90_{-0.43}^{+0.67}$ &
  $5.53_{-1.65}^{+0.51}$ \\
\makecell{kT\\(keV)} &
  $1.23_{-0.18}^{+0.15}$ &
  $1.41_{-0.33}^{+0.27}$ &
  $1.85_{-0.36}^{+0.43}$ &
  $1.16_{-0.17}^{+0.15}$ &
  $2.18_{-0.58}^{+2.38}$ &
  $1.48_{-0.33}^{+0.46}$ &
  $1.53_{-0.40}^{+0.34}$ &
  $1.55_{-0.18}^{+0.25}$ \\
Mg &
  $0.64_{-0.39}^{+1.44}$ &
  - &
  - &
  - &
  - &
  - &
  - &
  - \\
Si &
  $1.87_{-0.30}^{+0.61}$ &
  $2.38_{-0.63}^{+0.69}$ &
  $3.01_{-0.63}^{+0.78}$ &
  $1.96_{-0.35}^{+0.43}$ &
  $3.03_{-0.61}^{+1.25}$ &
  $2.50_{-0.64}^{+0.92}$ &
  $2.53_{-0.69}^{+0.68}$ &
  $3.09_{-0.67}^{+0.95}$ \\
S &
  $1.93_{-0.29}^{+0.44}$ &
  $2.36_{-0.56}^{+0.67}$ &
  $3.67_{-0.62}^{+0.91}$ &
  $2.20_{-0.34}^{+0.44}$ &
  $3.25_{-0.64}^{+1.25}$ &
  $3.15_{-0.69}^{+1.06}$ &
  $2.78_{-0.68}^{+0.76}$ &
  $3.57_{-0.62}^{+2.87}$ \\
Ar &
  $2.25_{-0.78}^{+1.01}$ &
  $1.97_{-0.89}^{+1.15}$ &
  $2.78_{-0.98}^{+1.24}$ &
  $3.32_{-1.09}^{+1.68}$ &
  $2.02_{-1.39}^{+2.00}$ &
  $4.74_{-1.47}^{+2.21}$ &
  $2.81_{-1.11}^{+1.49}$ &
  $3.36_{-1.20}^{+2.27}$ \\
Ca &
  $2.86_{-1.82}^{+2.71}$ &
  $4.60_{-2.46}^{+4.41}$ &
  $5.91_{-2.31}^{+3.28}$ &
  $4.27_{-3.18}^{+6.02}$ &
  - &
  - &
  $2.51_{-1.87}^{+4.28}$ &
  $5.13_{-2.56}^{+3.81}$ \\
\makecell{$\tau$\\($10^{10}$ cm$^{-3}$ s )} &
  $8.61_{-2.55}^{+3.25}$ &
  $6.35_{-1.67}^{+3.57}$ &
  $5.50_{-1.19}^{+1.85}$ &
  $5.26_{-1.36}^{+2.21}$ &
  $2.99_{-0.77}^{+1.20}$ &
  $6.02_{-2.00}^{+3.47}$ &
  $6.35_{-1.84}^{+3.77}$ &
  $6.64_{-4.02}^{+3.36}$ \\
$\chi_{\nu}^{2}$ (DoF) &
  $1.13$ ($219$) &
  $1.35$ ($213$) &
  $1.14$ ($243$) &
  $0.91$ ($227$) &
  $0.95$ ($205$) &
  $0.97$ ($200$) &
  $1.29$ ($221$) &
  $1.27$ ($211$)
\enddata
\end{deluxetable*}

\begin{deluxetable*}{ccccc}
\centering
\tablecaption{\textbf{G352.7--0.1}: Best fit parameters per region \label{tab:g352_results}}
\tablehead{
\colhead{Parameter} & 
\colhead{R00} & 
\colhead{R01} & 
\colhead{R02} & 
\colhead{R03} 
}
\startdata
\makecell{N$_{\rm H}$\\($10^{22}$ cm$^{-2}$)} &
  $4.75_{-0.07}^{+0.77}$ &
  $3.36_{-0.31}^{+0.70}$ &
  $5.08_{-0.61}^{+1.65}$ &
  $4.50_{-0.09}^{+2.42}$ \\
\makecell{kT\\(keV)} &
  $1.76_{-0.36}^{+0.08}$ &
  $2.34_{-0.55}^{+0.54}$ &
  $1.98_{-0.86}^{+0.15}$ &
  $3.89_{-2.24}^{+0.93}$ \\
Mg                     & $3.00_{-1.18}^{+1.18}$ & $0.99_{-0.29}^{+0.60}$ & $4.55_{-1.36}^{+1.53}$ & $1.93_{-1.91}^{+2.59}$ \\
Si                     & $8.68_{-0.88}^{+0.89}$ & $3.60_{-0.56}^{+1.08}$ & $8.67_{-1.15}^{+1.36}$ & $8.80_{-1.68}^{+1.93}$ \\
S                      & $7.68_{-0.76}^{+0.80}$ & $2.76_{-0.45}^{+0.56}$ & $6.18_{-0.85}^{+0.94}$ & $7.58_{-1.38}^{+1.65}$ \\
Ar                     & $6.87_{-2.40}^{+2.52}$ & $2.82_{-1.50}^{+1.75}$ & $6.31_{-3.25}^{+3.58}$ & $3.76_{-3.76}^{+4.45}$ \\
Ca                     & $13.2_{-6.6}^{+7.4}$   & $6.02_{-4.20}^{+5.90}$ & -                      & $12.1_{-10.5}^{+11.8}$ \\
Fe		& $15.5_{-4.4}^{+5.8}$ & $1.40_{-0.75}^{+2.27}$ & $11.2_{-5.4}^{+6.2}$ & $8.08_{-5.28}^{+11.9}$ \\
\makecell{$\tau$\\($10^{10}$ cm$^{-3}$ s )} &
  $3.92_{-3.77}^{+1.68}$ &
  $2.62_{-0.33}^{+0.66}$ &
  $2.59_{-0.55}^{+0.86}$ &
  $2.52_{-0.21}^{+3.08}$ \\
$\chi_{\nu}^{2}$ (DoF) & $1.10$ ($251$)         & $1.43$ ($242$)         & $1.05$ ($246$)         & $0.98$ ($221$)       
\enddata
\end{deluxetable*}

\begin{deluxetable*}{cccccccc}
\centering
\tablecaption{\textbf{0505--67.9}: Best fit parameters per region \label{tab:0505_results}}
\tablehead{
\colhead{Parameter} & 
\colhead{R00} & 
\colhead{R01} & 
\colhead{R02} & 
\colhead{R03} &
\colhead{R04} & 
\colhead{R05} & 
\colhead{R06} 
}
\startdata
\makecell{kT$_{\rm c}$\\(keV)} &
  $0.21_{-0.01}^{+0.01}$ &
  $0.22_{-0.01}^{+0.01}$ &
  $0.21_{-0.01}^{+0.01}$ &
  $0.20_{-0.01}^{+0.01}$ &
  $0.20_{-0.01}^{+0.01}$ &
  $0.20_{-0.01}^{+0.01}$ &
  $0.21_{-0.01}^{+0.01}$ \\
\makecell{$\tau_{\rm c}$\\($10^{13}$ cm$^{-3}$ s )} &
  \multicolumn{1}{l}{$0.62_{-0.18}^{+1.17}$} &
  \multicolumn{1}{l}{$0.26_{-0.02}^{+0.51}$} &
  \multicolumn{1}{l}{$4.79_{-0.62}^{+0.19}$} &
  \multicolumn{1}{l}{$4.91_{-1.32}^{+0.08}$} &
  \multicolumn{1}{l}{$4.95_{-0.63}^{+0.04}$} &
  \multicolumn{1}{l}{$4.91_{-1.72}^{+0.07}$} &
  \multicolumn{1}{l}{$4.62_{-0.89}^{+0.36}$} \\
\makecell{kT$_{\rm h}$\\(keV)} &
  $0.81_{-0.02}^{+0.02}$ &
  \multicolumn{1}{l}{$0.80_{-0.02}^{+0.01}$} &
  \multicolumn{1}{l}{$0.79_{-0.01}^{+0.03}$} &
  \multicolumn{1}{l}{$0.78_{-0.01}^{+0.05}$} &
  \multicolumn{1}{l}{$0.80_{-0.02}^{+0.01}$} &
  \multicolumn{1}{l}{$0.82_{-0.01}^{+0.01}$} &
  \multicolumn{1}{l}{$0.84_{-0.02}^{+0.02}$} \\
O &
  $0.50_{-0.13}^{+0.29}$ &
  - &
  $0.31_{-0.08}^{+0.09}$ &
  $0.61_{-0.04}^{+0.06}$ &
  $0.20_{-0.11}^{+0.14}$ &
  $0.56_{-0.03}^{+0.16}$ &
  $0.39_{-0.04}^{+0.18}$ \\
Ne &
  $0.84_{-0.07}^{+0.38}$ &
  $1.27_{-0.14}^{+0.43}$ &
  $0.95_{-0.19}^{+0.10}$ &
  $0.99_{-0.05}^{+0.02}$ &
  $0.92_{-0.07}^{+0.11}$ &
  $1.18_{-0.04}^{+0.18}$ &
  $1.01_{-0.06}^{+0.28}$ \\
Mg &
  $1.08_{-0.04}^{+0.33}$ &
  $1.42_{-0.09}^{+0.32}$ &
  $0.91_{-0.10}^{+0.13}$ &
  $0.96_{-0.06}^{+0.11}$ &
  $0.97_{-0.02}^{+0.13}$ &
  $1.26_{-0.07}^{+0.22}$ &
  $1.12_{-0.09}^{+0.13}$ \\
Si &
  $1.30_{-0.03}^{+0.37}$ &
  $1.45_{-0.12}^{+0.29}$ &
  $1.10_{-0.07}^{+0.05}$ &
  $1.01_{-0.06}^{+0.10}$ &
  $1.27_{-0.06}^{+0.09}$ &
  $1.35_{-0.06}^{+0.28}$ &
  $1.18_{-0.05}^{+0.13}$ \\
S &
  $1.53_{-0.08}^{+0.59}$ &
  $1.31_{-0.16}^{+0.34}$ &
  $1.34_{-0.12}^{+0.24}$ &
  $1.12_{-0.21}^{+0.19}$ &
  $1.21_{-0.10}^{+0.12}$ &
  $1.23_{-0.09}^{+0.19}$ &
  $0.96_{-0.16}^{+0.12}$ \\
Fe &
  $2.26_{-0.03}^{+0.46}$ &
  $1.77_{-0.08}^{+0.22}$ &
  $1.37_{-0.05}^{+0.14}$ &
  $0.91_{-0.04}^{+0.16}$ &
  $1.29_{-0.04}^{+0.13}$ &
  $1.48_{-0.06}^{+0.22}$ &
  $1.30_{-0.08}^{+0.12}$ \\
\makecell{$\tau_{\rm h}$\\($10^{11}$ s cm$^{-3}$)} &
  $3.70_{-0.42}^{+1.58}$ &
  $6.17_{-0.60}^{+2.18}$ &
  $4.28_{-1.05}^{+1.11}$ &
  $2.61_{-0.32}^{+0.10}$ &
  $3.12_{-0.36}^{+1.02}$ &
  $3.98_{-0.50}^{+0.62}$ &
  $4.11_{-0.44}^{+1.59}$ \\
\makecell{Line Centroid\\(keV)} &
  $1.24_{-0.01}^{+0.01}$ &
  $1.25_{-0.01}^{+0.01}$ &
  $1.24_{-0.01}^{+0.1}$ &
  $1.24_{-0.01}^{+0.01}$ &
  $1.25_{-0.01}^{+0.01}$ &
  $1.23_{-0.01}^{+0.01}$ &
  $1.25_{-0.01}^{+0.02}$ \\
$\chi_{\nu}^{2}$ (DoF) &
  $1.34$ ($782$) &
  $1.52$ ($813$) &
  $1.22$ ($801$) &
  $1.41$ ($803$) &
  $1.42$ ($776$) &
  $1.33$ ($839$) &
  $1.39$ ($827$)
\enddata
\end{deluxetable*}

\begin{deluxetable*}{ccccc}
\centering
\tablecaption{\textbf{0509--67.5}: Best fit parameters per region \label{tab:0509_results}}
\tablehead{
\colhead{Parameter} & 
\colhead{R00} & 
\colhead{R01} & 
\colhead{R02} & 
\colhead{R03} 
}
\startdata
\makecell{kT$_{\rm c}$\\(keV)} &
  $0.41_{-0.02}^{+0.02}$ &
  \multicolumn{1}{l}{$0.38_{-0.02}^{+0.02}$} &
  \multicolumn{1}{l}{$0.37_{-0.02}^{+0.02}$} &
  \multicolumn{1}{l}{$0.40_{-0.02}^{+0.02}$} \\
Ne                     & $1.72_{-0.13}^{+0.18}$ & $1.64_{-0.20}^{+0.17}$ & $1.38_{-0.13}^{+0.14}$ & $1.33_{-0.10}^{+0.15}$ \\
Mg                     & $0.82_{-0.49}^{+0.39}$ & $1.05_{-0.40}^{+0.38}$ & $0.76_{-0.37}^{+0.36}$ & $0.92_{-0.43}^{+0.32}$ \\
Si                     & $44.0_{-5.3}^{+7.7}$   & $52.4_{-7.0}^{+9.2}$   & $52.2_{-6.7}^{+9.3}$   & $51.0_{-6.1}^{+9.3}$   \\
S                      & $66.4_{-8.1}^{+10.4}$  & $74.9_{-8.7}^{+10.4}$  & $77.9_{-9.2}^{+10.7}$  & $70.5_{-8.9}^{+11.1}$  \\
Fe                     & $2.94_{-0.39}^{+0.42}$ & $2.67_{-0.43}^{+0.45}$ & $2.69_{-0.38}^{+0.45}$ & $2.58_{-0.37}^{+0.39}$ \\
\makecell{$\tau_{\rm c}$\\($10^{11}$ cm$^{-3}$ s )} &
  $3.81_{-0.91}^{+0.15}$ &
  $5.31_{-1.42}^{+2.75}$ &
  $4.91_{-1.44}^{+2.06}$ &
  $3.86_{-0.99}^{+1.94}$ \\
\makecell{kT$_{\rm h}$\\(keV)} &
  \multicolumn{1}{l}{$3.41_{-0.68}^{+2.01}$} &
  \multicolumn{1}{l}{$4.24_{-1.26}^{+4.63}$} &
  \multicolumn{1}{l}{$4.86_{-1.47}^{+4.53}$} &
  \multicolumn{1}{l}{$3.63_{-0.65}^{+1.76}$} \\
\makecell{Line Centroid\\(keV)} &
  $1.21_{-0.02}^{+0.02}$ &
  $1.19_{-0.04}^{+0.03}$ &
  $1.16_{-0.06}^{+0.04}$ &
  $1.21_{-0.02}^{+0.02}$ \\
$\chi_{\nu}^{2}$ (DoF) & $1.10$ ($496$)         & $1.27$ ($498$)         & $1.42$ ($501$)         & $1.27$ ($513$)
\enddata
\end{deluxetable*}

\begin{deluxetable*}{ccccc}
\centering
\tablecaption{\textbf{0509--68.7}: Best fit parameters per region \label{tab:n103b_results}}
\tablehead{
\colhead{Parameter} & 
\colhead{R00} & 
\colhead{R01} & 
\colhead{R02} & 
\colhead{R03} 
}
\startdata
\makecell{kT$_{\rm c}$\\(keV)} & $0.74_{-0.02}^{+0.01}$ & $0.74_{-0.01}^{+0.02}$ & $0.75_{-0.02}^{+0.02}$ & $0.76_{-0.01}^{+0.02}$ \\
\makecell{kT$_{\rm h}$\\(keV)} & $3.03_{-0.42}^{+0.68}$ & $3.08_{-0.50}^{+0.97}$ & $3.44_{-0.60}^{+0.47}$ & $2.89_{-0.47}^{+0.32}$ \\
O                                                        & $1.34_{-0.21}^{+0.52}$ & $1.39_{-0.19}^{+0.63}$ & $1.27_{-0.23}^{+0.56}$ & -                      \\
Ne                                                       & $3.11_{-0.88}^{+1.85}$ & $2.76_{-0.72}^{+2.42}$ & $1.84_{-0.96}^{+1.97}$ & $2.05_{-0.92}^{+1.09}$ \\
Mg                                                       & $0.93_{-0.63}^{+0.86}$ & $1.70_{-0.42}^{+1.05}$ & $2.31_{-0.40}^{+1.04}$ & $0.85_{-0.65}^{+0.72}$ \\
Si                                                       & $7.79_{-1.20}^{+2.90}$ & $8.34_{-1.41}^{+3.96}$ & $11.3_{-1.7}^{+3.3}$   & $8.06_{-0.91}^{+1.38}$ \\
S                                                        & $8.20_{-1.38}^{+2.47}$ & $9.33_{-1.62}^{+3.98}$ & $11.5_{-1.6}^{+3.7}$   & $8.63_{-1.15}^{+1.53}$ \\
Ar                                                       & $10.5_{-2.8}^{+4.8}$   & $12.1_{-2.9}^{+5.4}$   & $12.7_{-1.4}^{+7.2}$   & $8.75_{-2.67}^{+4.39}$ \\
Ca                                                       & $7.32_{-4.35}^{+6.24}$ & $10.8_{-4.0}^{+14.1}$  & $18.9_{-5.3}^{+9.0}$   & $7.28_{-4.80}^{+7.80}$ \\
Fe                                                       & $2.42_{-0.58}^{+1.15}$ & $3.23_{-0.83}^{+1.67}$ & $5.12_{-0.91}^{+1.61}$ & $3.24_{-0.58}^{+0.66}$ \\
\makecell{$\tau_{\rm h}$\\($\times 10^{11}$ cm$^{-3}$ s )} &
  $1.20_{-0.19}^{+0.30}$ &
  $1.09_{-0.17}^{+0.29}$ &
  $1.10_{-0.15}^{+0.26}$ &
  $1.34_{-0.21}^{+0.31}$ \\
\makecell{Line Centroid\\(keV)} &
  $1.29_{-0.03}^{+0.03}$ &
  $1.25_{-0.01}^{+0.01}$ &
  $1.24_{-0.01}^{+0.01}$ &
  $1.27_{-0.03}^{+0.02}$ \\
$\chi_{\nu}^{2}$ (DoF)                                   & 1.20 (1080)            & 1.07 (1060)            & 1.11 (1071)            & 1.09 (1070)          
\enddata
\end{deluxetable*}

\begin{deluxetable*}{cccccc}
\centering
\tablecaption{\textbf{0519--69.0}: Best fit parameters per region \label{tab:0519_results}}
\tablehead{
\colhead{Parameter} & 
\colhead{R00} & 
\colhead{R01} & 
\colhead{R02} & 
\colhead{R03} &
\colhead{R04} 
}
\startdata
\makecell{kT$_{\rm c}$\\(keV)} &
  $0.40_{-0.05}^{+0.03}$ &
  $0.41_{-0.04}^{+0.05}$ &
  $0.38_{-0.03}^{+0.02}$ &
  $0.36_{-0.10}^{+0.01}$ &
  $0.66_{-0.03}^{+0.04}$ \\
\makecell{kT$_{\rm h}$\\(keV)} &
  \multicolumn{1}{l}{$0.84_{-0.01}^{+0.01}$} &
  \multicolumn{1}{l}{$0.83_{-0.01}^{+0.02}$} &
  \multicolumn{1}{l}{$0.83_{-0.01}^{+0.01}$} &
  \multicolumn{1}{l}{$0.88_{-0.07}^{+0.01}$} &
  \multicolumn{1}{l}{$2.19_{-0.31}^{+0.37}$} \\
Ne &
  - &
  $1.07_{-0.74}^{+0.60}$ &
  - &
  $0.61_{-0.01}^{+4.23}$ &
  $0.70_{-0.31}^{+0.39}$ \\
Mg &
  $0.46_{-0.13}^{+0.27}$ &
  $0.29_{-0.20}^{+0.19}$ &
  $0.71_{-0.18}^{+0.26}$ &
  $0.59_{-0.46}^{+0.71}$ &
  $0.77_{-0.76}^{+0.18}$ \\
Si &
  $2.64_{-0.26}^{+0.56}$ &
  $2.56_{-0.32}^{+0.37}$ &
  $3.40_{-0.28}^{+0.52}$ &
  $3.52_{-0.02}^{+2.88}$ &
  $5.65_{-0.60}^{+0.66}$ \\
S &
  $4.24_{-0.51}^{+0.78}$ &
  $3.81_{-0.57}^{+0.65}$ &
  $4.72_{-0.78}^{+0.76}$ &
  $5.33_{-0.15}^{+4.15}$ &
  $9.16_{-1.07}^{+1.64}$ \\
Fe &
  $1.64_{-0.14}^{+0.24}$ &
  $1.47_{-0.14}^{+0.22}$ &
  $1.80_{-0.14}^{+0.19}$ &
  $1.92_{-0.18}^{+0.86}$ &
  $5.85_{-0.82}^{+0.53}$ \\
\makecell{$\tau_{\rm h}$\\($10^{12}$ cm$^{-3}$ s )} &
  $3.77_{-1.07}^{+7.47}$ &
  $3.84_{-1.64}^{+4.10}$ &
  $4.80_{-1.51}^{+11.3}$ &
  $1.66_{-0.37}^{+1.79}$ &
  $0.06_{-0.01}^{+0.01}$ \\
\makecell{Line Centroid\\(keV)} &
  $1.25_{-0.01}^{+0.01}$ &
  $1.23_{-0.01}^{+0.01}$ &
  $1.24_{-0.01}^{+0.01}$ &
  $1.22_{-0.17}^{+0.02}$ &
  $1.24_{-0.01}^{+0.06}$ \\
$\chi_{\nu}^{2}$ (DoF) &
  $1.40$ ($775$) &
  $1.26$ ($766$) &
  $1.25$ ($762$) &
  $1.21$ ($769$) &
  $1.28$ ($769$)
\enddata
\end{deluxetable*}

\begin{deluxetable*}{ccccccc}
\centering
\tablecaption{\textbf{0534--69.9}: Best fit parameters per region \label{tab:0534_results}}
\tablehead{
\colhead{Parameter} & 
\colhead{R00} & 
\colhead{R01} & 
\colhead{R02} & 
\colhead{R03} &
\colhead{R04} &
\colhead{R05} 
}
\startdata
\makecell{kT\\(keV)} &
  \multicolumn{1}{l}{$0.54_{-0.08}^{+0.12}$} &
  \multicolumn{1}{l}{$0.69_{-0.03}^{+0.11}$} &
  \multicolumn{1}{l}{$0.68_{-0.10}^{+0.05}$} &
  \multicolumn{1}{l}{$0.72_{-0.06}^{+0.13}$} &
  \multicolumn{1}{l}{$0.88_{-0.15}^{+0.18}$} &
  $0.66_{-0.07}^{+0.09}$ \\
O &
  $0.38_{-0.06}^{+0.02}$ &
  $0.47_{-0.05}^{+0.15}$ &
  $0.56_{-0.12}^{+0.11}$ &
  $0.62_{-0.08}^{+0.29}$ &
  $0.66_{-0.15}^{+0.24}$ &
  $0.90_{-0.17}^{+0.31}$ \\
Ne &
  $0.58_{-0.08}^{+0.82}$ &
  $1.08_{-0.15}^{+0.27}$ &
  $1.03_{-0.19}^{+0.21}$ &
  $1.21_{-0.13}^{+0.53}$ &
  $1.03_{-0.21}^{+0.30}$ &
  $1.30_{-0.23}^{+0.43}$ \\
Mg &
  $0.54_{-0.12}^{+0.61}$ &
  $0.66_{-0.09}^{+0.09}$ &
  $0.66_{-0.63}^{+0.10}$ &
  $0.83_{-0.17}^{+0.44}$ &
  $0.64_{-0.24}^{+0.35}$ &
  $0.83_{-0.30}^{+0.46}$ \\
Si &
  $0.32_{-0.19}^{+0.29}$ &
  $0.90_{-0.27}^{+0.55}$ &
  $0.78_{-0.30}^{+0.45}$ &
  $1.27_{-0.28}^{+0.79}$ &
  $0.98_{-0.37}^{+0.57}$ &
  $1.54_{-0.58}^{+0.89}$ \\
Fe &
  $0.54_{-0.13}^{+0.40}$ &
  $1.42_{-0.17}^{+0.57}$ &
  $1.30_{-0.36}^{+0.31}$ &
  $1.77_{-0.24}^{+0.96}$ &
  $1.73_{-0.45}^{+0.69}$ &
  $1.54_{-0.32}^{+0.63}$ \\
\makecell{$\tau$\\($10^{10}$ cm$^{-3}$ s )} &
  $10.9_{-10.9}^{+9.4}$ &
  $6.95_{-2.25}^{+1.29}$ &
  $4.20_{-7.64}^{+2.50}$ &
  $4.05_{-1.23}^{+0.88}$ &
  $2.90_{-0.76}^{+1.26}$ &
  $3.56_{-0.81}^{+1.03}$ \\
\makecell{Line Centroid\\(keV)} &
  - &
  $1.25_{-0.01}^{+0.01}$ &
  $1.25_{-0.01}^{+0.08}$ &
  $1.26_{-0.02}^{+0.02}$ &
  $1.26_{-0.02}^{+0.02}$ &
  $1.27_{-0.02}^{+0.03}$ \\
$\chi_{\nu}^{2}$ (DoF) &
  $1.15$ ($349$) &
  $1.19$ ($429$) &
  $1.05$ ($408$) &
  $1.40$ ($415$) &
  $1.31$ ($404$) &
  $1.25$ ($395$)
\enddata
\end{deluxetable*}

\begin{deluxetable*}{ccccc}
\centering
\tablecaption{\textbf{0548--70.4}: Best fit parameters per region \label{tab:0548_results}}
\tablehead{
\colhead{Parameter} & 
\colhead{R00} & 
\colhead{R01} & 
\colhead{R02} & 
\colhead{R03} 
}
\startdata
\makecell{kT$_{\rm c}$\\(keV)} & $0.29_{-0.03}^{+0.13}$ & $0.25_{-0.04}^{+0.01}$ & $0.26_{-0.04}^{+0.04}$ & $0.31_{-0.22}^{+0.11}$ \\
\makecell{kT$_{\rm h}$\\(keV)} & $0.69_{-0.02}^{+0.09}$ & $0.69_{-0.04}^{+0.01}$ & $0.68_{-0.05}^{+0.03}$ & $0.68_{-0.29}^{+0.13}$ \\
O                                                        & -                      & -                      & -                      & $0.61_{-0.07}^{+0.17}$ \\
Ne                                                       & $1.16_{-0.24}^{+0.84}$ & $1.01_{-0.26}^{+2.54}$ & $1.05_{-0.21}^{+0.55}$ & $0.86_{-0.12}^{+0.08}$ \\
Mg                                                       & $1.13_{-0.56}^{+0.58}$ & $0.87_{-0.40}^{+1.37}$ & $0.43_{-0.36}^{+0.31}$ & $0.66_{-0.09}^{+0.05}$ \\
Si                                                       & $1.94_{-0.02}^{+2.89}$ & $2.41_{-0.20}^{+2.53}$ & $2.76_{-0.55}^{+1.04}$ & $2.41_{-0.55}^{+0.29}$ \\
S                                                        & $1.22_{-0.59}^{+2.93}$ & $2.49_{-1.05}^{+3.13}$ & $2.87_{-1.09}^{+2.12}$ & $2.17_{-1.43}^{+0.88}$ \\
Fe                                                       & $1.40_{-0.02}^{+1.36}$ & $1.44_{-0.09}^{+1.29}$ & $1.35_{-0.18}^{+0.30}$ & $0.88_{-0.30}^{+0.17}$ \\
\makecell{$\tau_{\rm h}$\\($10^{11}$ cm$^{-3}$ s )} &
  $3.54_{-1.58}^{+3.09}$ &
  $6.18_{-2.48}^{+40.3}$ &
  $5.08_{-2.08}^{+3.15}$ &
  $1.38_{-0.37}^{+1.72}$ \\
\makecell{Line Centroid\\(keV)} &
  $1.21_{-0.07}^{+0.07}$ &
  $1.25_{-0.05}^{+0.03}$ &
  $1.26_{-0.03}^{+0.03}$ &
  $1.24_{-0.01}^{+0.28}$ \\
$\chi_{\nu}^{2}$ (DoF)                                   & $1.06$ ($363$)         & $0.90$ ($368$)         & $1.02$ ($373$)         & $1.01$ ($350$)
\enddata
\end{deluxetable*}

\clearpage
\appendix
\section{Nucleosynthesis Plots}
In the interest of space, we have opted not to include the entirety of our resultant data in the body of this work. Instead, we have opted to include them electronically, via the Center for Open Science's Open Science Framework (OSF). Below one will find a link to the complete collection of plots comparing the entirety of our suite of nucleosynthesis models to our observational results for each of the objects included in this study.

\href{https://doi.org/10.17605/OSF.IO/MZNCE}{doi:10.17605/OSF.IO/MZNCE}

\bibliographystyle{aasjournalv7}
\bibliography{bibliography}{}

\end{document}